\documentclass[final,5p,times]{elsarticle}
\usepackage{hyperref}
\usepackage{amsmath}
\usepackage{amssymb}
\usepackage{tabularx}
\usepackage{bm}
\usepackage[T1]{fontenc}
\usepackage{refcount}
\usepackage{enumitem}
\usepackage{upgreek}
\usepackage{csquotes}
\usepackage{mathrsfs}
\usepackage{soul}
\usepackage{comment}
\usepackage[usenames,dvipsnames,table]{xcolor}
\definecolor{jgreen}{HTML}{90bbbb}
\definecolor{myblue}{HTML}{7db3e8}
\definecolor{mybblue}{HTML}{93d6ec}
\definecolor{mydblue}{HTML}{6787e4}
\definecolor{mypurple}{HTML}{baa8f0}
\definecolor{mygreen}{HTML}{8cd9ad}
\definecolor{hyellow}{HTML}{ffe44d}
\definecolor{lred}{HTML}{8af48a}
\definecolor{Hred}{HTML}{f48a8a}
\definecolor{pastelpink}{HTML}{F8C8DC}
\definecolor{myorange}{HTML}{F7B787}
\definecolor{mypinkt}{HTML}{ec93bf}
\definecolor{raspberry}{HTML}{ee5db9}
\definecolor{retroorange}{HTML}{ffbf00}
\definecolor{retroyellow}{HTML}{ffd700}

\newcommand{\iu}{{i\mkern1mu}}

\begin{document}
\begin{frontmatter}
\ead{lidiajgomesdasilva.io, lidiajoana@pm.me}
\title{ \texttt{DiscoTEX 1.0}: \underline{Dis}continuous \underline{co}llocation and implicit-\underline{t}urned-\underline{ex}plicit (\texttt{IM\underline{TEX}}) integration \underline{symplectic}, symmetric numerical algorithms with higher order jumps for differential equations I: Numerical black hole perturbation theory applications}

\author{Lidia J. Gomes Da Silva}

\affiliation{organization={School of Mathematical Sciences, Queen Mary University of London},
            city={London},
            postcode={E1 4NS}, 
            country={UK}}
\begin{abstract}
Dirac $\delta-$distributionally sourced differential equations emerge in many dynamical physical systems from machine learning, finance, neuroscience, and seismology to black hole perturbation theory. Most of these systems lack exact analytical solutions and are thus best tackled numerically. In this work, we describe a generic numerical algorithm which constructs discontinuous spatial and temporal discretisations by operating on discontinuous Lagrange and Hermite interpolation formulae, respectively, recovering higher-order accuracy. By solving the distributionally sourced wave equation, which possesses analytical solutions, we demonstrate that numerical weak-form solutions can be recovered to high-order accuracy by solving a first-order reduced system of ordinary differential equations. The method-of-lines framework is applied to the \texttt{DiscoTEX} algorithm i.e through \underline{dis}continuous \underline{co}llocation with implicit\underline{-turned-explicit} (\texttt{IM\underline{TEX}}) integration methods which are symmetric and conserve symplectic structure. Furthermore, the main application of the algorithm is proved, for the first time, by calculating the amplitude at any desired location within the numerical grid, including at the position (and at its right and left limit) where the wave- (or wave-like) equation is discontinuous via interpolation using the \texttt{DiscoTEX} algorithm. This is demonstrated, firstly by solving the wave- (or wave-like) equation and comparing the numerical weak-form solution to the exact solution. We further demonstrate how to reconstruct the scalar and gravitational metric perturbations from weak-form numerical solutions of a non-rotating black hole, which do not have known exact analytical solutions, and compare them against state-of-the-art frequency domain results. We conclude by motivating how \texttt{DiscoTEX}, and related numerical algorithms, not only open a promising new alternative waveform generation route for the modelling of E(X)tremely and (E)xtreme-Mass-Ratio-Inspirals ((X)/(E)MRI)s via a self-consistent evolution in the time-domain, but also provides a highly-accurate necessary complementary framework to the current Fourier domain approach relying on a two-timescale expansion. 
\end{abstract}

\end{frontmatter}

\section{Introduction}\label{Sec1_intro}
Dirac $\delta-$distributionally-sourced wave and (wave-like) equations are present in many dynamical systems from machine learning \cite{huang2021solving}, finance \cite{lipton2021semi}, neuroscience \cite{lewis2003dynamics,caceres2011analysis,carrillo2015qualitative,caceres2017towards}, seismology \cite{johnson1974green, komatitsch1999introduction, dumbser2018efficient}, probability \cite{dembo2019criticality, cuchiero2023propagation} to gravitational physics \cite{barack2009gravitational, poisson2011motion, wardell2015self, barack2018self, pound2022black}.
In this work a new algorithm will be presented with application to gravitational physics and designed to optimally address the difficulties faced by previous numerical methods applied within the radiation-reaction community when modelling E(X)tremely and (E)xtreme-Mass-Ratio-Inspirals ((X)/(E)MRIs). These systems governing field equations describe the binary motion of a supermassive black hole (BH), of mass $M$, orbited by a much smaller compact object such as a black hole, of mass $\mu$. For XMRIs the compact object could further be a neutron-star, brown dwarf \cite{amaro2019extremely, vazquez2022revised, vazquez2023sgr} or primordial black hole \cite{barsanti2022detecting, huang2024primordial}. The significant mass discrepancy naturally leads to a perturbative treatment, through the black hole perturbation theory (BHPT) machinery, where the binary motion is systematically expanded in terms of the ratio between the two bodies' masses, $\epsilon = \mu/M$. The smaller BH is approximated as a point-particle where at zeroth-order in $\mu$ it follows the geodesic of the massive BH background, at first-order it deviates from the geodesic due to the interaction with its self-field. This deviation from its self-field is the self-force. The metric is perturbed as, 
\begin{equation}
    g^{\textrm{Exact}}_{\alpha\beta} = \bar{g}_{\alpha\beta} + \epsilon h^{(1)}_{\alpha\beta} +  \epsilon^{2} h^{(2)}_{\alpha\beta} + \mathcal{O}(\epsilon^{3}),
    \label{sec1_pert_metric}
\end{equation}
where the bar notation denotes the metric given with respect to the background BH. Extensive theoretical work \cite{hinderer2008two} has shown that expansions are only necessary up to second-order for accurate EMRI waveform models \cite{pound2022black, afshordi2023waveform}. The starting point in numerical BHPT is expanding the linearized Einstein equations,
\begin{equation}
    G_{\alpha\beta} = \frac{8 \pi G}{c^{4}}T_{\alpha\beta}, 
    \label{sec1_linEFEs}
\end{equation}
to the required orders, in this work, the focus is on expansions up to first-order given as, 
\begin{align}
      &G_{\alpha\beta} = \bar{G}_{\alpha\beta} + \delta G_{\alpha\beta}/2, \\
      &\delta G_{\alpha \beta} = \frac{16\pi G}{c^{4}} T_{\alpha \beta}. 
    \label{sec1_lineari_EinsteinEqu_pert}
\end{align}
To note here, going from the first to the second line, $\bar{G}_{\alpha\beta} = 0$, because the background BHs under study are solutions to the Einstein vacuum equations, specifically here we assume a Schwarzschild BH background as a reasonable first step.\footnote{Additionally $c$ is the speed of light and $G$ is the Gravitational constant which reduces to a unit due to the convention adopted, i.e. $G=c=1$.} The tensor, $T_{\alpha \beta}$, is the external matter perturbing the background BH written in terms of Dirac $\delta-$distributions reflecting the point-particle model of the compact object in BHPT, 
\begin{align}
         &T_{\alpha\beta} = \mu \int^{\infty}_{-\infty} \frac{u_{\alpha}u_{\beta} }{\sqrt{-g}}\delta^{4}(x^{\alpha} - z^{\alpha}(s)) \ ds  \nonumber \\ 
         &= \mu \frac{u_{\alpha}u_{\beta}}{u^{t}  r_{p}(t)^{2} \sin{\theta} } \delta(r - r_{p}(t))\delta(\theta - \theta_{p}(t))\delta(\phi - \phi_{p}(t)).
    \label{cha2_sem_pointparticle}
\end{align}
where $s$ is the particle's proper time, the point-particle moves on a geodesic with worldline $z^{\alpha}(s)$, the four-velocity is defined by $u^{\alpha} = dz^{\alpha}/ds$ and the $-p$ subscript denotes it is with respect to the point-particle. The natural next step for the numerical BH perturbation theorist is to take advantage of the spherical symmetry of the Schwarzschild background re-writing $h_{\alpha\beta}$ as a composition of radial and angular dependencies decomposed into a basis of spherical harmonics. Generically we have
\begin{equation}
    h_{\alpha \beta}(t,r,\theta,\phi) = \sum^{\infty}_{a}\sum_{lm}h^{a,lm}(t,r)(\textbf{t}^{a,lm})_{\alpha\beta}(\theta,\phi)
    \label{cha2_metricperturbation_ansataz}
\end{equation}
where $a = \{L0,T0,E1,E2,B1,B2,tt,Rt,Et,Bt\}$ denote the different spherical harmonic polarisation modes, for example $L0$ denotes the longitudinal scalar mode and $B2$ denotes the spin-2 mode with magnetic parity, see refs. \cite{thorne1980multipole,regge1957stability,zerilli1970gravitational, thompson2017, thompson2019gravitational}. The perturbations equations in equation \eqref{sec1_lineari_EinsteinEqu_pert} can then be obtained by varying the Einstein tensor, which in the Regge-Wheeler \cite{regge1957stability, zerilli1970gravitational, hopper2010gravitational, thompson2017gauge, thompson2019gravitational} (or radiation gauges \cite{teukolsky1972rotating, teukolsky1973perturbations, press1973perturbations, teukolsky1974perturbations}\footnote{If we started with a rotating Kerr background BH as it will be the case for modelling realistic X/EMRIs, see Section \ref{Sec4}.}) simplifies to a partial-differential-equation (PDE) of the type
\begin{align}
   &\bigg[ -\frac{\partial^{2}}{\partial t^{2}} + \frac{\partial^{2}}{\partial x^{2}} - V_{l}(r)\bigg] \Psi(t,r) = \square \Psi(t,r)  = S_{lm}(t,r)  \nonumber \\
   &= F(t,r) \delta(r-r_{p}) + G(t,r) \delta'(r-r_{p}),
   \label{ch1_bhpt_wave}
\end{align}
where $x$ is the tortoise coordinate, $x = r + 2M \log(r/2M -1)$, $V_{l}(r)$ is some potential remaining from the harmonic decomposition of the background Schwarzschild's or Kerr's backgrounds and $S_{lm}(t,r)$ describes the source term emerging from the scalar or gravitational perturbations. The numerical recipe for the necessary gravitational self-force (GSF) computations can then be summarised in three main phases: Phase 1 - requires one to accurately solve for the master functions $\Psi(t,r)$ given by the PDE \eqref{ch1_bhpt_wave}; Phase 2 - uses the numerical solution to reconstruct the metric perturbations as described by equation \eqref{sec1_pert_metric} and finally Phase 3 - uses the amplitudes of the metric perturbations to compute the gravitational self-force. As it stands it is highly unlikely that equations such as equation \eqref{ch1_bhpt_wave} possess exact known analytical solutions \cite{pound2022black}, one must therefore resort to numerical approaches. Given the distributional nature of the governing wave-like equation, a jump discontinuity arises at the point-particle location, $r=r_{p}(t)$, the strategy here is to let eq.~\eqref{ch1_bhpt_wave} admit weak-form solutions to the inhomogeneous master functions $\Psi(t,r)$ \cite{field2009discontinuous, hopper2010gravitational, thompson2017}:
\begin{equation}
    \Psi(t,r)=  \Psi^{+}(t,r) \Theta[r-r_{p}(t)] + \Psi^{-}(t,r)\Theta[r_{p}(t)-r],
    \label{ch1_weakform_master}
\end{equation}
where here the $\Psi^{+/-}$ are extensions of the smooth, homogeneous solutions of $\Psi(t,r)$ in equation \eqref{ch1_bhpt_wave} as they approach the particle position from the right and left, respectively, and $\Theta[r-r_{p}(t)]$ is the Heaviside step-function. As it has been shown by \cite{hopper2010gravitational, thompson2017} the metric perturbation in equation \eqref{sec1_pert_metric} can be rewritten in functional form as
\begin{align}
    &h(t,r) = h^{+}(t,r)\Theta[r-r_{p}(t)] +  h^{-}(t,r)\Theta[r_{p}(t)-r] \nonumber \\
    &+ h^{S}(t) \delta(r-r_{p}(t)), 
    \label{ch1_metric_weakform}
\end{align}
where $h^{+/-}(t,r)$ represents a smooth function in the region around the particle $(r>r_{p}/r<r_{p})$ and $h^{S}(t)$ is a smooth function only time-dependent as demonstrated by \cite{hopper2010gravitational} which gives the magnitude of the singularity. Schematically, Phase 3, \underline{numerically} simply amounts to using equation \eqref{ch1_metric_weakform} to compute the gravitational self-force as
\begin{align}
    &\mathcal{F}^{\alpha,lm}_{\rm{ret}}  = \lim_{r \rightarrow r_{p}^{\pm}} \sum^{m}_{l=-m} Y_{lm}(\theta, \phi) \int^{2\pi}_{0}\int^{\pi}_{0} \mathcal{F}^{\alpha}[\textbf{h}] Y^{*}(\theta',\phi') d\Omega', 
    \label{ch1_selfforce}
\end{align}
where this has been loosely defined. For a thorough regularised presentation please refer to state-of-the-art machinery of \cite{thompson2019gravitational}. Given the physical requirements from black hole perturbation theory and the gravitational self-force programme, it is clear we must pick a: \\
\begin{itemize}
    \item $[\texttt{REQ 1}]$: Suitable numerical solver - to be able to produce a highly accurate numerical solution to distributional wave-like equations of the type illustrated by equation \eqref{ch1_bhpt_wave} on any discretised spatial grid and time chart; 
    \item $[\texttt{REQ 2}]$: Suitable \enquote{interpolator} - from the numerical solution in  \texttt{[REQ 1]}, one must be able to accurately compute the amplitude of the metric perturbation at the particle/smaller BH location (i.e. the discontinuity, $r_{p}$) and in its neighbourhood/limit as radiation propagates towards the horizon $\mathcal{H}$/infinity $\mathscr{I}^{+}$. 
\end{itemize}

\begin{table*}
\begin{tabular}{||c|| c c c c c||  }
\hline
\textrm{Numerical Integrator}& 
\textrm{Integrator type}& 
\textrm{Energy conservation}&
\textrm{Symplectic}&
\textrm{Symmetric}&
\textrm{Stability} \\ \hline \hline 
Traditional RK & \texttt{EX} & yes, numerically & partial, numerically & no & CFL constrained \ \  \\ 
\cite{muller2010geodesicviewer,vincent2011gyoto, baubock2012ray, psaltis2011ray, chan2018gray2,gear1971numerical}  &  &  &  &  &  \  \\ \hline 
Gauss-Legendre RK & \texttt{IM} & yes, numerically & yes & yes & long-term stability \ \  \\ 
\cite{kopavcek2010transition, seyrich2012symmetric} &  &long-term  &  &  &  \  \\ \hline 
Mixed RK & \texttt{IMEX} & yes, numerically & yes & no & long-term stability \ \  \\ 
\cite{ascher1995implicit, ascher1997implicit} &  &long-term  &  &  &  \  \\ \hline 
\rowcolor{pastelpink}Horner-Form RK& \texttt{EX} & yes, numerically & partial, numerically & no & CFL constrained   \\ 
\rowcolor{pastelpink} &  &  &  &  &  \  \\ \hline 
\enquote{Slimplectic} & \texttt{IM} & yes, numerically & yes, but\footnotemark[4] & yes & long-term stability\ \  \\ 
\cite{tsang2015slimplectic}&  &long-term  &  &  &  \  \\ \hline 
Mixed symplectic & \texttt{IMEX} & yes & yes  & no & long-term stability \ \  \\ 
\cite{lubich2010symplectic, zhong2010global, mei2013dynamics}  &  &  &  &  &  \  \\ \hline 
Traditional CN & \texttt{IM} & yes, numerically & yes, numerically & yes & CFL constrained  \\ 
\cite{crank1947practical}  &  &  &  &  &  \  \\ \hline 
\rowcolor{myorange}\texttt{IMTEX} SCI ICN &  \texttt{IMTEX} & yes, numerically & yes, numerically & yes & CFL constrained \\ 
\rowcolor{myorange} \cite{teukolsky2000stability, 25thCapraTalk, phdthesis-lidia}  &  &  &  &  &  \  \\ \hline 
\rowcolor{pastelpink}Hermite &  \texttt{IM} & yes & yes & yes & unconditionally stable \\ 
\rowcolor{pastelpink}\cite{o2022conservative, da2023hyperboloidal} &  &  &  &  &  \  \\ \hline 
\rowcolor{pastelpink}Hermite \texttt{IMTEX} & \texttt{IMTEX} & yes & yes & yes & \\ 
\rowcolor{pastelpink}  &  &  &  &  &  \  \\   \hline 
\rowcolor{myorange}Spectral \texttt{SDIRK} & \texttt{IM} & yes & yes & -\footnotemark[5] & unconditionally stable\\ 
\rowcolor{myorange} \cite{macedo2014axisymmetric}  &  &  &  &  &  \  \\   \hline
\end{tabular}
\caption{Quick overview of different time-integrator schemes considered in this manuscript. Highlighted in pink are the integration schemes used in this work, and in orange one highlights upcoming work using different time-integrators. } \label{tab_my_geointegrators}    
\end{table*}
\footnotetext[4]{It is \enquote{slimpletic}! - phase space slims down as expected for dissipative systems \cite{tsang2015slimplectic}.}
\footnotetext[5]{Under consideration \cite{paper2}.}
As we reviewed in \cite{da2023hyperboloidal} there are four main difficulties associated with numerically solving these equations to accurately capture the orbital motion of EMRIs: 
\begin{itemize}
    \item Difficulty $(1)$ - how to treat the distributional source emerging from the point-particle model of the small BH; Numerically this means how can one resolve the distributional nature of the problem in both the space and time dimension; 
    \item Difficulty $(2)$ - what boundary conditions to impose such that radiation can be accurately extracted at the background BH limits; 
    \item Difficulty$(3)$ - time-integrators must ensure highly accurate long-term simulations that can maintain sufficient accuracy for signals potentially lasting the entire 3 $\pm$ year LISA mission; 
    \item Difficulty $(4)$ - how to choose initial conditions and understand their effect in the full radiative process. 
\end{itemize}

In this work, the focus will be mostly on difficulty $(1)$ and $(3)$. Difficulty's $(2)$  resolution amounts to a coordinate chart transformation of the problem in question with minimal changes to difficulties $(1,3)$. In Table \ref{tab_emris_timedomain} all the previous numerical methods used within the radiation-reaction community, to this author's knowledge are reviewed. The aim is to address important numerical algorithm details that were out of the scope of \cite{da2023hyperboloidal}. To resolve difficult $(1)$ we will use \underline{dis}continuous \underline{co}llocation methods to represent the Dirac $\delta$ distribution. These were initially introduced by \cite{2014arXiv1406.4865M} and further refined to the hyperboloidal wave equation and generic cases by \cite{reviews-lidia, 24thCapraTalk, 25thCapraTalk}. Here, given the local structure of the discontinuity at the point-particle is known a priori, we can adapt the Lagrange interpolation scheme by adding these known amplitudes, such that it holds for the case where the problem is smooth everywhere except at the point-particle’s location. In this work, it is demonstrated how these can be numerically optimised for any function in both space and time. For spatial discontinuous discretisations, both finite-difference and pseudospectral methods are considered generically against exact functions; for functions that are discontinuous in time, it is further shown how we can incorporate these discontinuities through Hermite interpolation formulae at higher orders, demonstrating orders $6^{th}$ to $12^{th}$ for the first-time and justifying our choices in \cite{da2023hyperboloidal}. Please refer to the first column of Table \ref{tab_emris_timedomain} for alternative solutions implemented by the community to resolve this difficulty. This work is further complemented by the companion paper \cite{discotexII}. \\

\texttt{EX} explicit time-integration methods have been the standard choice when handling difficulty $(3)$, particularly the implementation of the \texttt{Runge-Kutta}, \texttt{RK}, order-four algorithm \cite{press2007numerical, hairer2010geometric}. This scheme not only results in an \texttt{nth + 1} order error it is also riddled with errors from the real dynamical quantities, in its common \texttt{EX RK} form it preserves symplectic structure neither time-symmetry, energy is conserved with unbounded error numerically, and it also limited by a Courant-Friedrichs-Lewy (CFL) condition (i.e a limitation on the time discretization with respect to the spatial grid which ensures the numerical algorithm’s velocity does not surpass the physical velocity). These errors can usually be mitigated by further increasing the order of the algorithm \cite{muller2010geodesicviewer, vincent2011gyoto, baubock2012ray, psaltis2011ray, chan2018gray2} at the cost of increasing its complexity or by using adaptive step control \cite{gear1971numerical} which is more computationally expensive. \texttt{IM} implicit methods on the other hand are unconditionally stable allowing for larger step sizes and thus suitable for physical problems which have long-term evolutions, as in the case of (X)/(E)MRIs. A well-known example of these schemes is the Crank-Nicholson (CN) integrator \cite{crank1947practical}, which proved difficult to implement due to the complexity of the resulting algebraic equations. One solution came from Choptuik who suggested solving the CN scheme by self-consistent iterations (SCI) effectively turning it into an explicit scheme. To this author, within the numerical relativity community this was the first instance, where an \underline{im}plicit-\underline{t}urned-\underline{ex}plicit, here called \texttt{IMTEX}, the scheme was considered. However in the past, when attempting to use this \texttt {IMTEX} ICN scheme it was expected that iterating more than twice would not lead to improvements in both accuracy and stability \cite{teukolsky2000stability}, however, whereas the latter statement holds, as we showed in \cite{25thCapraTalk, lidiaICN, phdthesis-lidia} \footnote{See slides 9-10 of \cite{25thCapraTalk} for results. Full results are to be included in \cite{phdthesis-lidia, lidiaICN}, the noise in the plot on Slide 10 has been resolved, to be presented in \cite{phdthesis-lidia}. It is further noted ref. \cite{liu2024iterated} have recently independently verified some of the results, namely that increasing the number of iterations does result in improved accuracy.} accuracy, symmetry, symplectic structure preservation and energy conservation can be recovered with an increasing number of iterations. These results are the key motivation to introducing the term \texttt{IMTEX} in \cite{da2023hyperboloidal, phdthesis-lidia} as even though, while effectively explicit, for the \texttt{IMTEX} ICN scheme case it can preserve proprieties which were associated with the implicit scheme from which they are derived, and that should be kept in mind while implementing and studying its proprieties. Given the benefits of implicit schemes there have been significant efforts towards for example adapting the \texttt{Runge-Kutta} scheme to its \texttt{IMEX} form showing significant improvements in stability and long-term numerical energy conservation \cite{ascher1995implicit, ascher1997implicit}. Symplectic integrators \cite{hairer2010geometric} are characterised by the exact preservation of the symplectic structure, i.e. the phase-space trajectories and volume are conserved resulting in long-term stability and bounded energy conservation. One recent example is the completely symplectic scheme based on the implicit midpoint rule for conservative \cite{brown2006midpoint} and non-conservative \cite{tsang2015slimplectic} systems. Furthermore, recently mixed sympletic \texttt{IMEX} integrators have been explored \cite{lubich2010symplectic, zhong2010global, mei2013dynamics}. In this work difficulty $(3)$ will be handled by using both implicit and the implicit-turned-explicit \texttt{IM\underline{TEX}} Hermite integration schemes derived from implicit \texttt{IM} Hermite integration methods which show long-term stability, are energy-preserving, symmetric (that is they are both time-symmetric and \underline{$\rho-$ reversible}) and preserve symplectic structure. This was partially demonstrated by this author in \cite{o2022conservative}. We will present results in optimised Horner-form and up to the \texttt{12th}- order, justifying the choices used in previous and ongoing gravitational physics work \cite{da2023hyperboloidal}. This work is further complemented by the companion paper \cite{discotexII}. Additionally, to highlight the singular proprieties \texttt{IMTEX} possesses and distinguish from schemes which are \textit{exclusively} explicit \texttt{EX} we will compare its results to \texttt{Runge-Kutta} schemes which have also been written in optimal Horner-form. It will be shown to achieve comparable accuracies even to the lowest \texttt{2nd}- order \texttt{IMTEX} Hermite scheme one needs to go as far as \texttt{7th}- orders with the \texttt{EX RK} scheme. Finally, we complement these results with an explicit proof showing in \ref{app_imtex_props} that our \texttt{IMTEX} Hermite integration schemes conserve exactly and numerically symplectic structure and energy while being symmetric. In Table \ref{tab_emris_timedomain} column 4, entitled \enquote{Difficulty 3}, the numerical time-steppers used by the (X)/(E)MRI radiation reaction community can be found, the vast majority of the implementations were conditionally stable and hence minimised the chances of a successful long-term EMRI evolution. \\

Finally difficulty $(2)$ will be studied first by implementing radiation boundary conditions as previously done by \cite{field2009discontinuous, 2014arXiv1406.4865M}, followed by implementing a hyperboloidal slice \cite{penrose1963asymptotic, friedrich1983cauchy, zenginouglu2008hyperboloidal, zenginouglu2009gravitational} as an alternative framework that ensures the governing equation automatically enforce outflow at the boundaries. Here the spacetime is parametrised by a compact radial coordinate defined on a hyperboloidal time hypersurface, allowing for measurements directly at the BH horizon and future null infinity. In this work, the hyperboloidal slice known as the minimal gauge developed by \cite{ansorg2016spectral, jaramillo2021pseudospectrum, macedo2018hyperboloidal, macedo2020hyperboloidal, macedo2022hyperboloidal} is used. \\

Altogether we will demonstrate how to solve both wave and (wave-like) equations as given in equation \eqref{ch1_bhpt_wave} through the means of the method-of-lines recipe by using the \texttt{DiscoTEX} algorithm: we implement \underline{dis}continuous \underline{co}llocation methods to resolve the distributional point-source in both space and time direction and then integrate in time with a Hermite based \texttt{IM\underline{TEX}} numerical method, effectively resolving difficulty $(3)$. Furthermore, these numerical weak-form solutions will be used to obtain the required reconstructed metric amplitudes through a generic discontinuous interpolation formula possible via \texttt{DiscoTEX}. \texttt{DiscoTEX}'s validity will be verified firstly against the results obtained via exact known solutions of the wave equation \cite{field2009discontinuous, field2010persistent, field2023} this allows us to prove and validate the algorithm. Finally, we further compare the results and performance of \texttt{DiscoTEX} with its closest relatives: \texttt{DiscoIMP} where a Hermite implicit time-integration method is used and \texttt{DiscoREX} which uses a purely explicit \texttt{Runge-Kutta} time-stepper. \\

The paper is organised as follows. In Section 2 we review the distributional source wave-equation by highlighting the proprieties of both their exact solutions, as originally derived by Field \cite{field2009discontinuous, field2010persistent, field2023}, and their \textit{weak-form} solutions \cite{field2009discontinuous, field2010persistent, hopper2010gravitational}. In Section 3 it will be explained, in a step-by-step fashion, all the ingredients that makeup \texttt{DiscoTEX} algorithm, and related algorithms, and explain the numerical optimisation factors necessary for accurate implementation. This will be done by computing the solution to both the homogeneous and distributional sourced wave-like equation and by directly comparing it against its exact solution. Three different types of wave-like equations, as derived by Field \cite{field2009discontinuous, field2010persistent, field2023}, are studied in exact and weak form. This work is complemented by the companion paper \cite{discotexII} which computes these solutions up to the \texttt{12th}- order. In Section 4 we will demonstrate \texttt{DiscoTEX}'s applicability to numerical BHPT by solving the equations governing both the scalar and gravitational perturbations by a point-particle on a circular geodesic in Schwarzschild BH background.\footnote{To note here we model the gravitational perturbations by using the convention of \cite{martel2005gravitational, hopper2010gravitational}. These equations are different to the ones solved in \cite{da2023hyperboloidal} as it will be demonstrated. Hence, the work in this manuscript allows us to check both the consistency of \texttt{DiscoTEX} and the different formalisms used to model gravitational perturbations in Schwarzschild.} Furthermore, we will show  \texttt{DiscoTEX}'s main application, for the first time, by calculating both the scalar and gravitational self-force from the numerical weak-form amplitudes of the numerically reconstructed metric perturbations. All of our numerical BHPT results will be compared against state-of-the-art frequency domain work \cite{BHPToolkit, macedo2022hyperboloidal, thompson2019gravitational}. In addition, we will finish Section 4 by reviewing the key proprieties of (X)/(E)MRIs and their influence on numerical method choices and review the numerical techniques previously used within the community in Table \ref{tab_emris_timedomain}. Finally, we review the current state of (X)/(E)MRI modelling and conclude in Section 5.

\section{Distributionally sourced differential equations: exact \emph{vs} weak-form solutions}\label{sec2}
Distributionally sourced linear wave equations are a valuable toy model as their underlying structure is similar to the equations we wish to solve, as seen in equation \eqref{ch1_bhpt_wave}, but more importantly, they possess exact solutions. These solutions were first demonstrated in the literature by Field et al \cite{field2009discontinuous, field2010persistent, field2022discontinuous, field2023}, within the context of developing their novel Discontinuous Galerkin scheme for the modelling of EMRIs. In this work, their formalism is adapted and three types of exact solutions to the distributionally sourced equation are considered, 
\begin{align}
    \label{ch2_distr_wavequation_delta1_plus_delta0}
    & \textrm{Type I:} \ \ \square \Psi(t,x) = G(t) \delta(x - r_{p}) + F(t)\delta'(x - r_{p}), \\
    & \Psi(t,x) = -\frac{1}{2} \sin \mathcal{\theta} + \frac{1}{2} \iu \gamma^{2} [\dot{r}_{p} + \text{sgn}(x-r_{p})] \cos \mathcal{\theta},\label{ch2_distr_wavequation_delta1_plus_delta0_sol}
\end{align}
\begin{align}
    \label{ch2_distr_wavequation_delta0}
     &\textrm{Type II:} \ \ \square \Psi(t,x) = G(t) \delta(x - r_{p}),\\
    &\Psi(t,x) = -\frac{1}{2} \sin \mathcal{\theta}, 
    \label{ch2_distr_wavequation_delta0_sol}
\end{align}
and, 
\begin{align}
    \label{ch2_distr_wavequation_delta1}
     &\textrm{Type III:} \ \ \ \square \Psi(t,x) = G(t) \delta'(x - r_{p}),  \\
    &\Psi(t,x) = \frac{1}{2} \gamma^{2} [\dot{r}_p + \text{sgn}(x-r_{p})] \cos \mathcal{\theta}.
    \label{ch2_distr_wavequation_delta0_sol}
\end{align}

Furthermore in all these equations $\mathcal{\theta} = \gamma^{2}(t- x \, \dot{r}_{p} - |x - r_{p}|)$, $G(t) = \cos(t)= -\iu F(t)$ and $\rm{sgn}$ refers to the sigmun function. The aim of this manuscript is to give the generic framework of the \texttt{DiscoTEX} algorithm with black hole perturbation theory in mind, thus only equations of the type given in equation \eqref{ch1_bhpt_wave} are considered.

\subsection{Weak-form solutions to the distributionally sourced linear wave equation}
Given the presence of distributional term $\delta(x-r_{p})$ (or $\delta(r-r_{p}(t))$, as observed in equation ~\eqref{ch1_bhpt_wave} and its radial derivative, we extend the homogeneous solutions for the master functions $\Psi(t,r)$ to a \textit{weak-form solution} \cite{hopper2010gravitational} \footnote{Here, by \textit{weak-form solution} one refers to a general weak solution to the PDE, which is smooth but may have a set of measure zero non-differentiable or limited differentiable points \cite{da2023hyperboloidal}.} as they approach the particle position, $r_{p}$, from the left $(-)$ and right $(+)$ limits,   
\begin{equation} 
    \Psi(t,r)=  \Psi^{+}(t,r) \Theta[x-r_{p}(t)] + \Psi^{-}(t,r)\Theta[r_{p}(t)-x], 
    \label{ch2_weakformsolution_wave}
\end{equation}
where $\Theta(z)$ is the Heaviside function defined as
\begin{equation}\Theta(z)= \left\{ 
\begin{array}{ccc}
1 &\text{for}& z>0, \\
\frac{1}{2}&\text{for}& z=0, \\
0 &\text{for}& z<0.  
\end{array}
\right. 
\label{ch2_HeavisideStepFunction}
\end{equation}
A time-dependent jump is,
\begin{align}
     &[[\Psi]](t) =\Psi^{+}(t,r_{p}(t)) - \Psi^{-}(t,r_{p}(t)).
     \label{ch2_timedependent_jump}
\end{align}
Here, the jump formalism of ref.~\cite{hopper2010gravitational} is adapted to facilitate future work and comparisons, a derivative is given by a subscript. The reader is referred to \ref{app2} where a full derivation of the following jump conditions is given. We have that,  
\begin{align}
&(f_{p}^{2} - \dot{r}_{p}^{2} ) [[\Psi]] = F(t, r_{p}), \label{ch2_der_jump0} \\ 
&(f_{p}^{2} - \dot{r}_{p}^{2} ) [[\Psi_{r}]] + [\ddot{r}_{p} -  f_{p} f'_{p}]  [[\Psi]] + 2 \ \dot{r}_{p}\partial_{t}[[\Psi]]  = \nonumber \\
& =G(t, r_{p}) - \partial_{r_{p}} F(t, r_{p}), \label{ch2_der_jump1}    
\end{align}
where the jumps are given by, 
\begin{equation}    
    [[\Psi]] = \frac{F(t, r_{p})}{(f_{p}^{2} - \dot{r}_{p}^{2} ) }, 
    \label{ch2_gen_jump0}
\end{equation}
\begin{equation}    
    [[\Psi_{r}]] = \frac{G(t,r_{p})  - \partial_{r_{p}} F(t, r_{p})  -  2 \dot{r}_{p}\partial_{t}[[\Psi]]  - [\ddot{r}_{p} -f_{p}f'_{p} ][[\Psi]]    }{(f_{p}^{2} - \dot{r}_{p}^{2} )  }, 
    \label{ch2_gen_jump1}
\end{equation}
and finally, 
\begin{equation}    
    \partial_{t}[[\Psi]] = \frac{2 \dot{r}_{p}[f_{p} - \partial_{r_{p}}f_{p} - \ddot{r}_{p}    ]  F(t, r_{p})}{(f_{p}^{2} - \dot{r}_{p}^{2} )^{2}} +  \frac{\partial_{r_{p}} F(t, r_{p}) - \dot{r}_{p}F(t,r_{p}) }{(f_{p}^{2} - \dot{r}_{p}^{2} )}. 
    \label{ch2_gen_dt_jump0}
\end{equation}
Furthermore the following chain-rules can be used to obtain $[[\Psi_{t}]]$, 
\begin{eqnarray}
    \partial_{t} [[\Psi]] = [[\Psi_{r}]] \dot{r}_{p} + [[\Psi_{t}]], \nonumber \\ \label{ch2_chainrule_dtj0}
    [[\Psi_{x}]] = f_{p} [[\Psi_{r}]].
    \label{ch2_tortoise_r_jumprelation}
\end{eqnarray}
Higher-order jumps can be derived by applying higher-order chain rules, 
\begin{eqnarray}
    \label{ch2_chain_rules_dt_jt}
    \partial_{t} [[\Psi_{t}]] = \dot{r} [[\Psi_{rt}]]  + [[\Psi_{tt}]], \\ 
    \partial_{t} [[\Psi_{r}]] = \dot{r} [[\Psi_{rr}]]  + [[\Psi_{rt}]]. 
    \label{ch2_chain_rules_dt_jr}
\end{eqnarray}
With eq.~\eqref{ch2_chain_rules_dt_jr} $-$ $\dot{r}$ eq.~\eqref{ch2_chain_rules_dt_jt} and, because here we are dealing with the distributionally sourced wave-equation as aforementioned, it holds that $[[\Psi_{xx}]] = [[\Psi_{rr}]] = [[\Psi_{tt}]]$ and thus we have
\begin{eqnarray}
    [[\Psi_{rr}]] = \frac{1}{(1 - \dot{r}_{p}^{2})}\bigg(\partial_{t} [[\Psi_{t}]] - \dot{r} \ \partial_{t} [[\Psi_{r}]] \bigg). 
    \label{ch2_chain_rules_dt_first}
\end{eqnarray}
For an example of the application of these derivations one refers the reader to \cite{long2021time} where higher-order jumps are derived via the chain-rule method and to Chapter 5 of \cite{Dempsey:2017igj} for another application in black hole perturbation theory of Type II.

\section{\underline{Numerical weak-form} solutions to distributionally sourced linear differential equations via \texttt{DiscoTEX}}\label{Sec3}
\subsection{Reduction to first-order}\label{ch31_red_1order}

Numerically, the problem will be tackled through the method-of-lines (MoL) framework which has four main steps: picking boundary and initial conditions, followed by reducing the system to first-order with spatial discretisation and then integrating in time. 

The \texttt{1+1D} wave-like inhomogeneous equation is a hyperbolic PDE of the type 
\begin{equation}
    \partial_{t} \textbf{U} = \textbf{L} \cdot \textbf{U} + \mathcal{S}, \ \ \ \ \textbf{U} = \begin{pmatrix}\Psi\\\Pi
    \end{pmatrix},
    \label{ch3_general_pdes}
\end{equation}
where $\textbf{L}$ is a spatial differential operator and $\Psi(t,x)$, $\Pi(t,x)$ are the field variables and their partial derivatives with respect to time. We will show in this section how to construct discontinuous spatial and temporal discretisations by building upon well-established Lagrange and Hermite interpolation methods on multiple coordinate charts $(t,r)$ and $(\tau,\sigma)$. This work takes full advantage of the fact that we know all the information associated with the distributional part of the problem represented by the coefficients on the RHS of equation \eqref{ch1_bhpt_wave}, here represented generically by $\mathcal{S}$. 

\subsection{Discontinuous spatial discretisation via higher order jumps - resolving difficulty $(1)$ in the spatial dimension}\label{sec_3_2_discojumps}
\subsubsection{Spatial discretisation through Lagrangian interpolation}\label{sec_3_1}
We can discretise in space wave-like \texttt{1+1D} equations such as those in equation \eqref{ch1_bhpt_wave}, with governing field $\textbf{U}(t,x)$, such that $\textbf{U}(t,x) \rightarrow \textbf{U}(t, x_{i}) := \textbf{U}_{i}(t)  = \textbf{U}(t)$ where $a\leq x_{i} \leq b $ with the collocation nodes ranging from $0<i<N$. Essentially, we build the collocation polynomial of degree $N$, 
\begin{equation}
    p(x) = \sum^{N}_{j=0} c_{j} x^{j}, 
    \label{smooth_interpol_gen}
\end{equation}
determined by solving the linear algebraic system of conditions specifically given as 
\begin{equation}
    p(x_{i}) = U_{i}, 
    \label{smooth_interpol_gen_conditions}
\end{equation}
for the coefficients $c_{j}$. 
Rewriting it in Lagrangian formulae, the Lagrange interpolating polynomial (LIP) is retrieved, 
\begin{equation}
    p(x) = \sum^{N}_{j=0} U_{j} \pi_{j}(x),
    \label{smooth_interpol_LIP}
\end{equation}
where $\pi_{j}(x)$ is the Lagrange basis polynomial (LBP) given as
\begin{equation}
 \pi_{j}(x) = \prod^{N}_{\substack{k=0,\\ k\neq j}} \frac{x - x_{k}}{x_{j} - x_{k}}.
  \label{ch3_sb2_LBP}
\end{equation}
By acting on the LIP as given in eq.~\eqref{smooth_interpol_LIP},  we can differentiate (or integrate) the field $U$ any \texttt{n-th} times, explicitly, 
\begin{equation}
    U^{(n)}(x_{i}) \approx p^{(n)}(x_{i}) = \sum^{N}_{j=0} D^{(n)}_{ij} U_{j},  
    \label{spatial_disc}
\end{equation}
where, 
\begin{equation}
    D^{(n)}_{ij} = \pi^{(n)}_{j} (x_{i}). 
    \label{diffmatrix_gen}
\end{equation}

The explicit form of the spatial differential operators given in eq.~\eqref{ch1_bhpt_wave} is now given, and later discretised and included in the effective wave operator $\textbf{L}$, by using both spectral and finite-difference collocation methods \cite{o2022conservative, da2023hyperboloidal, phdthesis-lidia}. For the spectral method, the Chebyshev-Gauss-Lobatto collocation nodes are given by, 
\begin{eqnarray}
    x_{i} = \frac{a + b}{2} + \frac{b-a}{2}z_{i}, \nonumber \\ 
     z_{i} = -\cos{\theta_{i}}, \ \ \theta_{i} = \frac{i\pi}{N}, \ i = 0,1,...,N
    \label{chebyshev_lobatto_nodes} 
\end{eqnarray}
yielding, 
\begin{equation}
    {D}^{(1)}_{i j} = 
\frac{2}{b-a}   \begin{cases}
        \frac{c_i (-1)^{i+j}}{c_j (z_i-z_j)} & i \neq j\\
        -\frac{z_j}{2(1-z_j^2)} & i = j \neq 0,N\\
        -\frac{2 N^2 + 1}{6} & i = j = 0\\
        \frac{2 N^2 + 1}{6} & i = j = N
    \end{cases}
    \label{cha3_firstorder_diffmatrix}
\end{equation}
for the first-derivative matrix, and
\begin{small}
\begin{equation}
D_{ij}^{(2)} = {\left( {\frac{2}{{b - a}}} \right)^2}\left\{ {\begin{array}{*{20}{l}}
{\frac{{{{( - 1)}^{i + j}}}}{{{c_j}}}\frac{{z_i^2 + {z_i}{z_j} - 2}}{{(1 - z_i^2){{({z_i} - {z_j})}^2}}}}&{i \ne j,{\; }i \ne 0,N }\\
{\frac{2}{3}\frac{{{{( - 1)}^j}}}{{{c_j}}}\frac{{(2{N^2} + 1)(1 + {z_j}) - 6}}{{{{(1 + {z_j})}^2}}}}&{i \ne j,{\;}i = 0}\\
{\frac{2}{3}\frac{{{{( - 1)}^{j + N}}}}{{{c_j}}}\frac{{(2{N^2} + 1)(1 - {z_j}) - 6}}{{{{(1 - {z_j})}^2}}}}&{i \ne j,{\;}i = N}\\
{ - \frac{{({N^2} - 1)(1 - z_j^2) + 3}}{{3{{(1 - z_j^2)}^2}}}}&{i = j,{\; }i \ne 0,N}\\
{\frac{{{N^4} - 1}}{{15}}}&{i = j = 0{\; \rm{ or }\; }N}
\end{array}} \right.
\label{cha3_secondorder_diffmatrix}
\end{equation}
\end{small}
for the second derivative matrix. Alternatively, we further note the dot product from equation \eqref{cha3_firstorder_diffmatrix}, could be computed as $D^{(2)}_{ij} = D^{(1)}_{ij} \cdot D^{(1)}_{ij}$. For finite-difference methods, we use equidistant nodes, 
\begin{equation}
    x_{i} = a + i\frac{b-a}{N}, \ \ \ \ i=0,1,..,N
    \label{ch3_sb2_EquiNodes}
\end{equation}
and resort to the fast library of the \texttt{Wolfram Language} which uses Fornberg's algorithm to compute higher-order derivatives, $D^{(n)}_{ij}$, obtained through the command, 
\begin{verbatim}
Dn = 
NDSolve‘FiniteDifferenceDerivative[Derivative[n],X,
"DifferenceOrder" -> "4"]@
"DifferentiationMatrix"//Normal
//Developer`ToPackedArray//SparsedArray.
\end{verbatim}
 
\subsubsection{Discontinuous generalisation of the Lagrange interpolation method by incorporating higher order jumps}
\begin{figure*}
\includegraphics[width=94mm]{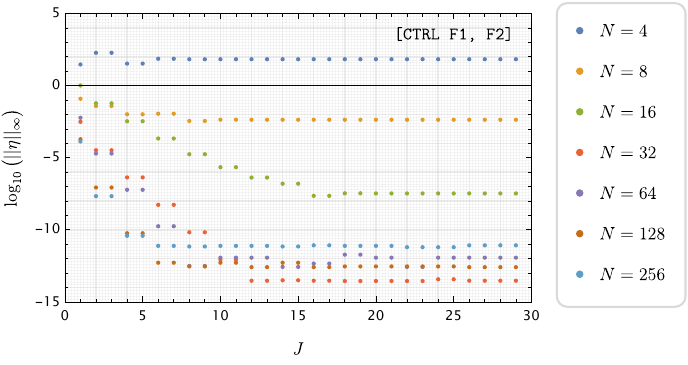}
\quad
\includegraphics[width=76mm]{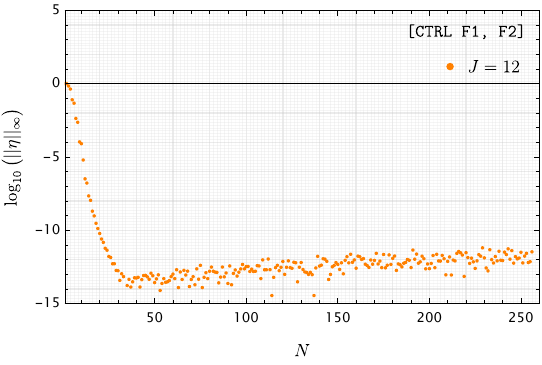}
\caption{\textbf{Top:} Plot showing the $l_{\infty}$ of the relative error from the exact and numerical solution with the number of Chebyshev nodes $N$ versus the number of jumps $J$. It is readily clear that for optimal implementations of the algorithm, we need to study both these numerical algorithm optimisation control factors. 
\textbf{Bottom:} From the plot on the top it is clear the optimal number of nodes is $N=32$ Chebyshev nodes with about $J=12$ jumps. \label{ch3_SimpSpaceFunc_convtest}} 
\end{figure*} 

The discontinuous generalisation to the Lagrange interpolation briefly reviewed above, can then be constructed. This was put forward generically in \cite{2014arXiv1406.4865M} and later improved by \cite{24thCapraTalk,25thCapraTalk, phdthesis-lidia, da2023hyperboloidal}. This method uses higher-order jumps as input, where here we leverage the fact we know the location of the particle, $x_{p}$, \textit{a priori}, such that it holds to write the master field variables a combination of jumps in their fields and derivatives. Hence we define a jump as: 
\begin{equation}
    \Psi^{(m)}(x_{p}^{+}) -  \Psi^{(m)}(x_{p}^{-}) = J_{m}(t), \ \ \  m=0,1, ..., \infty
    \label{ch3_b2_higherorderjumps}
\end{equation}
where explicitly, albeit in $(t,r)$ coordinates, this has been given for the first $m=\{0,1\}$ orders in equations (\eqref{ch2_gen_jump0}, \eqref{ch2_gen_jump1}) and will be later specified in hyperboloidal coordinates after it has been adequately incorporated within the Lagrangian method described above. 

Essentially we take the weak form of the solution to the wave-like functions as given in eq.~\eqref{ch1_bhpt_wave} and rewrite it as a \textit{generic} collocation polynomial
\begin{equation}
    p(x) =  p_{+}(x) \Theta(x-x_{p}) + p_{-}(x)\Theta(x_{p}-x),  
    \label{ch3_sb2_gen_p_poly}
\end{equation}
where the right/left interpolating polynomials are given respectively as
\begin{equation}
    p_{+}(x) = \sum^{N}_{j=0} c^{+}_{j}(x_{p})x^{j}, \ \ \ 
    p_{-}(x) = \sum^{N}_{j=0} c^{-}_{j}(x_{p})x^{j}.
    \label{ch3_sb2_gen_IPsLR}
\end{equation}

By solving equations \eqref{ch3_sb2_gen_IPsLR} as a system of algebraic equations with the collocation conditions given by, 
\begin{equation}  \Psi_i=\left\{ 
\begin{array}{ccc}
p_{+}(x_{i}), & \qquad x_{i} > x_{p}, \\
p_{-}(x_{i}), & \qquad x_{i} < x_{p},
\label{ch3_sb2_collcond_gen}
\end{array}
\right. 
\end{equation}
one determines half of the $(2N+2)$ polynomial coefficients, $c^{\pm}_{j}$.
The remaining coefficients are then determined by imposing the jump conditions in equation \eqref{ch3_sb2_gen_p_poly} as,
\begin{equation} 
p_{+}^{(m)}(x_{p}) - p_{-}^{(m)}(x_{p}) = \left\{ 
\begin{array}{ccc}
J_{m}, & m =0,1, ..., M \\
0,  & m = M+1, ..., N 
\end{array}
\right.
\label{ch3_sb2_coeffFR_JC_gen}
\end{equation}
where here $M$ ranges from $[-1,...,N]$ and the $M=N+1$ jumps are left unspecified until when studying the number of jumps optimal to the algorithm's implementation for the particular physical model here studied. 
Everything in the LBP is then rewritten as, $\pi_{j}(x)$, as given in eq.~\eqref{ch3_sb2_LBP}, which varies depending on whether we choose to work with spectral eq.~\eqref{chebyshev_lobatto_nodes} or a finite-difference eq.~\eqref{ch3_sb2_EquiNodes} collocation methods. It then suffices to solve algebraically the \textit{interpolating piecewise polynomial}
\begin{equation}
    p^{\pm}(x) = \sum^{N}_{j=0} C^{\pm}_{j}(x_{p}) \pi_{j}(x).
    \label{ch3_sb2_IPP}
\end{equation}
Specifically, we have the algebraic conditions,
\begin{eqnarray}
    C^{+}_{j}(x_{p}) = \Psi_{j} + \Theta(x_{p} - x_{j}) g(x_{j} - x_{p}), \\
    \label{ch3_sb2_ipp_collcright}
    C^{-}_{j}(x_{p}) = \Psi_{j} - \Theta(x_{j} - x_{p}) g(x_{j} - x_{p}), 
    \label{ch3_sb2_ipp_collcleft}
\end{eqnarray}
where, 
\begin{equation}
    g(x_{j} - x) = \sum^{M}_{m=0} \frac{J_{m}}{m!}(x_{j} - x_{p})^{m}
    \label{ch3_sb2_gVector_expSum}
\end{equation}
are the weights computed from the jump conditions derived at the discontinuity $x_{p}$. Finally, substituting equations (\eqref{ch3_sb2_IPP}-\eqref{ch3_sb2_ipp_collcleft}) into equation \eqref{ch3_sb2_gen_p_poly} we get the \textit{generic} interpolating piecewise polynomial, 
\begin{equation}
    p(x) =  \sum^{N}_{j=0} \bigg[ \Psi_{j} + \Delta_{\Psi}(x_{j} - x_{p}; x- x_{p}) \bigg] \pi_{j}(x)
\end{equation}
where the $\Delta_{\Psi}$ function is given by
\begin{small}
\begin{align}
    &\Delta_{\Psi}(x_{j} - x_{p}; x -x_{p}) = \nonumber \\
    &=\bigg[ \Theta(x-x_{p})\Theta(x_{p} -x_{j})
     -  \Theta(x_{p}-x)\Theta(x_{j} - x_{p})\bigg]  g(x_{j} - x_{p}) \nonumber \ \ \ \ \ \ \  \\
    &= \bigg[\Theta(x_{i} - x_{p}) - \Theta(x_{j} - x_{p})\bigg]g(x_{j} - x_{p})  \; \rm{when} \; x = x_{i}. \ \ \ \ \ 
    \label{ch3_sb2_delta}    
\end{align}
\end{small}
\noindent In the end we approximate our master field variables as given in equation \eqref{ch1_bhpt_wave} by 
\begin{equation}
    \Psi(t,x) \approx \sum^{N}_{j=0} \bigg[ \Psi_{j}(t) + \Delta_{\Psi}(x_{j} - x_{p}(t); x - x_{p}(t))  \bigg] \pi_{j}(x).
    \label{ch3_sb2_WFsolution}
\end{equation}
To be precise, all the differential operators in eq.~\eqref{ch1_bhpt_wave} and further specified in eq.~\eqref{ch34_waveTR_l_op_boundaryCondos} and eqs.~(\eqref{ch34_sb2_diffOperators_gamma}-\eqref{ch34_sb2_diffOperators_iota}), are discretised as, 
\begin{eqnarray}
  \partial_{x}^{n}(t,x)\Psi|_{x = x_{i}} = p^{(n)}(x) = \sum^{N}_{j=0}  D^{(n)}_{ij}  \Psi_{j} + s_{i}^{(n)}(t), \ \ \ \ 
\label{cha2_spatial_disc_generic}
\end{eqnarray}
where $  s^{(n)}_{i}(t)$ is given as 
\begin{equation}
    s^{(n)}_{i}(t) = \sum^{N}_{j=0}D^{(n)}_{ij} \Delta_{\Psi}\big(x_{j} - x_{p}(t); x_{i} - x_{p}(t)\big)  
    \label{ch3_sb2_SpaceSpource}
\end{equation}
and the user-specifiable high-order jumps in eq.~\eqref{ch3_sb2_gVector_expSum} obtained through the computation of the higher order recurrence relation are given as \footnote{We note in our previous work \cite{da2023hyperboloidal} this equation is given for circular orbits as thus all terms reducing to zero in this regime are omitted. The derivation was nevertheless given explicitly, for the first time, and thus reproducible to the reader.},
\begin{align}
    &J_{m+2}(t) =  -\frac{1}{(f^{2}_{p}- \dot{r}^{2}_{p})} 
    \bigg( \bigg( \ddot{J}_{m}(t) - 2 \dot{r}_{p} \dot{J}_{m+1}(t) \nonumber \\ 
    & - \ddot{r}_{p} J_{m+1}(t) + \dot{r}^{2}_{p} J_{m+2}(t)   \bigg) \nonumber \\
    &+  \sum^{m}_{k=0} {m \choose k} \bigg[ - \rho^{(k)}(r_{p}) J_{m+1-k}(t) +  V^{(k)}(r_{p}) J_{m-k}(t) \bigg]   \nonumber \\
&- \sum^{m}_{k=1} {m \choose k} \eta^{(k)}(r_{p}) J_{m+2-k}(t)    \bigg).
\label{cha3_rec_relation_mjumps_PhysicalChart_RWZ}
\end{align}
Furthermore the initialising jumps $J_{0}(t), J_{1}(t)$ are generically given as in equations (\eqref{ch2_gen_jump0}, \eqref{ch2_gen_jump1}) for when the wave equation is, for example, as given in \eqref{aa_ugly_pde} for the $\Psi(t,r)$ field. Shortly the recurrence relation for the field in $\Psi(t,x)$ will be specified. In what follows, we will also apply the machinery just explained when solving wave and wave-like equations in a hyperboloidal coordinate chart where $(t \rightarrow t(\tau,\sigma), r \rightarrow r(\sigma))$. We note this implementation has thoroughly been explained in previous work \cite{da2023hyperboloidal}. 

\subsubsection{Example of discontinuous differentiation of a simple spatial function}\label{example_disco_space}
To illustrate the usefulness of the above scheme as well as understand the required numerical optimisation factors associated with resolving functions with discontinuities in space, the following discontinuous function
\begin{equation}
f(x) = \sin(\pi |x - x_{p}|)
\label{ch3_spatial_function}
\end{equation}
is considered in the spatial domain $x \in [-1,1]$ and where $x_{p}$ is the discontinuity at say $x_{p} = 0.001$. Exact higher-order derivatives of this function can be found via, 
\begin{equation}
    f^{n}_{\textbf{E}}(x) =  \left\{ 
    \begin{array}{ccc}
    f^{(n)}(x)^{+} &\text{for}& x>x_{p}, \\
    f^{(n)}(x)^{-} &\text{for}& x<x_{p}, \\
    \frac{f^{(n)}(x)^{+} - f^{(n)}(x)^{-}}{2} & \text{else}, 
    \end{array}
    \right. 
    \label{ch3_simplefunction_exact}
\end{equation}
where the subscript $\textbf{E}$ indicates the solution is in exact form. The analytic jumps associated with this function and its derivatives are given by, 
\begin{equation}    
    J_{m} = \frac{1}{m!} \bigg[ \lim_{x \rightarrow x_{p}^{+}} f^{(m)}(x) - \lim_{x \rightarrow x_{p}^{-}} f^{(m)}(x) \bigg], \hspace{0.9cm} m=0,1,2,...,M
    \label{ch3_simplefunction_space_AnalJumps}
\end{equation}
where $M$ is the user-specifiable number of jumps. 
\begin{figure}
\includegraphics[width=75mm]{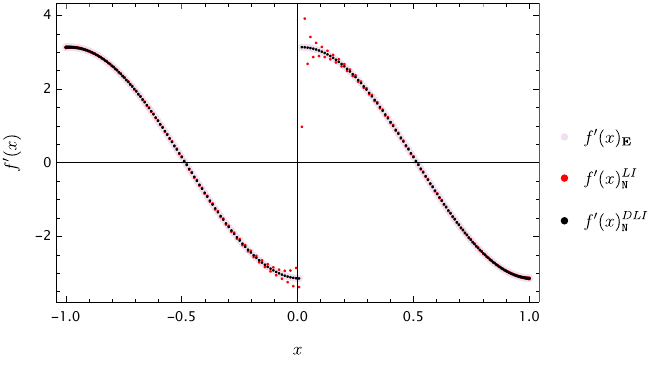}
\quad
\includegraphics[width=75mm]{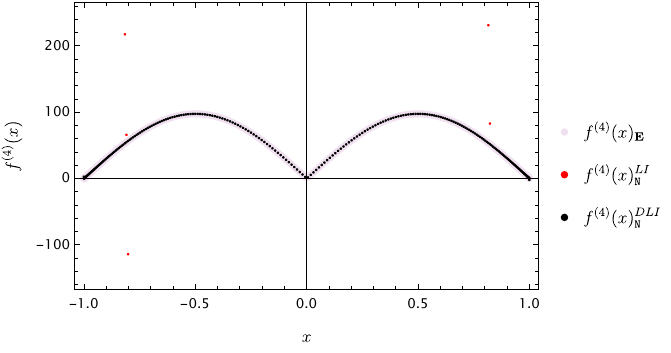}
\caption{Plots of the exact spatial discontinuous function as given by equation \eqref{ch3_simplefunction_exact} against the numerical solution is obtained with traditional Lagrangian smooth interpolation methods and followed with the correction introduced by the discontinuous algorithm. We demonstrate the accuracy of the algorithm by considering both the first- and fourth-order spatial derivatives. \label{ch3_SimpSpaceFunc_AllMethods}} 
\end{figure}   
To construct the numerical solution $f^{(n)}_{\texttt{N}}$ we make use of eq.~\eqref{ch3_sb2_delta}, now given as, 
\begin{equation}
    f^{(n)}(x_{i})_{\texttt{N}} \approx p^{(n)}(x_{i}) = \sum^{N}_{j=0} D^{(n)}_{ij} [f_{j} + \Delta(x_{j} - x_{p}; x_{i}-x_{p})], 
    \label{ch3_simpFunc_num_disco_sol}
\end{equation} 
and here $N$ highlights this is the numerical solution, $f_{j} = f(x_{j})$ and we now substitute equation \eqref{ch3_sb2_delta} here to get, 
\begin{equation}
    f^{(n)}(x_{i})_{\texttt{N}} \approx \sum^{N}_{j=0} D^{(n)}_{ij} [f_{j} + g_{j}\Theta_{i} - g_{j}\Theta_{i}], 
    \label{ch3_simpFunc_num_disco_sol_2}
\end{equation}
where $g_{j} = g(x_{j} - x_{p})$ and $\Theta_{i}= \Theta(x_{i} - x_{p})$. This amounts to simple vector-matrix and matrix-matrix multiplication obtained in \texttt{Mathematica} for example as, 
\begin{verbatim}
fn = Dn.f + (Dn.g)*h - Dn.(g*h)
\end{verbatim}
where \verb$fn$ $= f(x_{i})_{\texttt{N}}$, \verb $f$ $=f(x_{j})$ \verb$Dn$ $ = D^{(n)}_{ij}$, $h = \Theta_{i/j}$ and $g= g_{i/j}$ vectors. It is important to emphasise here that the numerical accuracy associated with discontinuous collocation Lagrange interpolation methods depends on two main factors which are \textbf{user-specifiable}: 
\begin{enumerate}
   \item \texttt{[CTRL F1]} - Number of nodes; \label{controlfactor1}
\item \texttt{[CTRL F2]} - Number of jumps.\label{controlfactor2}
\end{enumerate}
From the numerical experiments shown in Figure \ref{ch3_SimpSpaceFunc_convtest} it's clear high accuracies are possible for $N=32$ Chebyshev nodes, using equation \eqref{chebyshev_lobatto_nodes}, and with $J=12$ nodes. In Figure \ref{ch3_SimpSpaceFunc_AllMethods} we compare the numerical solution, $f'(x)^{DLI}$,  for the first-order derivative against the exact solution $f'(x)^{E}$ and the solution that would be obtained if the discontinuous corrections, given by $\Delta$ in equation \eqref{ch3_simpFunc_num_disco_sol}, were not included \cite{reviews-lidia, phdthesis-lidia}. 

\subsection{High order discontinuous time integration}\label{sec_discotex_justime}
\subsubsection{The \texttt{IM}plicit-\texttt{T}urned-\texttt{EX}plicit \texttt{IMTEX} Hermite geometric integrators and relatives - addressing difficulty $(3)$}\label{sec_imtex}
\begin{figure}
\includegraphics[width=89mm]{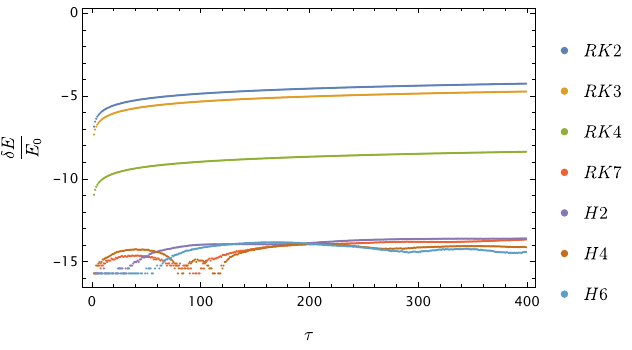}
\caption{Fractional energy error from the evolution of equation (25) as given in Ref.\cite{da2023hyperboloidal} with no RHS terms i.e $\mathcal{S} = S_{lm}(\tau,\sigma) = 0$ and potential on the LHS  is $V_{l}= 0$. We observe both \texttt{HF EX RK} and \texttt{IMTEX} Hermite integrators errors are bounded, with lowest order \texttt{IMTEX} integrators significantly outperforming even the highest \texttt{HF EX RK} schemes. Furthermore, it is observed at this juncture, that there is no point in using a Hermite \texttt{IMTEX} scheme of order higher than 4, this is likely due to the build-up of round-off error and the fact most of the residual error comes from the spatial discretisation which becomes more significant as the order of Hermite \texttt{IMTEX} increases. \label{ch3_RK_HER}} 
\end{figure}   
Even though the above resolves difficulty $(1)$ in the spatial direction, one still needs to address how the discontinuity is resolved in the time direction to fully accurately solve for difficulty $(1)$. However, before we do so, the work done in \cite{o2022conservative, da2023hyperboloidal} will be revisited and difficulty $(3)$ will be addressed. Our motivation here is two-fold. Firstly, before the distributionally sourced wave-equations of the type eq.~\eqref{ch1_bhpt_wave} are numerically solved, the effect that increasing the order of integration of these numerical schemes has on the numerical accuracy and speed of the numerical solutions must be clearly understood.  Secondly, we wish to highlight the differences between a \textbf{purely explicit} \texttt{EX} non-geometric time integrator such as Runge-Kutta \texttt{EX RK} scheme and the novel \texttt{IM}plicit-\texttt{T}urned-\texttt{EX}plicit, i.e. the \texttt{IMTEX}, Hermite integration scheme for both homogeneous and distributionally-sourced wave-like equations.

In Section \ref{sec_3_1} it was shown how the field variable $\textbf{U}(t,x)$ is discretised in space such that equation \eqref{ch3_general_pdes}, is effectively reduced to first-order form and solved as a system of coupled ordinary differential equations (ODEs) given as, 
\begin{equation}
    \frac{d \textbf{U}}{dt} = \textbf{L} \cdot \textbf{U} + \mathcal{S}, 
    \label{ch33_reduction_odes}
\end{equation}
where $\textbf{U}(t,x) \rightarrow \textbf{U}(t)$ with $\textbf{U}(t,x_{i}) := \textbf{U}_{i}(t)$. For now, the source term $\mathcal{S}$ is dropped and we consider only the homogeneous problem. As demonstrated in \cite{o2022conservative}, one starts by applying the fundamental theorem of calculus to solve for the field variable $\textbf{U}$ in time in the form 
\begin{equation}
    \textbf{U}(t_{n+1}) = \textbf{U}(t_{n}) + \int^{t_{n+1}}_{t_{n}} \textbf{U}(t) \ dt. 
    \label{ch33_fundamentalCalculus}
\end{equation}
Then approximate $\textbf{U}(t)$ by applying a two-point Taylor expansion, (or a 2-point Hermite interpolant) where an osculating polynomial is constructed such that it matches the values of $\textbf{U}$ and its derivatives at the endpoints $t_{n/n+1}$ approximating the integral in \eqref{ch33_fundamentalCalculus} \cite{lanczos1956applied, dyche1956multiple, markakis2019time}. We then get the \textit{generalised Hermite rule} to order $l-1$, 
\begin{equation}
    \int^{t_{n+1}}_{t_{n}} \textbf{U}(t) dt = \sum^{l}_{k = 1} c_{lk} \Delta t^{k} \bigg( 
    \textbf{U}^{(k-1)}_{n} + (-1)^{(k-1)} \textbf{U}^{(k-1)}_{n+1} \bigg) + \textbf{R}_{l},
    \label{ch33_gen_Hermite_rule}
\end{equation}
where $k$ denotes the order of the time-derivative of our function at any $t_{n}$ such that, 
\begin{equation}
    \textbf{U}^{(k)}_{n} = \frac{d^{k} \textbf{U}(t)}{dt^{k}}\bigg|_{t=t_{n}}. 
    \label{ch33_timederivativesfield}
\end{equation}
Furthermore, the expansion coefficients $c_{lk}$ are given as, 
\begin{equation}
    c_{lk} = \frac{l!(2l-k)!}{k!(2l)!(l-k)!}, 
    \label{ch33_exp_coefficients}
\end{equation}
and the remainder $\textbf{R}_{l}$ as, 
\begin{equation}
    \textbf{R}_{l} = (-1)^{l} \frac{(l!)^{2}}{(2l+1)!(2l)!} \Delta t^{2l+1} \textbf{U}^{(2l)}(t), \ \ \ \ \ \ t \in [t_{n}, t_{n+1}]. 
    \label{ch33_remainder}
\end{equation}
Choosing $l=3$ yields the Lotkin's rule \cite{lotkin1952new} in \textit{implicit}, \texttt{(IM)} form, namely
\begin{align}
   &\int^{t_{n+1}}_{t_{n}} \textbf{U}(t) dt = \frac{\Delta t}{2}(\textbf{U}_{n} + \textbf{U}_{n+1}) + \nonumber \\ 
   &\frac{\Delta t^{2}}{10}(\dot{\textbf{U}}_{n} - \dot{\textbf{U}}_{n+1})
   + \frac{\Delta t^{3}}{120}(\ddot{\textbf{U}}_{n}+ \ddot{\textbf{U}}_{n+1})  + \mathcal{O}(\Delta t^{7}), 
    \label{ch33_hermiteintegration_hermite6rule}
\end{align}

This approximates to, 
\begin{small}
\begin{align}
   &\textbf{U}_{n+1}  - \textbf{U}_{n} = \frac{\Delta t}{2}\textbf{L} \cdot (\textbf{U}_{n} + \textbf{U}_{n+1}) + \nonumber \\ 
    &\frac{\Delta t^{2}}{10}\textbf{L} \cdot (\dot{\textbf{U}}_{n} - \dot{\textbf{U}}_{n+1}) + \frac{\Delta t^{3}}{120}\textbf{L} \cdot (\ddot{\textbf{U}}_{n}+ \ddot{\textbf{U}}_{n+1}). \ \ \ \  \ \ \ \  
    \label{ch33_hermiteintegration_hermite6rule_matrix}
\end{align}
\end{small}
Now, we transform the above equation by matrix-inversion to its \texttt{IM}plicit-\texttt{T}urned-\texttt{EX}plicit, \texttt{IMTEX} form and solve as, 
\begin{small}
\begin{align}
&\bigg[ \textbf{I} - \frac{\Delta t}{2}\textbf{L} + \frac{\Delta t^{2}}{10}\textbf{L}\cdot \textbf{L}  - \frac{\Delta t^{3}}{120}\textbf{L}\cdot\textbf{L}\cdot\textbf{L} \bigg] \cdot  \textbf{U}_{n+1}   = \nonumber \\ 
&\bigg[ \textbf{I} + \frac{\Delta t}{2}\textbf{L} 
+ \frac{\Delta t^{2}}{10}\textbf{L}\cdot \textbf{L}  + \frac{\Delta t^{3}}{120}\textbf{L}\cdot\textbf{L}\cdot\textbf{L} \bigg] \cdot \textbf{U}_{n},  \ \ \ 
\label{ch33_imtex_order6}
\end{align}
\end{small}
where equation \eqref{ch33_reduction_odes} is used to further simplify the \texttt{nth-} order time derivatives, for example,  $\dot{\textbf{U}} = \textbf{L} \cdot \textbf{U}$. Given the operator $\textbf{L}$ is linear, we can further optimise the expression above by writing it in its Horner-form \texttt{HF} and solve this as\footnote{We note an erroneous version of this equation appeared in \cite{Markakis:2023pfh}. It is not clear to this author if this affects their results as the order of the integrator they used to attain their physical results is not specified.}, 
\begin{align}
    &\textbf{U}_{n+1} = \textbf{U}_{n} \nonumber \\
    &+ (\Delta t \ \textbf{L}) \cdot \bigg( \textbf{I} + \frac{1}{60}(\Delta t \ \textbf{L}) \cdot (\Delta t \ \textbf{L} )\bigg) \cdot \textbf{HFH6}\cdot \textbf{U}_{n} 
    \label{ch3_hornerfor_sh6}
\end{align}
where the \texttt{HF} matrix is specifically given as, 
\begin{eqnarray}
    \textbf{HFH6} = \bigg( \textbf{I} - \frac{\Delta t}{2} \textbf{L}\cdot \bigg( \textbf{I} - \frac{\Delta t \ \textbf{L}}{5} \cdot \bigg( \textbf{I} - \frac{\Delta t}{12} \textbf{L} \bigg)\bigg)\bigg)^{-1}. 
    \label{ch3_hfh6}
\end{eqnarray}
We note in \ref{app_higherorder_imtex} all \texttt{IMTEX} evolution schemes are given up to \texttt{12th}-order. 
To obtain Runge-Kutta integration schemes one integrates a 1-point Taylor expansion of $\textbf{U}(t)$ about $t_{n}$, 
\begin{equation}
    \int^{t_{n+1}}_{t_{n}} \textbf{U}(t) dt = \sum^{l}_{m=1}\frac{\Delta t^{m}}{m!} \textbf{U}^{(m-1)}_{n} + \textbf{R}_{l}, 
    \label{ch3_1pointTaylor}
\end{equation}
where the remainder is given by, 
\begin{equation}
    \textbf{R}_{l} = \frac{\Delta t^{l+1}}{(l+1)!} \textbf{U}^{(l)}(t), \hspace{0.3cm} t \in [t_{n}, t_{n+1}],  
    \label{ch3_remainder_rk}
\end{equation}
and the $m-th$ derivative by, 
\begin{equation}
    \textbf{U}^{(m)}_{n} = \frac{d^{m}\textbf{U}(t)}{dt^{m}}\bigg|_{t=t_{n}}.  
    \label{ch3_rk_deri}
\end{equation}
To obtain the traditional \texttt{Runge-Kutta} \texttt{RK} scheme, we treat the derivatives in equation \eqref{ch3_rk_deri} as constant polynomial coefficients and eliminate them by evaluating the Taylor $\textbf{U}(t)$ at several points. For an order \texttt{7th} \texttt{RK} scheme the above yields the following \texttt{EX} scheme
\begin{eqnarray}
    \textbf{U}_{n+1} = \textbf{U}_{n} + (\textbf{HFRK7}-\textbf{I})\cdot \textbf{U}_{n}, 
    \label{ch3_hornerform_rk7}
\end{eqnarray}
where \texttt{HFRK7} matrix is given as, 
\begin{small}
\begin{align} &\textbf{HFRK7} = \textbf{I} + \textbf{A} \cdot \bigg( \textbf{I} + \textbf{A} \cdot \bigg(    \frac{1}{2}\textbf{I} +  \textbf{A} \cdot \bigg( \frac{1}{6}\textbf{I} + \nonumber \\ 
&\textbf{A} \cdot \bigg(\frac{1}{24}\textbf{I} + \textbf{A} \cdot \bigg( \frac{1}{120}\textbf{A} + \bigg( \frac{1}{720} \textbf{I} + \frac{1}{5040}\textbf{A}\bigg)\cdot \textbf{A} \bigg)  \bigg) \bigg) \bigg). 
    \label{ch3_hornerformer_rk7mat}
\end{align}
\end{small}
We note \texttt{EX RK} schemes can be found from orders \texttt{2nd} to \texttt{6th} in \ref{app_ex_rk}.  
\begin{table}
\begin{small}
\begin{tabular}{l|| c ||c  }
\textrm{Order of \texttt{HF} \texttt{IMTEX} \texttt{NH}}& 
\textrm{Accuracy, $\eta$}&
\textrm{Wall-clock times, s\footnotemark} \\ \hline
Order 2, \texttt{NH2} & $2.5 \times 10^{-14}$ & $5.1072372$ \ \  \\ 
Order 4, \texttt{NH4} & $7.3\times 10^{-15}$ & $4.9639754$ \ \  \\ 
Order 6, \texttt{NH6} & $3.8 \times 10^{-15}$ & $5.2465613$ \ \  \\ 
Order 8, \texttt{NH8} & $3.6 \times 10^{-14}$ & $5.4844739$ \ \  \\ 
Order 10, \texttt{NH10} & $2.2\times 10^{-14}$ & $6.1766433$ \ \  \\ 
Order 12, \texttt{NH12} & $4.0 \times 10^{-14}$ & $6.1136290$ \ \ \\ \hline
\textrm{Order of \texttt{HF} \texttt{NRK}}& 
\textrm{Accuracy, $\eta$}&
\textrm{Wall-clock times, s} \\ \hline
Order 2, \texttt{RK2} & $5.7\times 10^{-5}$ & $5.3512382$ \ \ \\ 
Order 3, \texttt{RK3} & $1.9\times 10^{-5}$ & $5.6920472$ \ \ \\ 
Order 4, \texttt{RK4} & $4.4\times 10^{-9}$ & $6.1857173$ \ \ \\ 
Order 5, \texttt{RK5} & $8.6 \times 10^{-10}$ & $6.0316653$ \ \ \\ 
Order 6, \texttt{RK6} & $ 1.9\times 10^{-13}$ & $6.2187127$ \ \ \\ 
Order 7, \texttt{RK7} & $ 2.2\times 10^{-14}$ & $6.0237433$ \ \ \\ \end{tabular}
\caption{Comparison between the accuracy and computational wall-clock times of the different explicit \texttt{EX} numerical integration schemes. Firstly, we give the results for the numerical \texttt{IMTEX} Hermite \texttt{NH} scheme optimised to Horner-Form \texttt{HF} and the optimised Runge-Kutta \texttt{RK} \texttt{EX}. It is observed that an order 7 \texttt{RK EX} scheme is required for competitive accuracy with even the lowest-order \texttt{IMTEX} Hermite scheme.} \label{tab_discontinuousTime-allorders}
\end{small}
\end{table}
\footnotetext{We note wall-clock times were recorded by measuring the time taken to perform the operations highlighted by numerical schemes in equations (\eqref{ch33_imtex_order6}, \eqref{ch3_hornerform_rk7}).}Remarkably, it is needed to extend the \texttt{Runge-Kutta} scheme to \texttt{7th} order to achieve the same results as with our lowest order integration scheme, as evidenced by Table \ref{tab_discontinuousTime-allorders} and Figure \ref{ch3_RK_HER}. Furthermore, it's observed that a plateau seems to be reached at about $10^{-15}$ accuracy for \texttt{IMTEX} Hermite schemes, this is because most of the residual error stems from the spatial discretisation which becomes more significant with increasing order of integration due to the increase in the number of operators which cumulative increases round-off error. This can potentially be mitigated with the implementation of an error minimisation algorithm such as compensation summation and performing the operations prior to the evolution with higher precision (higher than 16). At this juncture, we have not explored this but it is left as a subject for future work. The merits of \texttt{IMTEX} over traditionally \texttt{EX} explicit schemes are obvious, furthermore, \texttt{IMTEX} schemes preserve the symplectic structure and are energy conserving \textbf{both exactly and numerically} for quadratic Hamiltonian's. We refer the reader to \ref{app_imtex_props} for explicit analytical proof of these proprieties. It is with this conservation proprieties in mind that the \texttt{IMTEX} Hermite is selected as the time-stepper to resolve difficult $(3)$, implementing thus \texttt{DiscoTEX}. However, in the next sections, we will also show solutions where the implicit scheme is used, implementing \texttt{DiscoIMP}, this will be compared and discussed. Explicit schemes can also be constructed with the \texttt{EX RK HF} optimised schemes, and one refers the reader to \cite{phdthesis-lidia,discotexII,discoREX} where this has been implemented and compared, via the evolution algorithm \texttt{DiscoREX}. 

\subsubsection{Discontinuous time integration via Hermite interpolation higher-order formulas - resolving difficulty $(1)$ in the time dimension}\label{sec332_disco_time_int}

We now resume the problem raised by difficulty $(1)$, resolved in the spatial direction in Section \ref{sec_3_2_discojumps}, and address how to resolve the presence of a discontinuity in the time direction. This can be achieved by correcting the time-integration algorithms presented above for the non-smooth case. Here, it is shown, how once again, we can use the method of undetermined coefficients to accommodate for the discontinuous nature of the problem by writing the integrals as piecewise interpolating polynomials and using the \textbf{known} jump conditions to solve the algebraic system of equations. In this section, it is shown how to do this for \texttt{6th} order, as this was an important step when deciding on what order of algorithm to use when applying these methods to the self-force problem, such as in \cite{da2023hyperboloidal}. Furthermore, this work is complemented by a second paper \cite{discotexII} where the schemes from \texttt{2nd} to \texttt{12th} presented in this section are corrected to the non-smooth case, in that work we also compute the numerical solution to the distributionally sourced wave-equation up to order \texttt{12th}. The description of this procedure was done initially at second order by \cite{2014arXiv1406.4865M} and later improved to higher orders by this author and collaborators.
\begin{figure*}
    \centering
    \includegraphics{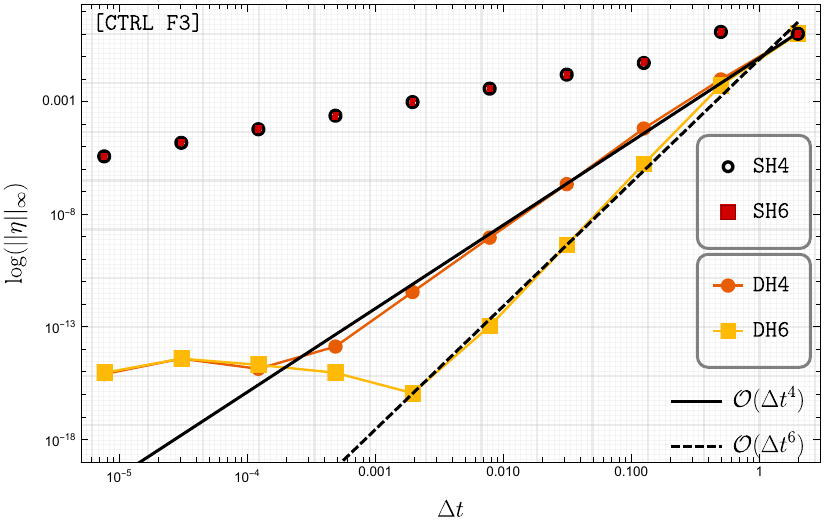}
    \caption{Numerical error associated with the evaluation of the integral in equation \eqref{ch3_legendreQP_time} and with the numerical scheme highlighted in equation \eqref{ch33_hermiteintegration_hermite6rule} corrected by incorporating the discontinuous behaviour through the discontinuous time-integration rule given in \eqref{sec2_disco_time_h6}. For an order-6 integration scheme, sixth-order convergence is observed as given by the \texttt{DH6} line. For a smooth integrator as given by line \texttt{SH6} inaccurate results are recorded. In a companion paper i.e. \cite{discotexII} we further compare results up to order-\texttt{12th}.}
    \label{ch3_DTh_h6}
\end{figure*}
We start by considering the first-order differential equation, 
\begin{equation}
    \frac{dy}{dt} = f(t, y(t)), 
    \label{ch3_disco_time_1st_ode}
\end{equation}
on a small time interval $[t_\nu, t_{\nu+1}]$. Applying the fundamental theorem of calculus we have, 
\begin{equation}
    y(t_{\nu+1}) - y(t_{\nu}) = \int_{t_{\nu}}^{t_{\nu+1}} f(t, y(t)) dt. 
    \label{ch3_disco_time_fundamental_theo_calculus}
\end{equation}
For the smooth case, as demonstrated in the previous section this amounts to a simple approximant such as that given in equation \eqref{ch33_hermiteintegration_hermite6rule}. For the non-smooth case, we incorporate the discontinuous behaviour by constructing the interpolant as a piecewise polynomial, 
\begin{equation}
    f(t, y(t)) \approx p(t) = p_{+}(t) \Theta(t - t_{\times}) +  p_{-}(t) \Theta(t_{\times} - t),  
    \label{ch3_disco_time_NS_interpo}
\end{equation}
where $t_{\times}$ is the point where the function is discontinuous such that $t_{\times} \in [t_{\nu}, t_{\nu+1}]$ and $f$ is approximated as $p_{+}$ in $[t_{\times}, t_{\nu + 1}]$ and $p_{-}$ in $[t_{\nu}, t_{\times}]$. Specifically one has
\begin{align}
    \label{ch3_disco_time_system_right}
    &p_{+}(t) = a_{0} + a_{1} t  + a_{2} t^{2}  + a_{3} t^{3}  + a_{4} t^{4} + a_{5} t^{5}, \\
    &p_{-}(t) = b_{0} + b_{1} t  + b_{2} t^{2}  + b_{3} t^{3}  + b_{4} t^{4} + b_{5} t^{5}. 
    \label{ch3_disco_time_system_left}
\end{align}
We have 12 unknown coefficients, however, as explained before and derived in Section \ref{sec2}, the jump conditions are known so we can impose the following collocation conditions:
\begin{align}
    &p_{-}(t_\nu) = f_{\nu},  p_{+}(t_{\nu+1}) = f_{\nu+1}, \\
    &p_{-}'(t_\nu) = df_{\nu},  \hspace{0.2cm} p_{+}'(t_{\nu+1}) = df_{\nu+1}, \\
    &p_{-}''(t_\nu) = ddf_{\nu},  \hspace{0.2cm} p_{+}''(t_{\nu+1}) = ddf_{\nu+1}, \\
    &p_{-}'''(t_\nu) = dddf_{\nu},  \hspace{0.2cm} p_{+}'''(t_{\nu+1}) = dddf_{\nu+1}, \\
    &p_{-}^{(4)}(t_\nu) = d4f_{\nu},  \hspace{0.2cm} p_{+}^{(4)}(t_{\nu+1}) = d4f_{\nu+1}, \\
    &p_{-}^{(5)}(t_\nu) = d5f_{\nu},  \hspace{0.2cm} p_{+}^{(5)}(t_{\nu+1}) = d5f_{\nu+1}, \\
    &p_{+}(t_\times)-  p_{-}(t_\times) = \textbf{J}_{0}, \\
    &p_{+}'(t_\times)-  p_{-}'(t_\times) = \textbf{J}_{1}, \\
    &p_{+}''(t_\times)-  p_{-}''(t_\times) = \textbf{J}_{2}, \\
    &p_{+}'''(t_\times)-  p_{-}'''(t_\times) = \textbf{J}_{3}, \\
    &p_{+}^{(4)}(t_\times)-  p_{-}^{(4)}(t_\times) = \textbf{J}_{4}, \\
    &p_{+}^{(5)}(t_\times)-  p_{-}^{(5)}(t_\times) = \textbf{J}_{5}, 
    \label{ch3_disco_time_12_collocation_condos}
\end{align}
where here the subscript notation $p^{(n)}$ denotes the \texttt{nth-} order derivative. With these 12 conditions, it is possible to solve for the 12 polynomial coefficients highlighted in equations (\eqref{ch3_disco_time_system_right}, \eqref{ch3_disco_time_system_left}) as a linear system of algebraic equations. Integrating both of the piecewise polynomials attained yields
\begin{small}
\begin{align}
&f(t)_{\texttt{NDH6}} = \frac{\Delta t }{2} \bigg(f(t_{\nu}) + f(t_{\nu+1} )  \bigg) + \nonumber \\
&\frac{\Delta t^{2} }{10}  \bigg(\dot{f}(t_{\nu}) - \dot{f}(t_{\nu+1} )  \bigg) + \frac{\Delta t^{3} }{120}  \bigg(\ddot{f}(t_{\nu}) + \ddot{f}(t_{\nu+1} )  \bigg) + \textbf{J}_{\texttt{H6}}(\Delta t_{\times}, \Delta t), \ \ \ \ \ \ \ \ \label{sec2_disco_time_h6}
\end{align}
\end{small}
where $\textbf{J}_{\texttt{H6}}(\Delta t_{\times}, \Delta t)$ is given by
\begin{align}
    &\textbf{J}_{\texttt{H6}}(\Delta t_{\times} , \Delta t) = \frac{1}{2}(\Delta t - 2 \Delta t_{\times} ) \textbf{J}_{0}  + \frac{1}{10} (\Delta t^{2} - 5 \Delta t \Delta t_{\times} + 5 \Delta t_{\times}^{2}) \textbf{J}_{1} + \nonumber\\
    &\frac{1}{120} (\Delta t - 2 \Delta t_{\times}) (\Delta t^{2} - 10 \Delta t \Delta t_{\times}  + 10 \Delta t_{\times} ^{2}) \textbf{J}_{2}  -  \nonumber \\
    &\frac{1}{120}(\Delta t - \Delta t_{\times}) \Delta t_{\times} (\Delta t^{2} - 5\Delta t \Delta t_{\times} + 5 \Delta t_{\times} ^{2}) \textbf{J}_{3} + \nonumber \\
    &\frac{1}{240} (\Delta t - 2\Delta t_{\times} ) (\Delta t - \Delta t_{\times} )^{2} \Delta t_{\times} ^{2}  \textbf{J}_{4} - \nonumber \\
    &\frac{1}{720} (\Delta t - \Delta t_{\times} )^{3}\Delta t_{\times}^{3}   \textbf{J}_{5}. \label{sec2_disco_time_Jh6}
\end{align}
Similarly to the previous section, the effects of increasing the order of the time-integration of the algorithm on the accuracy and computational times are thoroughly studied. As mentioned at the beginning of this section, we refer the reader to the companion paper \cite{discotexII} where all formulas can be found from  \texttt{2nd}- to \texttt{12th}-order. We start by considering the time-dependent Legendre polynomial
\begin{equation}
    f(t) = P_{5}(t)\Theta(t) + Q_{5}(t)\Theta(t), 
    \label{ch3_legendreQP_time}
\end{equation}
where $P_{5}(t), Q_{5}(t)$ are the fifth Legendre polynomials of the first and second kind respectively. Introducing a discontinuity at $t_{\times}=0$ in the interval $t \in [-0.55,0.45]$, equation \eqref{ch3_legendreQP_time} admits the following analytical jumps, 
\begin{align}
    &\textbf{J}_{0} = \frac{8}{15},  \ \textbf{J}_{1} = \frac{15}{8}, \ \textbf{J}_{2} = -16, \ \textbf{J}_{3} = -\frac{105}{2}, \\
    &\textbf{J}_{4} = 384, \ \textbf{J}_{5} = 945,  \ \textbf{J}_{6} = -3840, \ \textbf{J}_{7} =0, \\
    &\textbf{J}_{8} = -46080, \ \textbf{J}_{9} = 0,  \ \textbf{J}_{10} = -1935360, \ \textbf{J}_{11} = 0. 
    \label{ch3_legende_anal_jumps}
\end{align}
\begin{figure}
\includegraphics[width=84mm]{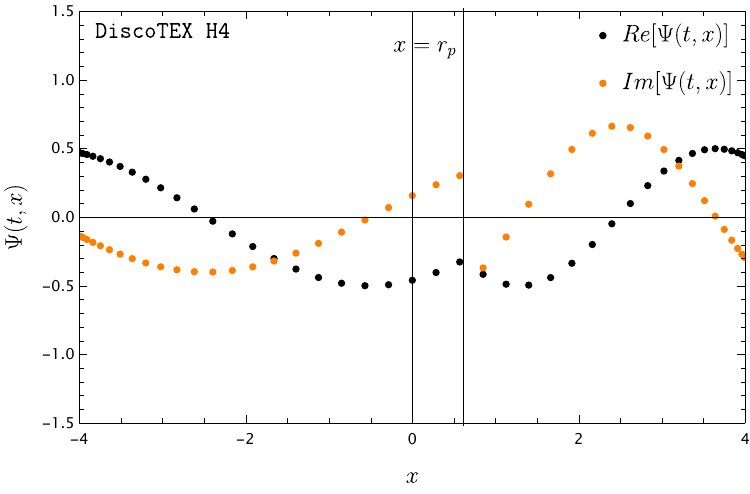}
\quad
\includegraphics[width=84mm]{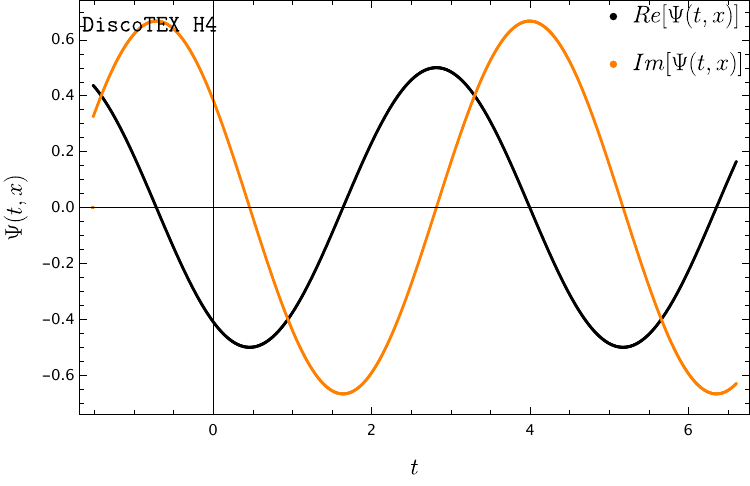}
\caption{Numerical weak-form solution to $\Psi(t,x)$ obtained via \texttt{DiscoTEX} with moving radiation boundary conditions. \textbf{Left:} Numerical field $\Psi(t,x)$ for a point-particle in time-dependent linear motion $r_{p}(t) = v t$, where $v$ is the particle's velocity. Specifically, here $r_{p} \approx 0.0634$ at the coordinate-time $t=0.1902$ and $v=1/4$. 
\textbf{Right:} Waveform for the point-particle computed on the numerical domain $x \in [-4,4]$ and $t \in [-1.521, 6.602]$. \label{ch3_discotex_tr_wave_snap}} 
\end{figure} 
Integrating analytically the above polynomial in the interval $t \in [-0.55,0.45]$ one gets $\approx 0.1125883303464025$. From Figure \ref{ch3_DTh_h6} it is observed that the error associated with the discontinuous time-integration rule retrieved in equation \eqref{sec2_disco_time_h6} scales with order 6 as expected. Furthermore, the results are complemented by the companion paper \cite{discotexII} showing the discontinuous time-integration rule scaling with the expected orders for up to \texttt{12th-} order. In all these plots we further include the result attained if we were to perform a smooth integration as given by equation \eqref{ch33_hermiteintegration_hermite6rule} without the discontinuous corrections. To further investigate the effect increasing the order of the integrator had on numerical results the numerical integrals at $\Delta t = 0.0001$ are evaluated at the final time $t = 0.45$ for all the \texttt{12th} order discontinuous Hermite interpolation schemes. As it is observed in Table \ref{tab_discontinuousTime-allorders} increasing after \texttt{4th} order for the same interval the scheme's accuracy 
\begin{table}
\begin{small}
\begin{tabular}{l|| c ||c  }
\textrm{Order of \texttt{NDH}}& 
\textrm{Accuracy, $\eta$}&
\textrm{Wall-clock times, s} \\ \hline
Order 2, \texttt{NDH2} & $3.1396164973385154\times 10^{-9}$ & $0.4410348$ \ \  \\ 
Order 4, \texttt{NDH4} & $3.733124920302089 \times 10^{-15}$ & $3.2397477$ \ \  \\ 
Order 6, \texttt{NDH6} & $3.733124920302089 \times 10^{-15}$ & $7.083297$ \ \  \\ 
Order 8, \texttt{NDH8} & $3.733124920302089 \times 10^{-15}$ & $11.21189$ \ \  \\ 
Order 10, \texttt{NDH10} & $3.733124920302089 \times 10^{-15}$ & $15.4928979$ \ \  \\ 
Order 12, \texttt{NDH12} & $3.733124920302089 \times 10^{-15}$ & $19.2839227$ \ \ \\ 
\end{tabular}
\caption{Here we give the accuracy and wall-clock computational times of the corrected Hermite integration schemes interpolated to accommodate for time functions which have a discontinuity. We evaluate the function at the same final time of $t_{f} =0.45 $ and a time-step of $\Delta t =0.0001$. After 4 orders the same accuracy is attained for this interval while the computational task-time significantly increases. These results are complemented by those in paper \cite{discotexII}.}\label{tab_discontinuousTime-allorders}
\end{small}
\end{table}
saturates with no improvement, while the computational time significantly increases. These results ultimately motivated our selection of an order-4 scheme when performing the numerics in \cite{da2023hyperboloidal}, and the rest of the work in this manuscript. 

Finally, it is paramount to highlight, that, for numerical implementations, another user-specifiable optimisation control factor emerges:
\begin{enumerate}
    \item \textcolor{gray}{\texttt{[CTRL F\ref{controlfactor1}]} Number of \texttt{N} nodes;}
    \item \textcolor{gray}{\texttt{[CTRL F\ref{controlfactor2}]} Number of \texttt{J} jumps;}
    \item \texttt{[CTRL F3]} the time $\Delta t $ step-size, \label{controlfactor3}
\end{enumerate}
which should be studied before conducting numerical simulations. We also highlight the jumps in the time direction, $\texttt{J}$ \textbf{are fixed and determined by the order of the discontinuous time-integration algorithm} as evidenced by equations ~(\eqref{sec2_disco_time_h6}, \eqref{sec2_disco_time_Jh6}) and the respective equations for the different order integrators given in the companion paper \cite{discotexII}.  

\subsection{Numerical weak-form solutions to the distributionally sourced wave equation via the \texttt{DiscoTEX} numerical solver }\label{Sec34_DiscoTEX_for_wave}
With the technology studied in in Sections \ref{sec_3_2_discojumps}-\ref{sec_discotex_justime} we have demonstrated: 
\begin{itemize}
    \item Difficulty $(1)$ - can be accurately resolved in both the space (Section \ref{sec_3_2_discojumps}) and time (Section \ref{sec_discotex_justime}) directions through \underline{dis}continuous \underline{co}llocation methods in both dimensions - \texttt{Disco}; 
    \item Difficulty $(3)$ - can be accurately resolved with suitable Hermite implicit \texttt{IM}, \texttt{-IMP}, or implicit-turned-explicit \texttt{IM\underline{TEX}}, \texttt{-TEX}, numerical integrators, Section \ref{sec_imtex}. 
    \item implement solver - \texttt{DiscoTEX} or related solvers such as - \texttt{DiscoIMP} \cite{phdthesis-lidia} or \texttt{DiscoREX} \cite{phdthesis-lidia, discoREX}. 
\end{itemize}
\begin{figure*}
\includegraphics[width=94mm]{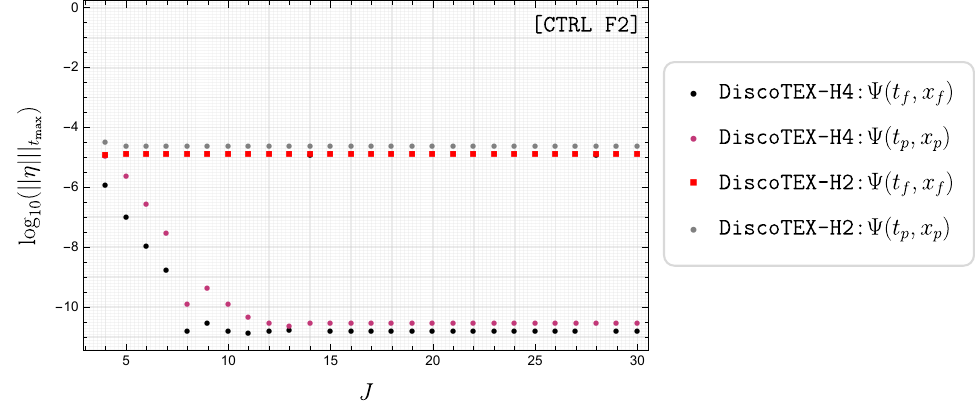}
\quad
\includegraphics[width=94mm]{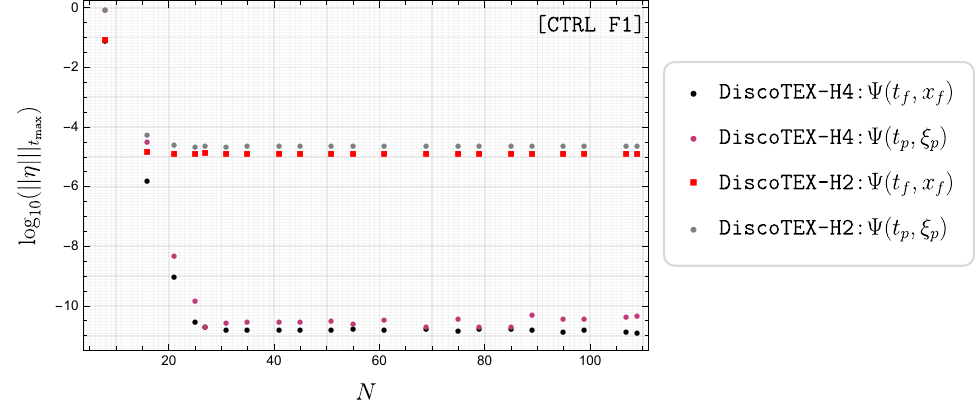}
\caption{Numerical convergence studies for optimal implementation of \texttt{DiscoTEX} with an order $2$ and $4$ \texttt{IMTEX} Hermite \texttt{H2/H4} time integrator to meet requirements \texttt{[REQ 1], [REQ 2]} with high accuracy. \texttt{[REQ 1]} is assessed by computing the error of the numerical solution $\Psi(t_{f}, x_{f})$ at $\sigma = 0$, i.e at infinity $\mathscr{I}^{+}$ against the exact solution as given by equation \eqref{ch34_distr_wavequation_delta1_plus_delta0_sol}. \texttt{[REQ 2]} is assessed by computing the error of the numerical solution $\Psi(t_{p}, x_{p})$ via \texttt{DiscoTEX}'s interpolator at the point-particle position $x_{p}(t) = x_{p}$ as given in equation \eqref{ch34_psi_discotex_interpol} against the  exact solution evaluate at the same $x_{p}$ particle location. \textbf{Left:} Convergence study determining the optimal number of jumps as $J=20$. \textbf{Right:} Convergence study determining the optimal number of Chebyshev nodes at around $N=34$. \label{ch3_discotex_wave_conv_test}} 
\end{figure*} 
The aim is now to show how we can adapt the numerical method-of-lines recipe framework to incorporate for these corrections which accurately resolve the discontinuous nature of the problem. We will show requirements \texttt{[REQ 1], [REQ 2]} can be accurately met by implementing \texttt{DiscoTEX:} i.e. using discontinuous collocation methods to resolve the particle discontinuity in space and time, followed by the use of an \texttt{IM\underline{TEX}} Hermite time-integrator. For the first time, it will be shown how \texttt{DiscoTEX} can be used to complete solutions to equations of Type I as given in equations (\eqref{ch2_distr_wavequation_delta1_plus_delta0}, \eqref{ch2_distr_wavequation_delta1_plus_delta0_sol}), i.e  
\begin{align}
    \label{ch34_distr_wavequation_delta1_plus_delta0}
    & \textrm{Type I:} \ \ \square \Psi(t,x) = F(t) \delta'(x - x_{p}) + G(t)\delta(x - x_{p}), \\
    & \Psi(t,x) = -\frac{1}{2} \sin \mathcal{\theta} + \frac{1}{2} \iu \gamma^{2} [\dot{x}_{p} + \text{sgn}(x-x_{p})] \cos \mathcal{\theta}.\label{ch34_distr_wavequation_delta1_plus_delta0_sol}
\end{align}
For simplicity, as stated in Section \ref{sec2}, we define $F(t) = -\iu G(t) = -\iu \cos{t}$. Additionally, we note we defined $r_{p} = x_{p}$ to avoid confusion with $r$, the polar coordinate. In the body of this work, as explained the method-of-lines framework is used, hence, to reduce it to a first-order system of equations the master function $\Pi(t,x) = \partial_{t} \Psi(t,x)$ is introduced, with the equation above, now given as 
\begin{align}
    \label{ch34_WaveTR_red_pde_sys_psi}
    &\partial_{t}\Psi = \Pi, \\
    &\partial_{t}\Pi = \partial_{x}^{2} \Psi -  F(t) \delta'(x - x_{p}) - G(t)\delta(x - x_{p}). 
    \label{ch34_WaveTR_red_pde_sys_pi}
\end{align}
As explained in \cite{da2023hyperboloidal, 2014arXiv1406.4865M} the idea is to use discontinuous collocation methods, as thoroughly explained in Section \ref{sec_3_1} through equations (\eqref{smooth_interpol_gen} - \eqref{cha3_rec_relation_mjumps_PhysicalChart_RWZ}), and approximate both master functions as
\begin{align}
    \label{ch34_psi_disco}
    & \Psi(t,x) \approx \sum^{N}_{j=0} \bigg[ \Psi_{j} + \Delta_{\Psi}(x_{j} - x_{p}(t) ; x - x_{p}(t)) \bigg] \pi_{j}(x), \\
    \label{ch34_pi_disco}
    & \Pi(t,x) \approx \sum^{N}_{j=0} \bigg[ \Pi_{j} + \Delta_{\Pi}(x_{j} - x_{p}(t) ; x - x_{p}(t)) \bigg] \pi_{j}(x),
\end{align}
and the differential operators as, 
\begin{align}
    &\partial_{(x)}^{n}(t,x)\Psi|_{x = x_{i}} =p^{(n)}(x)  \\
    &= \sum^{N}_{j=0}  D^{(n)}_{ij}  \Psi_{j} + s_{i}^{(n)}(t), \ \ \ \ 
    \label{ch34_diff_op_disco}
\end{align}
with $s^{(n)}_{i}(t)$ given as, 
\begin{equation}
    s^{(n)}_{i}(t) = \sum^{N}_{j=0}D^{(n)}_{ij} \Delta_{\Psi}\big(x_{j} - x_{p}(t); x_{i} - x_{p}(t)\big),  
    \label{ch34_sb2_SpaceSpource}
\end{equation}
where the user-specifiable higher order jumps are given by a recurrence relation of the type given in equation \eqref{cha3_rec_relation_mjumps_PhysicalChart_RWZ}. This equation, as explained in \ref{app2}, is derived for the case where the wave equation operator is in $(t,r)$ coordinates. Here, as explained, we work with the version transformed to $(t,x)$ coordinates through a tortoise coordinate transformation as defined in equation \eqref{ch1_bhpt_wave}. The methodology explained in \ref{app2} and originally given in \cite{da2023hyperboloidal}, in $(t,x)$ greatly simplifies, to, 
\begin{align}
    \label{ch34_wavetx_REC_j0}
    &J_{0}(t) = \gamma^{2}(t) F(t), \\
    \label{ch34_wavetx_REC_j1}
    &J_{1}(t) = \gamma^{2}(t)  \big[G(t) - 2 \dot{J}_{0}(t) - \ddot{x}_{p}(t) J_{0}(t)   \big], \\
    &J_{m+2}(t) = \gamma^{2}(t)  \big[ \ddot{J}_{m}(t) - 2\dot{x}_{p}\dot{J}_{m+1}(t) - \ddot{x}_{p}(t) J_{m+1}(t) \big], 
    \label{ch34_wavetx_REC_jmp2}
\end{align}
where, 
\begin{align}
    \gamma^{2}(t) = \frac{1}{(1- \dot{x}_{p}^{2})}.  
    \label{lorenz_factor}
\end{align}

This equation was first included in the second version of \cite{2014arXiv1406.4865M}.\footnote{The author derived this equation independently as part of PhD work and it appears in \cite{reviews-lidia}. This author started with the first version of \cite{2014arXiv1406.4865M} in 2019 \cite{markakis2019time, OBoyleTalk}.} Despite multiple efforts by multiple people throughout several years, \cite{2014arXiv1406.4865M,o2022timeQMUL, o2022time, o2023discontinuous} this equation will only now be fully demonstrated and proven here thanks to the existence of the exact equation as given in equation \eqref{ch34_distr_wavequation_delta1_plus_delta0_sol} originally derived by Field \cite{field2009discontinuous, field2010persistent} over a decade ago. 
As highlighted in Sections \ref{ch31_red_1order}, \ref{sec_discotex_justime} in equations (\eqref{ch3_general_pdes}, \eqref{ch33_reduction_odes}) we start by reducing the problem to a first-order system of differential equations, i.e., 
\begin{align}
    &\frac{d\textbf{U}}{dt} = \textbf{L} \cdot \textbf{U} + \mathcal{S} + \Upsilon, \ \ \  \textbf{U} = \begin{pmatrix}
        \Psi(t) \\ \Pi(t)\end{pmatrix},\nonumber \\
    &\mathcal{S} = \begin{pmatrix}  0 \\ \textbf{s}_{\Psi}(t) +  \textbf{s}_{\Pi}(t)\end{pmatrix},
    \ \ \Upsilon = \begin{pmatrix} \Upsilon_{\Psi}(t) \\ \Upsilon_{\Pi}(t)  \end{pmatrix}, \label{ch34_red_1ode} \
\end{align}
\begin{figure*}
\centering
\includegraphics[width=150mm]{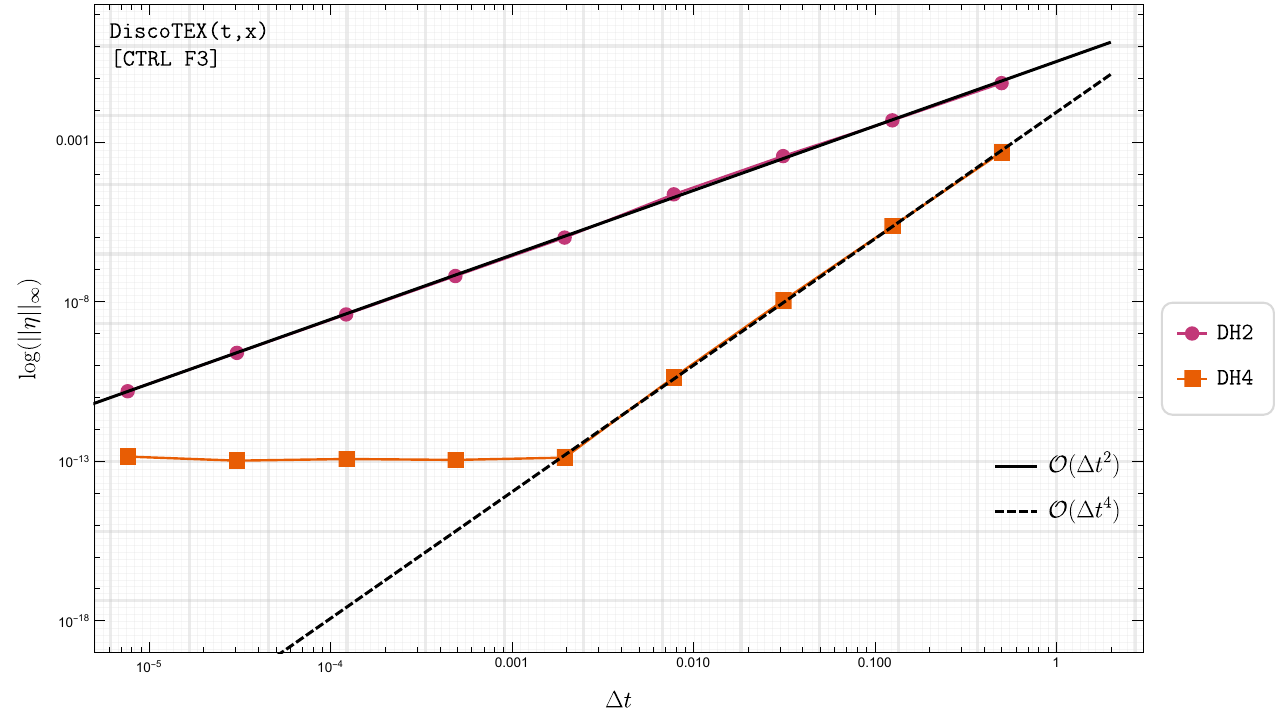}
\caption{Numerical error associated with computation of the numerical weak-form solution $\Psi(t,x_{f})$ against the exact solution in Eq.(\ref{ch34_distr_wavequation_delta1_plus_delta0_sol}) with both the discontinuous Hermite integrator of order-2 and order-4, respectively \texttt{DH2,DH4}. As expected, for an order-2 integration scheme second-order convergence is observed as given by the \texttt{DH2} line and for an order-4 integration scheme fourth-order convergence is observed as given by the \texttt{DH4} line. \label{ch3_discotex_wavetr_stepsize}} 
\end{figure*} 
where all functions were written with respect to time. The matrix evolution operator associated with equation \eqref{ch34_distr_wavequation_delta1_plus_delta0_sol} is, 
\begin{align}
    & \textbf{L} = \begin{pmatrix} 0 & I \\ \textbf{L}_{1} & \textbf{L}_{2} \end{pmatrix} = \hspace{0.2cm}  \begin{pmatrix} 0 & I \\ \textbf -\partial_{x}^{2} & 0 \end{pmatrix}. 
    \label{ch34_waveTR_l_op}
\end{align}
For convenience we define,   
\begin{align}
    \begin{pmatrix} \Upsilon_{\Psi}(t) =  - [[\Psi]]_{i}\Xi_{i}= -J_{0}(t)*\Xi_{i}\\ \Upsilon_{\Pi}(t) = - [[\Pi]]_{i}\Xi_{i} = - \mathbb{J}_{0}(t)*\Xi_{i}\end{pmatrix}
    \label{ch34_td_corrections_tofinal_sol}
\end{align}
with $J_{0}(t)$ is as defined in eq.~\eqref{ch34_wavetx_REC_j0}. $\mathbb{J}_{0}(t)$ is the jump associated with the time derivative of the master $\Psi(t,x)$ i.e $\Pi(t,x)$, this can be generically derived as
\begin{align}
    \label{ch34_rec_time_PI_jumps}
    \mathbb{J}(t) = \mathbb{J}_{m}(t) = \partial_{t}(J_{m}(t)) - \dot{x}_{p}(t) J_{m+1}(t), \\
    \mathbb{J}(t)|_{m=0} = \mathbb{J}_{0}(t) = \partial_{t}(J_{0}(t)) - \dot{x}_{p}(t) J_{1}(t). 
    \label{ch34_rec_time_PI_jump}
\end{align}
We refer the reader to \ref{app2} specifically equation \eqref{app_temp_approx} for a clarification on the derivation of this relation. $[[\Psi]]_{i}$ is as defined in equation \eqref{ch2_timedependent_jump}, and $[[\Pi]]_{i}$ is given by 
\begin{align}
    \label{ch3_time-dependent_pi_jump}
    &[[\Pi]]_{i} = \Pi^{+}(t,x_{p}(t)) - \Pi^{-}(t,x_{p}(t)), \hspace{0.2cm} t \rightarrow \frac{x_{i}}{v}\\
    &\Xi_{i} = \Theta(t_{n+1} - t_{i})\Theta(t_{i} - t_{n}),
    \label{ch34_switch}
\end{align}
where $\Theta$ is as defined in equation \eqref{ch2_HeavisideStepFunction} and $\Xi_{i}$ acts as a \texttt{switch} which turns on these corrections in the time direction for both the $\Psi(t)$ and $\Pi(t)$ master functions when the particle worldline $x_{p}(t)$ crosses the $i-$th grid-point as they are integrated through \texttt{DiscoTEX} at a time $t_{i} \in [t_{n}, t_{n+1}]: x_{p}(t)= x_{i}$. One emphasises here a minus sign has been introduced to highlight the correction needed when considering the jumps in the time-direction, this is crucial for the algorithm to work. In \cite{2014arXiv1406.4865M} eq.(C5), even though not necessarily wrong as there are no derivatives of the Dirac distribution as per their eq.(47), the term $\Upsilon_{\Psi}(t)$ it's just zero as $J_{0}(t) = 0$, and hence only equation \eqref{ch34_rec_time_PI_jump}.\footnote{In \cite{2014arXiv1406.4865M} this was only partially explained for problems of Type II as in equations (\eqref{ch2_distr_wavequation_delta0}, \eqref{ch2_distr_wavequation_delta0_sol}). The attempts at understanding this in a simpler toy model by those authors in \cite{o2022timeQMUL, o2022time, o2023discontinuous} are not satisfactory. See footnote 10 of this manuscript.}

As in equation \eqref{ch34_waveTR_l_op} $\textbf{L}$ only has a differential operator $-\partial_{x}^{2}$ and the source $\mathcal{S}$ simplifies to
\begin{align}
    \label{ch34_red_1ode} 
    &\mathcal{S} = \begin{pmatrix}  0 \\ \textbf{s}_{\Psi}(t) +  \textbf{s}_{\Pi}(t) = 0 \end{pmatrix}, \\
    &\mathcal{S} = \begin{pmatrix}  0 \\ \textbf{s}_{\Psi}(t) \end{pmatrix}, 
\end{align}
where $\textbf{s}_{\Psi}(t)$ is simply given by 
\begin{align}
    \label{ch34_disco_ope}
    &\textbf{s}_{\Psi}(t) =  \tilde{\textbf{s}}^{(2)}_{\Psi}(t), \\
    &\tilde{\textbf{s}}^{(2)}_{\Psi}(t) = \sum^{N}_{j=0} D^{(2)} _{ij} \big[ \Delta_{\Psi}(x_{j}-x_{p};x_{i} - x_{p}) \big]. 
    \label{s2psi_chi}
\end{align}

\subsubsection{Numerical weak-form solution via \texttt{DiscoTEX} with moving radiation boundary conditions}\label{Sec34_DiscoTEX_for_wave_tx}
\begin{figure*}
\includegraphics[width=90mm]{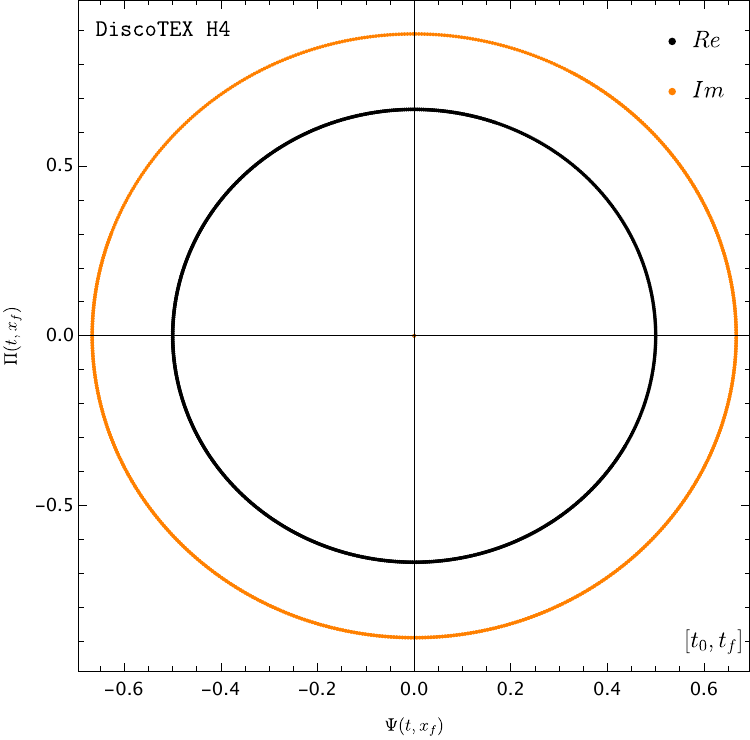}
\quad
\includegraphics[width=90mm]{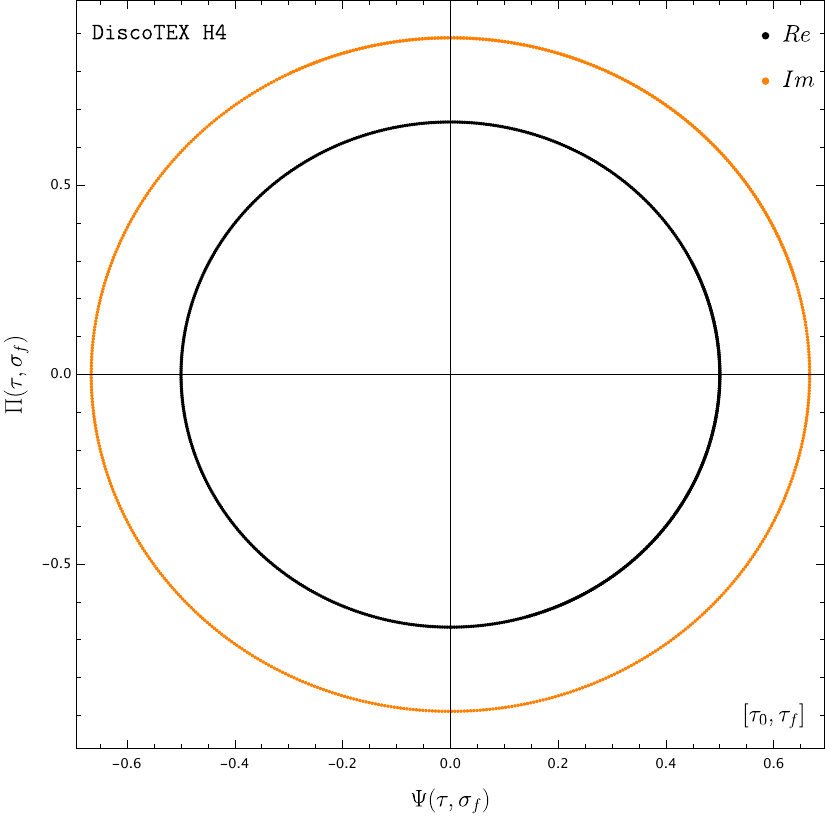}
\caption{Phase portraits for the numerical weak-form solutions obtained via \texttt{DiscoTEX} with an order $4$ \texttt{IMTEX} Hermite \texttt{H4} time integrator in both the $(t,r)$ and $(\tau,\sigma)$ coordinate charts. 
\textbf{Left:} Phase portrait of the numerical weak-form solutions to the field $\Psi(t,x_{f})$, $\Pi(t,x_{f})$ evaluated in the numerical time interval $t \in [-1.521,6.602] $ at the last grid point. 
\textbf{Right:} Phase portrait of the numerical weak-form solutions to the field $\Psi(\tau,\sigma_{f})$, $\Pi(\tau,\sigma_{f})$ evaluated in the numerical time interval $\tau \in [-1.52,4.50] $ at the last grid point. It is noteworthy to point out that unlike in \cite{da2023hyperboloidal}, Fig.(2),  here one uses the exact solutions as initial data, whereas in our previous work, we used trivial initial data. \label{ch3_discotex_wave_phaseps}} 
\end{figure*}
In this work \texttt{DiscoTEX} will be implemented with both moving radiation boundary conditions \cite{reviews-lidia} and with hyperboloidal slicing \cite{reviews-lidia} recently used in \cite{da2023hyperboloidal}. As demonstrated in \cite{field2009discontinuous, 2014arXiv1406.4865M}, we can impose purely outgoing solutions at the first $x_{i=0}$ and last $x_{i=N}$ grid-points, respectively as
\begin{align}
    & \Pi(t, x=a) = \partial_{x}\Psi(x=a), \\
    & \Pi(t, x=b) = -\partial_{x}\Psi(x=b).
    \label{ch34_mrcbs}
\end{align}
Specifically, the wave operator now corrects to
\begin{align}
    & \textbf{L} = \begin{pmatrix} 0 & I \\ \textbf{L}_{1} & \textbf{L}_{2} \end{pmatrix} = \hspace{0.2cm}  \begin{pmatrix} 0 & I \\ \textbf -\partial_{x}^{2} & M_{ij} \end{pmatrix}, 
    \label{ch34_waveTR_l_op_boundaryCondos}
\end{align}
where $M_{ij}$ and the source term $\mathcal{S}$ are defined as in equation (B9) of 
\cite{2014arXiv1406.4865M} and thus eq.~\eqref{ch34_disco_ope} now corrects to, 
\begin{align}
    \label{ch34_red_1ode} 
    &\mathcal{S} = \begin{pmatrix}  0 \\ \textbf{s}_{\Psi}(t) + \big( \textbf{s}_{\Pi}(t) \neq 0 \big)\end{pmatrix} 
   = \begin{pmatrix}  0 \\ \textbf{s}_{\Psi}(t) + \textbf{s}_{\Pi}(t) \end{pmatrix}, 
\end{align}
with $\textbf{s}_{\Psi}(t)$ and $\textbf{s}_{\Pi}(t)$ being, respectively defined as, 
\begin{align}
    \label{ch34_disco_ope_new}
    &\textbf{s}_{\Psi}(t) =   \textbf{s}^{(2)}_{\Psi}(t) + \bar{\textbf{s}}^{(1)}_{\Psi}(t) + 
    \bar{\textbf{s}}^{(2)}_{\Psi}(t) , \\
    \label{s2psi_chi}
    &\textbf{s}^{(2)}_{\Psi}(t) = \sum^{N-1}_{j=1}D^{(2)} _{ij} \big[ \Delta_{\Psi}(x_{j}-x_{p};x_{i} - x_{p}) \big],\\
    \label{s2psi_chi_bar}
    &\bar{\textbf{s}}^{(2)}_{\Psi}(t) = \sum^{N}_{j=0} \bar{D}^{(2)} _{ij} \big[ \Delta_{\Psi}(x_{j}-x_{p};x_{i} - x_{p}) \big],\\
    \label{s1psi_chi_bar}
    &\bar{\textbf{s}}^{(1)}_{\Psi}(t) = \sum^{N}_{j=0} \bar{D}^{(1)} _{ij} \big[ \Delta_{\Psi}(x_{j}-x_{p};x_{i} - x_{p}) \big],\\
    \label{ch34_disco_ope_l22_new}
    &\textbf{s}_{\Pi}(t) =  \bar{\textbf{s}}^{(1)}_{\Pi}(t), \\
    &\bar{\textbf{s}}^{(1)}_{\Pi}(t) = \sum^{N}_{j=0} \bar{D}^{(1)} _{ij} \big[ \Delta_{\Pi}(x_{j}-x_{p};x_{i} - x_{p}) \big], 
    \label{s1pi_bar}
\end{align} 
where the bar notation denotes the corrections to the differential operators introduced by the moving radiation boundary conditions framework adapted in \cite{2014arXiv1406.4865M}. Unlike in their work, where an implicit order-2 Hermite integration scheme was used, we know work with an order-4 Hermite time-stepper, thus the time-derivative of $s_{i}(t)$ naturally arises from the substitution $\dot{\textbf{U}} = \textbf{L} \cdot \textbf{U} + \mathcal{S}$ that is used. An analogous replacement has been done in the smooth case when we went from equation \eqref{ch33_hermiteintegration_hermite6rule_matrix} to \eqref{ch33_imtex_order6}, for example, the only difference now is the additional term $\mathcal{S}$ arising from the discontinuous machinery.
Explicitly, 
\begin{align}
\textbf{U}_{n+1} - \textbf{U}_{n} &= \int^{t_{n+1}}_{t_{n}} \bigg( \textbf{L} \cdot \textbf{U} + \mathcal{S} \bigg) dt \nonumber \\ 
&= \frac{\Delta t}{2} \textbf{L} \cdot (\textbf{U}_{n} + \textbf{U}_{n+1}) + \frac{\Delta t}{2} (\mathcal{S}_{n} + \mathcal{S}_{n+1}) \nonumber \\
&+\frac{\Delta t ^{2}}{12} \textbf{L} \cdot (\dot{\textbf{U}}_{n} - \dot{\textbf{U}}_{n+1}) + \frac{\Delta t^{2}}{12} (\dot{\mathcal{S}}_{n} - \dot{\mathcal{S}}_{n+1}) \nonumber \\ 
&= \frac{\Delta t}{2} (\textbf{U}_{n} + \textbf{U}_{n+1}) + \frac{\Delta t}{2} (\mathcal{S}_{n} + \mathcal{S}_{n+1}) \nonumber \\
&+\frac{\Delta t^{2}}{12} \textbf{L} \cdot (\textbf{L} \cdot \textbf{U}_{n} - \textbf{L} \cdot \textbf{U}_{n+1}) + \frac{\Delta t^{2}}{12} \textbf{L} \cdot (\mathcal{S}_{n} - \mathcal{S}_{n+1}) \nonumber \\ 
&+ \frac{\Delta t^{2}}{12} (\dot{\mathcal{S}}_{n} - \dot{\mathcal{S}}_{n+1}), 
    \label{discoHermitereduced}
\end{align}
thus the \texttt{4th}- order Hermite scheme introduces the new state vector, 
\begin{align}
    \label{ch34_dot_sourceh4} 
    &\dot{\mathcal{S}} = \begin{pmatrix}  0 \\ \textbf{st}_{\Psi}(t) + \textbf{st}_{\Pi}(t) \end{pmatrix}, 
\end{align}
\begin{figure*}
\includegraphics[width=84mm]{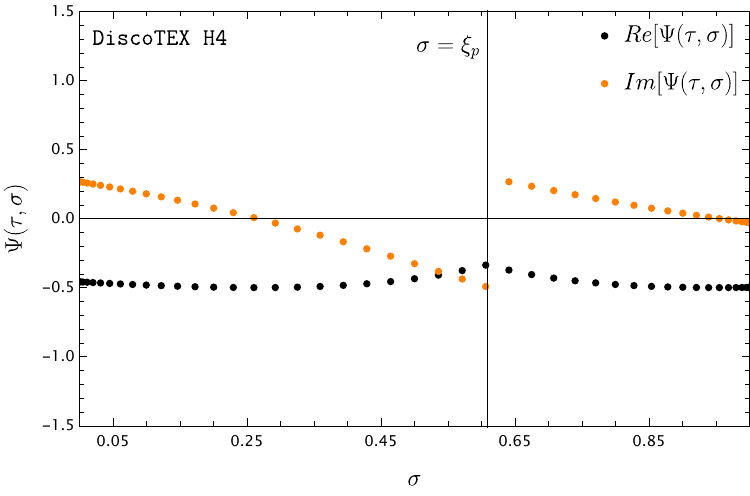}
\quad
\includegraphics[width=84mm]{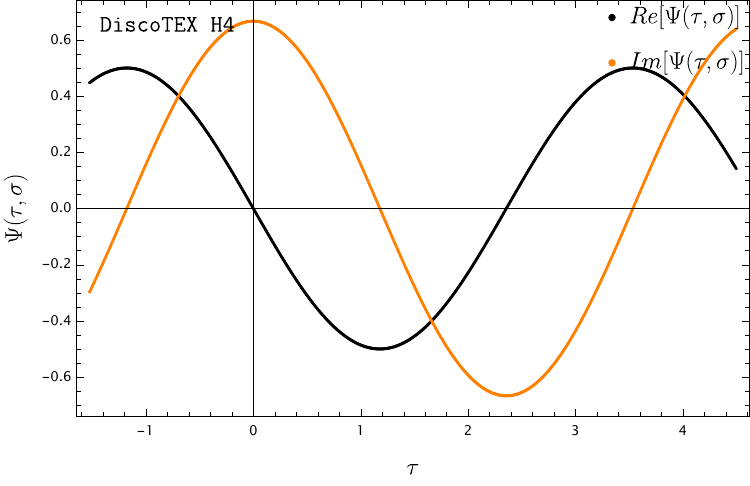}
\caption{Numerical weak-form solution to $\Psi(\tau,\sigma)$ obtained via \texttt{DiscoTEX}. \textbf{Left:} Numerical field $\Psi(\tau,\sigma)$ for a point-particle in time-dependent linear motion $\xi_{p}(\tau_{c}) $, where $v$ is the particle's velocity. Specifically, here $\xi_{p} \approx 0.608$ at the coordinate-time $\tau_{c} \approx0.876$ and $v=1/4$. 
\textbf{Right:} Waveform for the point-particle computed on the numerical domain $\sigma \in [0,1]$ and $\tau \in [-1.52, 4.50]$.\label{ch3_discotex_hyperboloidal_wave_snap}} 
\end{figure*} 
This is demonstrated further in the companion paper \cite{discotexII} where higher-order time-integration schemes are implemented. If we were to keep the notation they used for equation (B9) for the source term we would now have, 
\begin{align}
    &\dot{\textbf{s}}_{i}(t) = \dot{s}^{(2)}(t) + D^{(2)}_{i0} \bigg(A_{0}\dot{s}^{(1)}_{0}(t) + B_{0}\dot{s}^{(1)}_{N}(t) 
    + C_{0}\dot{s}^{(2)}_{0}(t)  + E_{0}\dot{s}^{(2)}_{N}(t) \nonumber \\
    & - C_{0}\dot{r}^{(1)}_{0}(t)  +  E_{0}\dot{r}^{(1)}_{N}(t) \bigg) +
    D^{(2)}_{iN} \bigg(A_{N}\dot{s}^{(1)}_{0}(t) + B_{N}\dot{s}^{(1)}_{N}(t) + C_{N}\dot{s}^{(2)}_{0}(t) \nonumber \\
   &   + E_{N}\dot{s}^{(2)}_{N}(t) - C_{N}\dot{r}^{(1)}_{0}(t)  +  E_{N}\dot{r}^{(1)}_{N}(t) \bigg), 
\end{align}
however, here as we have done before, we replaced their $\dot{s}^{(1,2)}(t)$ and $\dot{r}^{(1)}(t)$ terms, by
\begin{align}
     \label{ch34_disco_ope_new_dot}
    &\textbf{st}_{\Psi}(t) =   \textbf{st}^{(2)}_{\Psi}(t) + \bar{\textbf{st}}^{(1)}_{\Psi}(t) + 
    \bar{\textbf{st}}^{(2)}_{\Psi}(t) , \\
      \label{s2psi_chi_dot}
     &\textbf{st}^{(2)}_{\Psi}(t) = \sum^{N-1}_{j=1}D^{(2)} _{ij} \big[\partial_{t} \Delta_{\Psi}(x_{j}-x_{p};x_{i} - x_{p}) \big],\\
    \label{s2psi_chi_bar_dot}
    &\bar{\textbf{st}}^{(2)}_{\Psi}(t) = \sum^{N}_{j=0} \bar{D}^{(2)} _{ij} \big[ \partial_{t} \Delta_{\Psi}(x_{j}-x_{p};x_{i} - x_{p}) \big],\\
    \label{s1psi_chi_bar_dot}
    &\bar{\textbf{st}}^{(1)}_{\Psi}(t) = \sum^{N}_{j=0} \bar{D}^{(1)} _{ij} \big[ \partial_{t} \Delta_{\Psi}(x_{j}-x_{p};x_{i} - x_{p}) \big],\\
    &\bar{\textbf{st}}^{(1)}_{\Pi}(t) = \sum^{N}_{j=0} \bar{D}^{(1)} _{ij} \big[ \partial_{t} \Delta_{\Pi}(x_{j}-x_{p};x_{i} - x_{p}) \big]. 
    \label{s1pi_bar_dot}
\end{align}

The partial-derivative with respect to time explicitly acts on the $g$ vector as per equations (\eqref{ch3_b2_higherorderjumps}-\eqref{ch3_sb2_SpaceSpource}). Explicitly $\partial_{t}\Delta_{\Psi}$ is, 
\begin{align}
&\partial_{t}\Delta_{\Psi}(x_{j}-x_{p};x_{i} - x_{p})  = \nonumber \\
&\bigg[\Theta(x_{i} - x_{p}(t)) - \Theta(x_{j} - x_{p}(t))\bigg] \partial_{t}\big(g_{\Psi}(x_{j} - x_{p}(t))\big)  \; \rm{when} \; x = x_{i}, 
\end{align}
with the time-derivative of $g_{\Psi}(x-x_{p}(t))$, given in equation \eqref{ch3_sb2_gVector_expSum}, being obtained as, 
\begin{align}
    \partial_{t}\big(g_{\Psi}(x - x_{p}(t))\big) &= \frac{\partial g_{\Psi}(x-x_{p}(t))}{\partial t} = \nonumber \\ 
    &= \frac{\partial}{\partial t} \bigg[ \sum^{M}_{m=0}\frac{J(t)}{m!}(x_{j} - x_{p} (t))^{m}\bigg] \nonumber \\
    &= \sum^{M}_{m=0} \bigg[ \frac{1}{m!}\dot{J}(x_{j} - x_{p} )^{m} - \frac{m}{m!} \dot{x}_{p} J (x_{j} - x_{p} )^{m-1}  \bigg].  
    \label{gdt}
\end{align}
The term $\partial_{t} \Delta_{\Pi}(x_{j}-x_{p};x_{i} - x_{p})$ is obtained analogously, however, the jump is now on the master field $\Pi(t,x)$, as it has been defined in equation \eqref{ch34_rec_time_PI_jumps}, thus 
\begin{align}
    \partial_{t}\big(g_{\Pi}(x - x_{p}(t))\big) &= \frac{\partial g_{\Pi}(x-x_{p}(t))}{\partial t} = \nonumber \\ 
     &= \frac{\partial}{\partial t} \bigg[ \sum^{M}_{m=0}\frac{\mathbb{J}(t)}{m!}(x_{j} - x_{p} (t))^{m}\bigg] \nonumber \\
    &= \sum^{M}_{m=0} \bigg[ \frac{1}{m!}\dot{\mathbb{J}}(x_{j} - x_{p} )^{m} - \frac{m}{m!} \dot{x}_{p} \mathbb{J} (x_{j} - x_{p} )^{m-1}  \bigg].  
    \label{gpidt}
\end{align}
To emphasize the different jumps involved we have added the subscripts $\Psi/\Pi$ to the $g$ vector accordingly. All these equations alternatively re-define $\bar{s}_{i}(t)$ given in the last equation after equation (B9) of \cite{2014arXiv1406.4865M}. 

Furthermore, here we highlight that terms in equations (\eqref{s2psi_chi_bar}, \eqref{s1psi_chi_bar}, \eqref{s1pi_bar}, \eqref{s2psi_chi_bar_dot}, \eqref{s1psi_chi_bar_dot}, \eqref{s1pi_bar_dot}) are only evaluated at the end points as they emerge from corrections to the radiation boundary conditions as given after equation (B9) in \cite{2014arXiv1406.4865M}. 

Finally, combining the discontinuous collocation machinery through the spatial corrections described above and correcting the time-integration Hermite \texttt{IMTEX} order-4 scheme equations (\eqref{ch3_hornerfor_sh4}, \eqref{ch3_hfh4}) with the discontinuous collocation correction in time given by equations (\eqref{app3_disco_time_h4}, \eqref{app3_disco_time_Jh4}), we have the final evolution scheme, \texttt{DiscoTEX}, namely, 
\begin{align}
    &\textbf{U}_{n+1} = \textbf{U}_{n}+  \textbf{HFH4} \cdot  \bigg[ \textbf{A} \cdot  \bigg[ \textbf{U}_{n} + \frac{\Delta t}{12} \bigg(\textbf{s}_{n} - \textbf{s}_{n+1} \bigg)  \bigg] \nonumber  \\
    &+  \frac{\Delta t}{2} \bigg( \textbf{s}_{n} + \textbf{s}_{n+1} \bigg) + \frac{\Delta t^{2}}{12} \bigg(\textbf{st}_{n} - \textbf{st}_{n+1}  \bigg) \nonumber \\
    & + \Upsilon(t) +  \textbf{J}_{\texttt{H4}} (\Delta t_{\times}, \Delta t) \Xi  \bigg], 
    \label{ch3_discoTEX4_Wave_tr}
\end{align} 
in the $(t,x)$ coordinate chart. It is left to define the jumps in $\textbf{J}_{\texttt{H4}}(\Delta t_{\times}, \Delta t)$. As described in Section \ref{sec332_disco_time_int}, complemented by \ref{app_disco_time_h4}, a fourth-order discontinuous time-integration rule, described by equations (\eqref{app3_disco_time_h4}, \eqref{app3_disco_time_Jh4}) scheme will impose four jump conditions associated with the jumps of the field equation in the time direction, i.e. the time jumps. From equation \eqref{ch34_WaveTR_red_pde_sys_psi} it is clear there will be a time jump associated with the integration of $\Pi(t,x)$, and from equation \eqref{ch34_WaveTR_red_pde_sys_pi} another due to the integration of $\partial_{x}^{2} \Psi$. 
This prompts the effective vector, 
\begin{align}
    \textbf{J}_j(t) = 
    \begin{pmatrix}
    \mathbb{JJ}_j(t)\\
    \mathcal{JJ}_j(t)
\end{pmatrix} =  \begin{pmatrix}
    {\mathbb{J}_{0}(t),\mathbb{K}_{0}(t),\mathbb{L}_{0}(t),\mathbb{M}_{0}(t)} \\
    {\mathcal{J}_{0}(t), \mathcal{K}_{0}(t),\mathcal{L}_{0}(t),\mathcal{M}_{0}(t)}
\end{pmatrix}\bigg|_{t\rightarrow \frac{x_{i}}{v}},
    \label{ch34_wavetr_timejumps}
\end{align}
where here $j = 0,..3,$ and it depends on the order of the integration scheme used. We refer the reader to the companion paper \cite{discotexII} where time integration is used at higher orders. As explained just before Section \ref{sec_discotex_justime} ends and Section \ref{Sec34_DiscoTEX_for_wave} starts, we use an order-4 scheme so only 4 jumps will be necessary. Concretely $\mathbb{J}_{0}$ is as given in equations (\eqref{ch34_rec_time_PI_jumps}, \eqref{ch34_rec_time_PI_jump}), the other jumps are given as, 
\begin{align}
    &\mathbb{K}_{m}(t) =  \partial_{t}(\mathbb{J}_{m}(t)) - \dot{x}_{p}(t) \mathbb{J}_{m+1}(t),\\
    &\mathbb{K}(t)|_{m=0} = \mathbb{K}_{0}(t)= \partial_{t}(\mathbb{J}_{0}(t)) - \dot{x}_{p}(t) \mathbb{J}_{1}(t)= \mathbb{JJ}_1, \\
    &\mathbb{L}(t) =  \partial_{t}(\mathbb{K}_{m}(t)) - \dot{x}_{p}(t) \mathbb{K}_{m+1}(t),\\
    &\mathbb{L}_0(t)|_{m=0} = \mathbb{L}_{0}(t)= \partial_{t}(\mathbb{K}_{0}(t)) - \dot{x}_{p}(t) \mathbb{K}_{1}(t)= \mathbb{JJ}_2, \\
    &\mathbb{M}_m(t) = \partial_{t}(\mathbb{L}_{m}(t)) - \dot{x}_{p}(t) \mathbb{L}_{m+1}(t), \\
    &\mathbb{M}_0(t)|_{m=0} = \mathbb{M}_{0}(t)= \partial_{t}(\mathbb{L}_{0}(t)) - \dot{x}_{p}(t) \mathbb{L}_{1}(t) = \mathbb{JJ}_3, 
\end{align}
\begin{figure*}
\includegraphics[width=94mm]{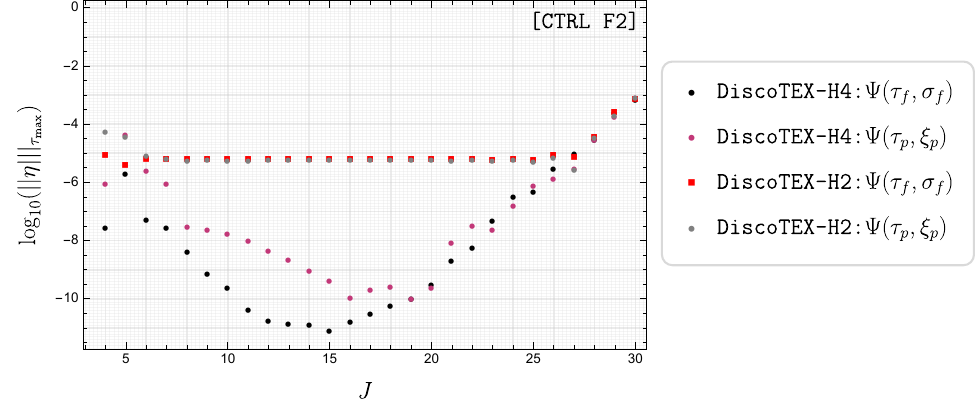}
\quad
\includegraphics[width=94mm]{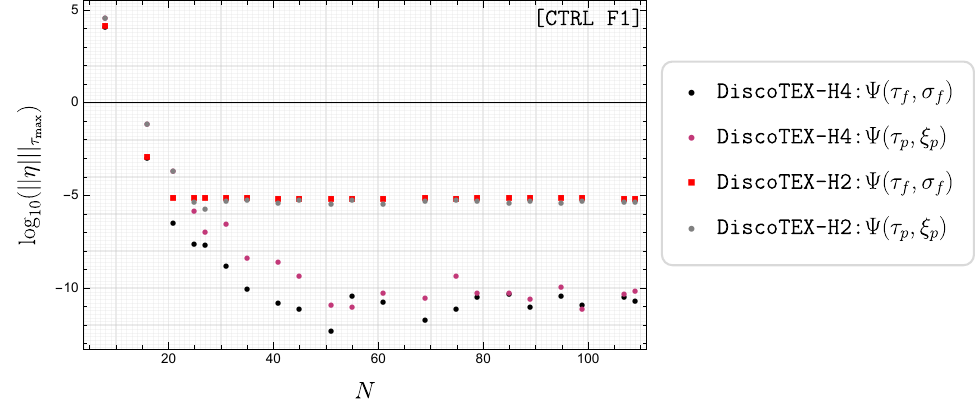}
\caption{Numerical convergence studies for optimal implementation of \texttt{DiscoTEX} with an order $2$ and $4$ \texttt{IMTEX} Hermite \texttt{H2/H4} time integrator to meet requirements \texttt{[REQ 1], [REQ 2]} with high accuracy. \texttt{[REQ 1]} is assessed by computing the error of the numerical solution $\Psi(\tau_{f}, \sigma_{f})$ at $\sigma = 1$, i.e. at infinity $\mathscr{I}^{+}$ against the exact solution as given by Eq.(\ref{ch34_distr_wavequation_delta1_plus_delta0_hyper}). \texttt{[REQ 2]} is assessed by computing the error of the numerical solution $\Psi(\tau_{p}, \xi_{p})$ via \texttt{DiscoTEX}'s interpolator at the point-particle position $\xi_{p}(\tau_{c}) = \xi_{p}$ as given in Eq.(\ref{ch3_discoTEX4_Wave_hyper}) against the  exact solution evaluate at the same $\xi_{p}$ particle location. \textbf{Left:} Convergence study determining the optimal number of jumps as $J=15$.
\textbf{Right:} Convergence study determining the optimal number of Chebyshev nodes at around $N=45$. \label{ch3_discotex_hyperboloidal_wave_conv_test}} 
\end{figure*} 
The time jump $\mathcal{J}_{0}(t)$ associated with the differential operator of $\textbf{L}$ on the master field $\Psi(t,x)$ i.e. the term $\partial^{2}_{x} \Psi$ is

\begin{align}
    \label{d2x_timejump_gen1}
    &\mathcal{J}(t) = - J_{\bar{m}}(t), \\
    &\mathcal{J}(t)|_{\bar{m} = 2} = \mathcal{J}_{0}(t) = - J_{2}(t) = \mathcal{JJ}_{0},
     \label{d2x_timejump_gen2}
\end{align}
where $\bar{m} = \{2, \cdots, m-2\}$ and $J_{2}(t)$ is given by equation \eqref{ch34_wavetx_REC_jmp2}, explicitly stated in equation \eqref{wave_tx_j2} of \ref{app_space_jumps}.  The terms $\mathcal{JJ}_{j=1,2,3}$ are then given as
\begin{align}
    \label{tx_k_time}
    &\mathcal{K}_0(t) = \partial_{t}(\mathcal{J}_{\bar{n}}(t)) - \dot{x}_{p}(t) \mathcal{J}_{\bar{n}+1}(t), \\
    &\mathcal{K}(t)|_{\bar{n}=0} = \mathcal{K}_{0}(t)= \partial_{t}(\mathcal{J}_{0}(t)) - \dot{x}_{p}(t) \mathcal{J}_{1}(t) = \mathcal{JJ}_1,\\
    &\mathcal{L}(t) =  \partial_{t}(\mathcal{K}_{\bar{n}}(t)) - \dot{x}_{p}(t) \mathcal{K}_{\bar{n}+1}(t),\\
    &\mathcal{L}_0(t)|_{\bar{n}=0} = \mathcal{L}_{0}(t)= \partial_{t}(\mathcal{K}_{0}(t)) - \dot{x}_{p}(t) \mathcal{K}_{1}(t) = \mathcal{JJ}_2, \\
    &\mathcal{M}(t) = \partial_{t}(\mathcal{L}_{\bar{n}}(t)) - \dot{x}_{p}(t) \mathcal{L}_{\bar{n}+1}(t), \\
    &\mathcal{M}_0(t)|_{\bar{n}=0} = \mathcal{M}_{0}(t)= \partial_{t}(\mathcal{L}_{0}(t)) - \dot{x}_{p}(t) \mathcal{L}_{1}(t) = \mathcal{JJ}_3, 
    \label{tx_m_time}
\end{align}
where we introduced the dummy variable $\bar{n}$ reflecting the new set of higher order jumps used as stated in equation \eqref{d2x_timejump_gen1}. Finally the vectors $\textbf{J}_{0}(t), \textbf{J}_{1}(t), \textbf{J}_{2}(t), \textbf{J}_{3}(t)$ define the necessary time jumps for successful implementation of the discontinuous time-integration rule described by  $\textbf{J}_{\texttt{H4}}(\Delta t_{\times}, \Delta t)$ in equations (\eqref{app3_disco_time_h4}, \eqref{app3_disco_time_Jh4}) and equations (C8-C32) of \cite{da2023hyperboloidal} (in that work $\textbf{J}_{\texttt{H4}}(\Delta t_{\times}, \Delta t)$ = 0, as explained, as we modelled a circular orbit).\footnote{The term $\textbf{J}_{\texttt{H2/H4}}(\Delta t_{\times}, \Delta t)$ in \cite{o2023discontinuous} which those authors call $K_{H2/H4}$ for example in Eq.(4.51)), is not correct in \cite{o2023discontinuous, o2022timeQMUL,o2022time}. Their first time jump \cite{o2023discontinuous}, $K_{0}$ should, match here $\mathbf{J}_{0}$ and it does not. Assuming that they were initialising their scheme from jumps start counting in $1,...,3$, their presentation could be correct, albeit with a few necessary modifications. However, these authors explicitly state that they use equations ((3.6)--(3.9)), in which case their implementation is indeed incorrect. In this section, this is discussed in detail and explicitly given in \ref{app_timejumps}. Furthermore, the implementation here is correct as evidenced by the highly accurate results and the verification, \textit{for the first time}, of the \textbf{full} recurrence relation given in equations (A5-A8) of \cite{2014arXiv1406.4865M} in various coordinates charts with \texttt{DiscoTEX} and \texttt{DiscoIMP}.}  Additionally, we explicitly give the first few jumps in both the spatial and time direction through \ref{app_space_jumps}-\ref{app_timejumps}. An alternative evolution algorithm is given in \ref{app_discoimp} using the purely implicit \texttt{IM} Hermite time-integration scheme, hereafter named \texttt{DiscoIMP}. Shortly these implementations will be discussed, in the following section. Before the numerical calibrations necessary for optimal implementation of \texttt{DiscoTEX}, and \texttt{DiscoIMP}, the problem will be considered in a new coordinate chart as an attempt at resolving difficulty $(2)$ emerging when modelling (X)/(E)MRIs. 

\subsubsection{Numerical weak-form solution via \texttt{DiscoTEX} with hyperboloidal slicing as an alternative to incorporation of boundary conditions}\label{Sec34_DiscoTEX_for_wave_hyperboloidal}

The development of \texttt{DiscoTEX} was mostly motivated with application to numerical black hole perturbation theory in mind and hence one of the main goals is the ability to accurately extract information at both the black hole horizon and infinity. As explained in \cite{da2023hyperboloidal}, deciding on what boundary conditions to implement has led to what is known as the outer radiation boundary problem, where it is not entirely clear if the extrapolated information at these boundaries will be contaminated from the implementation of boundary conditions such as those described above. In the past decade, the gravitational self-force numerical relativity \footnote{Here the term numerical relativity is loosely used, usually it is associated with the implementation of numerical methods to solving the non-linear Einstein equations in \texttt{3D+1}.} community has decided to use hyperboloidal methods to resolve this problem motivated by the work of \cite{zenginouglu2010asymptotics}. There have been numerous implementations of these slices, though here one highlights the works of \cite{bernuzzi2011binary, vishal2023towards} as they had previously attempted this problem by the implementation of boundary conditions described above with the same numerical strategies to resolve difficulties $(1)$ and $(3)$, also see Table \ref{tab_emris_timedomain}. The core idea of hyperboloidal slicing is to parameterise the spacetime with a compact radial coordinate defined on a hyperboloidal time surface, essentially the coordinate map $(t \rightarrow t(\tau, \sigma), x \rightarrow x(\sigma))$ \cite{zenginouglu2009gravitational} is applied. The technique here used to construct these slices is the \enquote{scri-fixing technique} \cite{zenginouglu2008hyperboloidal}. The coordinate chart is mapped as 
\begin{align}
    &t = \tau - H(\sigma), \hspace{0.2cm} x = \frac{1}{2} \bigg( \frac{1}{\sigma} + \ln{(1-\sigma)} - \ln{\sigma} \bigg). 
    \label{ch34_minimal_gauge_tx}
\end{align}

It is important to highlight that whereas the map of the time coordinate is exactly the same as we described in \cite{da2023hyperboloidal}, here the tortoise coordinate $x$ is directly mapped and not $r$. The height function $H(\sigma)$ is given by, 
\begin{align}
    &H(\sigma) = \frac{1}{2} \bigg[ \ln{(1-\sigma)} - \frac{1}{\sigma} + \ln{\sigma} \bigg]
    \label{ch34_heightfunction}
\end{align}
as originally introduced by \cite{ansorg2016spectral, jaramillo2021pseudospectrum, macedo2018hyperboloidal, macedo2020hyperboloidal}. As in equation \eqref{ch34_distr_wavequation_delta1_plus_delta0_sol}, we now have the following exact equation of Type I, 
\begin{align}
    & \textrm{Type I:} \ \ \square \Psi[\tau,\sigma] = F(\tau) \delta'(\sigma - \xi_{p}) + G(\tau)\delta(\sigma - \xi_{p}).
    \label{ch34_distr_wavequation_delta1_plus_delta0_hyper}
\end{align}
The master function and related functions transform from $\Psi(t,x) \rightarrow \Psi (t(\tau,\sigma), x(\sigma))$, which here are loosely simply denoted by $\Psi(\tau,\sigma)$, the final wave equation can be retrieved by applying the following equation as given originally in \cite{jaramillo2021pseudospectrum}, 
\begin{small}
\begin{align}
    &\bigg[  \bigg(1 - \bigg( \frac{H'}{x'}\bigg) \bigg) \partial_{\tau}^{2} - \frac{2}{x'}\bigg( \frac{H'}{x'}\bigg)\partial_{\tau}\partial_{\sigma} - \frac{1}{x'}\bigg(\frac{H'}{x'} \bigg)'\partial_{\tau}
    - \frac{1}{x'} \partial_{\sigma}\bigg( \frac{1}{x'} \partial_{\sigma} \bigg)\bigg] \Psi(\tau,\sigma) = 0, 
    \label{ch34_trans_equation}
\end{align}
\end{small}
where prime indicates derivative with respect to $\sigma$. The equation now reads,
\begin{align}
    \square \Psi(\tau,\sigma) = \mathcal{S}(\tau,\sigma) , 
    \label{ch34_hyper_wave_equation}
\end{align}
with, 
\begin{align}
    \square \Psi = \bigg( -\Gamma(\sigma) \partial^{2}_{\tau}  + \varepsilon(\sigma)\partial_{\sigma}\partial_{\tau} + \varrho(\sigma)\partial_{\tau} + \chi(\sigma)\partial^{2}_{\sigma} + \iota(\sigma)\partial_{\sigma}  \bigg) \Psi. 
    \label{hyper_dalembert_op}
\end{align}
The coefficients above read, 
\begin{align}
\label{ch34_sb2_diffOperators_gamma}
&\Gamma(\sigma) = - 4 \sigma^{2}(-1+\sigma^{2}),\\
\label{ch34_sb2_diffOperators_vareps}
&\varepsilon(\sigma) = -4(-1+\sigma) \sigma^{2}(-1 + 2\sigma^{2}),  \\
\label{ch34_sb2_diffOperators_varrho}
&\varrho(\sigma) =  -8(-1+\sigma)\sigma^{3}, \\
\label{ch34_sb2_diffOperators_chi}
&\chi(\sigma) =-4(-1 + \sigma)^{2}\sigma^{4},\\
&\iota(\sigma) = -4 (-1+\sigma)\sigma^{3}(-2+3\sigma),
\label{ch34_sb2_diffOperators_iota}
\end{align}
where, as in \cite{da2023hyperboloidal}, it's emphasised that the function $\chi(\sigma)$ vanishes at the boundaries where $\sigma = \{0,1\}$, reflecting the outflow behaviour automatically enforced by this hyperboloidal slicing at the future null infinity and the horizon and thus naturally imposes boundary conditions at $\sigma = 0$ and $\sigma=1$. The term $\mathcal{S(\tau)}$ is place-hold notation to denote the distributional source function as only a time-dependent function as before. We will shortly explain how this is handled. The first-order reduction of the PDE into a couple of differential equations goes as explained in equation \eqref{ch34_red_1ode} in the new coordinate chart, but now the operator $\textbf{L}$ is described as, 
\begin{align}
    \label{ch34_hyperboloidal_L1_p}
    &\textbf{L}_{1} = \frac{1}{\Gamma(\sigma)}\bigg( \chi(\sigma)\partial^{2}_{\sigma} + \iota(\sigma)\partial_{\sigma} \bigg), \\
    &\textbf{L}_{2} = \frac{1}{\Gamma(\sigma)}\bigg( \varepsilon(\sigma) \partial_{\sigma} - \varrho(\sigma) \bigg),
    \label{ch34_hyperboloidal_L2_p}
\end{align}
where for convenience as described in \cite{da2023hyperboloidal}, the tilde notation is introduced to denote the division of the coefficients in the operators by $\Gamma(\sigma)$, e.g., $\tilde{\varepsilon}(\sigma) =\varepsilon(\sigma)/\Gamma(\sigma)$. We also stress that the wave-equation here is different from the equation described in \cite{da2023hyperboloidal}. Here, we study just the flat spacetime and not the Schwarzschild's. The term $\Upsilon$ is defined as equation \eqref{ch34_td_corrections_tofinal_sol} but now with respect to the $\tau$ quantity. We simply have, 
\begin{align}
    \begin{pmatrix} \Upsilon_{\Psi}(\tau) =   [[\Psi]]_{i}\Xi_{i}= J_{0}(\tau)*\Xi_{i}\\ \Upsilon_{\Pi}(\tau) =  [[\Pi]]_{i}\Xi_{i} = \mathbb{J}_{0}(\tau)*\Xi_{i}\end{pmatrix}.
    \label{ch34_td_corrections_tofinal_sol_hyper}
\end{align}
The reader is directed to \ref{app_hyper_rec_hyper_jumps}, where it is explained how $J_{0}(\tau), J_{1}(\tau)$ are obtained in the new chart by the correct application of the Dirac delta-distribution proprieties. For clarity, the time jumps are similarly defined as, 
\begin{align}
    \label{ch34_rec_time_PI_jumps_hyper}
    \mathbb{J}(\tau) = \mathbb{J}_{m}(\tau) = \partial_{\tau}(J_{m}(\tau)) - \dot{\xi}_{p}(\tau) J_{m+1}(\tau), \\
    \mathbb{J}(\tau)|_{m=0} = \mathbb{J}_{0}(\tau) = \partial_{\tau}(J_{0}(\tau)) - \dot{\xi}_{p}(\tau) J_{1}(\tau),
    \label{ch34_rec_time_PI_jump_hyper}
\end{align}
\begin{figure*}
\centering
\includegraphics[width=150mm]{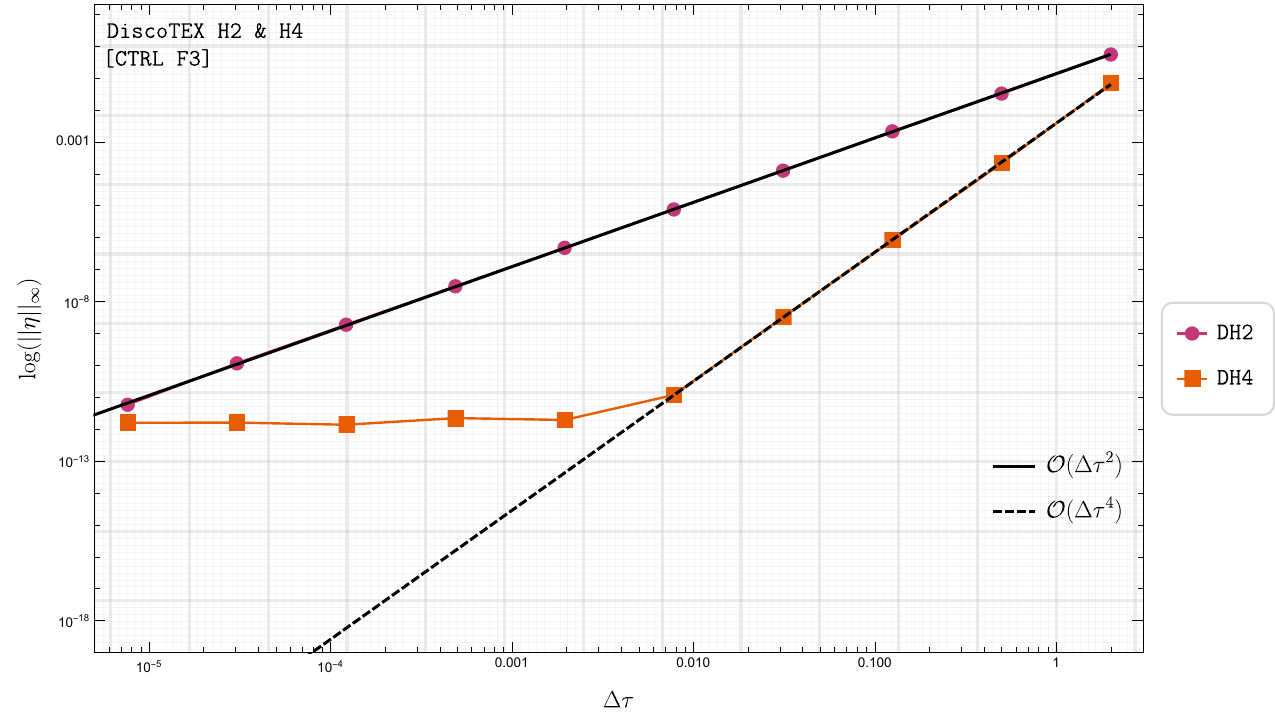}
\caption{Numerical error associated with computation of the numerical weak-form solution $\Psi(\tau,\sigma_{f})$ against the exact solution in equation \eqref{ch34_distr_wavequation_delta1_plus_delta0_hyper} with both the discontinuous Hermite integrator of order-2 and order-4, respectively \texttt{DH2,DH4}. As expected, for an order-2 integration scheme second-order convergence is observed as given by the \texttt{DH2} line and for an order-4 integration scheme fourth-order convergence is observed as given by the \texttt{DH4} line.\label{ch3_discotex_hyperboloidal_stepsize}} 
\end{figure*} 
where $\xi_{p}(\tau)$ denotes the new particle location in the hyperboloidal coordinate chart and is defined at the end of \ref{app_initialising_jumps}. The vector $\mathcal{S}$ is now described by, 
\begin{align}
    \tilde{\textbf{s}}(\tau) = \begin{pmatrix}
        0 \\
        \tilde{s}_{\Psi}^{(1)} + \tilde{s}_{\Psi}^{(2)} + \tilde{s}_{\Pi}^{(1)}
    \end{pmatrix}, 
    \label{s_spatial_disc_vector_hyper_Wave}
\end{align}
as described in \cite{da2023hyperboloidal} and $\dot{\mathcal{S}}$ by, 
\begin{align}
    \tilde{\textbf{st}}(\tau) = \begin{pmatrix}
        0 \\
        \tilde{st}_{\Psi}^{(1)} + \tilde{st}_{\Psi}^{(2)} + \tilde{st}_{\Pi}^{(1)}
    \end{pmatrix}. 
    \label{s_spatial_disc_vector_hyper_Wave}
\end{align}
In this work, however, as emphasised through equations (\eqref{ch34_minimal_gauge_tx} - \eqref{ch34_hyperboloidal_L2_p})) please note the operators are different, even though we use the same notation. Despite the different governing wave-like equation, here describing Minkowski's spacetime, the implementation of the discontinuous collocation machinery is similar due to the differential operators describing the evolution operator $\textbf{L}_{1}, \textbf{L}_{2}$ being the same i.e. we have in $\textbf{L}_{1}$, two spatial first and second-order differential operators, and $\textbf{L}_{2}$ we have one first-order spatial operator. For completion, we include here these operators, originally given in \cite{reviews-lidia, da2023hyperboloidal}. The differential operators corrected with the discontinuous collocation algorithm are given by, 
\begin{align}
  &\tilde{\upchi}(\sigma) \partial_{\sigma}^{2}\Psi\bigg|_{\sigma=\sigma_{i}} \nonumber \\
  &= \sum^{N}_{j=0} \bigg( \text{diag}(\tilde{\upchi}(\sigma_{i})) \cdot D^{(2)} \bigg)_{ij} \big[ \Psi_{j} + \Delta_{\Psi}(\sigma_{j}-\sigma_{p};\sigma_{i} - \sigma_{p}) \big], \nonumber   \\
  \label{chid2_discretization}
\end{align} 
\begin{align}
  &\tilde{\iota}(\sigma) \partial_{\sigma}\Psi\bigg|_{\sigma=\sigma_{i}} = \nonumber \\
  &\sum^{N}_{j=0} \bigg( \text{diag}(\tilde{\iota}(\sigma_{i})) \cdot D^{(1)} \bigg)_{ij} \big[ \Psi_{j} + \Delta_{\Psi}(\sigma_{j}-\sigma_{p};\sigma_{i} - \sigma_{p}) \big],  \nonumber   \\
  \label{iotad1_discretization}
\end{align}
\begin{align}
  &\tilde{ \upvarepsilon}(\sigma) \partial_{\sigma}\Pi\bigg|_{\sigma=\sigma_{i}} = \nonumber \\
  &\sum^{N}_{j=0} \bigg( \text{diag}(\tilde{ \upvarepsilon}(\sigma_{i})) \cdot D^{(1)} \bigg)_{ij} \big[ \Pi_{j} + \Delta_{\Pi}(\sigma_{j}-\sigma_{p};\sigma_{i} - \sigma_{p}) \big].  \nonumber   \\
  \label{alpha_d1_discretization}
\end{align}
The explicit form of $\tilde{\textbf{s}}(\tau)$ in equation \eqref{s_spatial_disc_vector_hyper_Wave} containing all the necessary corrections to the differential operators is
\begin{align}
    &\tilde{s}_{\Psi}^{(1)} = \sum^{N}_{j=0} \bigg( \text{diag}(\tilde{\iota}(\sigma_{i})) \cdot D^{(1)} \bigg)_{ij} \big[\Delta_{\Psi}(\sigma_{j}-\sigma_{p};\sigma_{i} - \sigma_{p}) \big], \ \ \ 
    \label{s1psi_iota}
\end{align}
\begin{align}
    &\tilde{s}_{\Psi}^{(2)} = \sum^{N}_{j=0} \bigg( \text{diag}(\tilde{\upchi}(\sigma_{i})) \cdot D^{(2)} \bigg)_{ij} \big[ \Delta_{\Psi}(\sigma_{j}-\sigma_{p};\sigma_{i} - \sigma_{p}) \big], \ \ \ 
    \label{s2psi_chi}
\end{align}
\begin{align}
    &\tilde{s}_{\Pi}^{(1)} = \sum^{N}_{j=0} \bigg( \text{diag}(\tilde{ \upvarepsilon}(\sigma_{i})) \cdot D^{(1)} \bigg)_{ij} \big[  \Delta_{\Pi}(\sigma_{j}-\sigma_{p};\sigma_{i} - \sigma_{p}) \big]. 
    \label{spi_epsilon}
\end{align}
To be precise equations (\eqref{s1psi_iota}, \eqref{s2psi_chi}) contribute to the corrected $\textbf{L}_{1}$ operator associated with the master function $\Psi(\tau,\sigma)$ whereas equation \eqref{spi_epsilon} contributes to the corrected $\textbf{L}_{2}$ operator. We would also like to emphasise, in the terms described above in equations (\eqref{chid2_discretization}-\eqref{spi_epsilon}), the numerical computation of these source terms was diagonalised followed by the dot product as highlighted by the term \texttt{diag()} acting on a given coefficient, to prevent a potential erroneous implementation of the hyperboloidal evolution operators described by equations (\eqref{ch34_sb2_diffOperators_gamma}-\eqref{ch34_hyperboloidal_L2_p}). There is no exceptional novelty here, from the technique described in equations (\eqref{cha2_spatial_disc_generic} - \eqref{ch3_sb2_SpaceSpource}), it is just a cautionary step and not necessary, provided the order of operations is carefully implemented and as explicitly stated in this work. The latter may not be obvious to a newcomer numerical relativist. The order these operations are performed is paramount to ensure a correct implementation, specifically, and for example,  $\varepsilon(\sigma) \partial_{\sigma} \neq  \partial_{\sigma} \varepsilon(\sigma)$. Carefully diagonalisation of both the evolution and source terms is more likely to prevent such typos, albeit, as stressed, not necessary. Finally, as in the previous section, the final algorithm describing \texttt{DiscoTEX}'s implementation in hyperboloidal coordinates reads
\begin{align}
    &\textbf{U}_{n+1} = \textbf{U}_{n}+  \textbf{HFH4} \cdot  \bigg[ \textbf{A} \cdot  \bigg[ \textbf{U}_{n} + \frac{\Delta \tau}{12} \bigg(\tilde{\textbf{s}}_{n} - \tilde{\textbf{s}}_{n+1} \bigg)  \bigg] \nonumber  \\
    &+  \frac{\Delta \tau}{2} \bigg( \tilde{\textbf{s}}_{n} + \tilde{\textbf{s}}_{n+1} \bigg) + \frac{\Delta \tau^{2}}{12} \bigg(\tilde{\textbf{st}}_{n} - \tilde{\textbf{st}}_{n+1}  \bigg) \nonumber \\
    &  + \Upsilon +  \textbf{J}_{\texttt{H4}} (\Delta \tau_{\times}, \Delta \tau) \Xi \bigg]. 
    \label{ch3_discoTEX4_Wave_hyper}
\end{align} 
The term pertaining to the discontinuous integration order-4 correction, $\textbf{J}_{\texttt{H4}} (\Delta \tau_{\times}, \Delta \tau)$ , is defined similarly as in equations (\eqref{app3_disco_time_h4}- \eqref{app3_disco_time_Jh4}) but is now transformed in the new chart. The time jumps as in the previous section read
\begin{align}
    \textbf{J}_j(\tau) = 
    \begin{pmatrix}
    \mathbb{JJ}_j(\tau)\\
    \mathcal{JJ}_j(\tau)
\end{pmatrix} =  \begin{pmatrix}
    {\mathbb{J}_{0}(\tau),\mathbb{K}_{0}(\tau),\mathbb{L}_{0}(\tau),\mathbb{M}_{0}(\tau)} \\
    {\mathcal{J}_{0}(\tau), \mathcal{K}_{0}(\tau),\mathcal{L}_{0}(t),\mathcal{M}_{0}(\tau)}
\end{pmatrix}\bigg|_{\tau \rightarrow \frac{\sigma_{p}(\tau)}{v}} , 
    \label{ch34_wavetr_timejumps_hyper}
\end{align}
We refer the reader to \ref{app_hyper_timejumps}, equations (\eqref{hyper_wave_jb_0}-\eqref{wave_Mc_0_exp}) for the explicit $\textbf{J}_{0}(\tau),\textbf{J}_{1}(\tau),\textbf{J}_{2}(\tau),\textbf{J}_{3}(\tau)$ terms. It is also worth highlighting it is reassuring to see \underline{the match} between the expressions in $(t,r)$ coordinates as given by equations (\eqref{wave_jb_0} - \eqref{wave_Mc_0}). 

\subsubsection{Numerical weak-form solutions to the distributionally sourced wave equation via \texttt{DiscoTEX} in the $(t,x)$, $(\tau,\sigma)$ coordinate charts - Optimisation of the numerical simulations and discussion }  

The main goal of this work is to be capable of computing highly accurate numerical solutions for distributionally sourced wave-equations via discontinuous collocation methods with implicit-turned-explicit \texttt{IMTEX} Hermite time-integrator i.e via \texttt{DiscoTEX}, as summarised by equations (\eqref{ch3_discoTEX4_Wave_tr}, \eqref{ch3_discoTEX4_Wave_hyper}) and prove it can successfully address the two main requirements, \texttt{[REQ 1,2]}, for numerical black perturbation theory applications in the time-domain. As explained through Sections \ref{sec_3_2_discojumps}-\ref{sec_discotex_justime} we need to consider three optimisation numerical factors for convergence, 
\begin{itemize}
    \item \texttt{[CTRL F1]} - Number of nodes;
    \item \texttt{[CTRL F2]} - Number of jumps; 
    \item \texttt{[CTRL F3]} - Time step-size.
\end{itemize}
The convergence tests must be done with the two requirements in mind as discussed in the introduction:
\texttt{[REQ 1]-} we must be able to compute the numerical weak-form solution accurately within a given spatial domain where $x/(\sigma) = [a,b]$; \texttt{[REQ 2]-} we must be able to accurately interpolate the numerical solution at any wanted position, particularly at the point-particle position and its right and left limits. To gauge this, the relative error of these solutions against the exact solutions given by equations (\eqref{ch34_distr_wavequation_delta1_plus_delta0_sol}, \eqref{ch34_td_corrections_tofinal_sol_hyper}) will be computed for both the coordinate charts under study, i.e $(t,x)$ and $(\tau,\sigma)$ respectively, 
\begin{equation}
    \eta = \bigg| 1- \frac{\Psi_{\rm{Numerical}}}{\Psi_{\rm{Exact}}} \bigg|.
    \label{ch3_relativeerror}
\end{equation}
As we have done in \cite{da2023hyperboloidal}, we start the numerical studies by checking if the orbital behaviour is that of what is expected from a point-particle distributionally sourcing a wave-equation in linear motion. From Figure \ref{ch3_discotex_tr_wave_snap} and Figure \ref{ch3_discotex_hyperboloidal_wave_snap} it is observed the discontinuous behaviour for when a particle is at $x_{p}(t)$ in an interval $t \in [-1.52,6.60]$ and $\xi_{p}(\tau) = $ in an interval $\tau \in [-1.52, 4.50]$, respectively. Given the solution to equations of Type I, as given in exact form equation \eqref{ch34_distr_wavequation_delta1_plus_delta0_sol}, we expect to observe sinusoidal behaviour of complex nature as evidenced on the right-plots of these figures. Additionally, in these plots we note that in Figure \ref{ch3_discotex_tr_wave_snap}, the spatial domain is restricted to $x = [-4,4]$ because we of the use of radiation boundary conditions, whereas in Figure \ref{ch3_discotex_hyperboloidal_wave_snap} the minimal gauge automatically introduces a compact domain of $\sigma = [0,1]$ where infinity $\mathscr{I}^{+}$ is located at $\sigma=0$ and the horizon at $\sigma=1$. Another simple, yet, important test is to assess if sympletic structure is preserved with \texttt{DiscoTEX} as expected when implementing \texttt{IMTEX} Hermite integrators schemes. As seen in Figure \ref{ch3_discotex_wave_phaseps}, the symplectic structure is preserved for the numerical solutions in either coordinate chart. Particularly here, because the exact solutions are available \cite{field2009discontinuous, field2010persistent, field2023}, we choose to use them as initial data unlike in \cite{da2023hyperboloidal} where trivial initial data of the type $(\Psi=\Pi=0)$ is used and a new optimisation numerical control factor is introduced. This difference is highlighted by contrasting Figure 2 in \cite{da2023hyperboloidal} with Figure \ref{ch3_discotex_wave_phaseps} where the solution is stable for the entirety of the temporal domains used in the simulations. \\

Before discussing the convergence studies for numerical weak-form solutions in either coordinate charts, it is left to explain the required numerical machinery for meeting \texttt{[REQ 2]}. After successfully computing numerical weak-form solutions, one needs to design an \enquote{interpolator} which is capable of handling the discontinuous nature of the problem given wanted solutions at any desired locations within the numerical spatial domains under study. Generic numerical evaluations as explained in previous sections can be achieved via equations (\eqref{ch34_psi_disco} - \eqref{ch34_sb2_SpaceSpource}), 

\texttt{-generic DiscoTEX interpolator:}
\begin{align}
    \label{ch34_psi_discotex_interpol}
    & \Psi(t,x) \approx \sum^{N}_{j=0} \bigg[ \Psi_{j} + \Delta_{\Psi}(x_{j} - x_{p}(t) ; x - x_{p} t) \bigg] \pi_{j}(x), \\
    \label{ch34_pi_discotex_interpol}
    & \Pi(t,x) \approx \sum^{N}_{j=0} \bigg[ \Pi_{j} + \Delta_{\Pi}(x_{j} - x_{p}(t) ; x - x_{p} t) \bigg] \pi_{j}(x),
\end{align}
with the numerical solutions, here $\Psi_{j}/\Pi_{j}$ obtained via \texttt{DiscoTEX} from the numerical evolution of equations (\eqref{ch3_discoTEX4_Wave_tr}, \eqref{ch3_discoTEX4_Wave_hyper}) in the aforementioned numerical space and temporal domains. Furthermore, it is also important, as it will be necessary for equations that emerge in black hole perturbation theory such as in equation \eqref{ch1_metric_weakform}, to determine if one can accurately compute generic amplitudes for higher order derivatives of the master functions. These are computed, for example, via, 

\texttt{- generic DiscoTEX interpolator for higher order derivatives:}
\begin{align}
    & \partial_{x}^{(n)}\Psi(t,x) \approx \sum^{N}_{j=0} \bigg[ \Psi_{j} \ \partial_{x}^{(n)}\pi_{j}(x) + \Delta_{\Psi}(x_{j} - x_{p} ; x - x_{p}) \partial_{x}^{(n)}\pi_{j}(x) \bigg]. 
      \label{ch34_psi_discotex_interpol_gen_deriva}
\end{align}
Equations (\eqref{ch3_discoTEX4_Wave_tr}, \eqref{ch3_discoTEX4_Wave_hyper}) along with equations (\eqref{ch34_psi_discotex_interpol}-\eqref{ch34_psi_discotex_interpol_gen_deriva}) describe all the machinery of the \texttt{DiscoTEX} algorithms necessary for the for computation of requirements \texttt{[REQ 1,2]}. The numerical error of the numerical weak-form solution at the particle and its right and left limits is gauged by computing the relative error as given in equation \eqref{ch3_relativeerror}.

It is now left to study the optimal numerical control factors associated with the accurate implementation of numerical weak-form solutions as described by the evolution algorithms in equations (\eqref{ch3_discoTEX4_Wave_tr}, \eqref{ch3_discoTEX4_Wave_hyper}). For both the solutions here, it was decided to use Chebyshev collocation nodes as described in equations \eqref{chebyshev_lobatto_nodes}, however, the discontinuous collocation scheme is generic and as demonstrated in \cite{da2023hyperboloidal} solutions with finite-difference schemes are also possible however tend to require a significantly higher number of nodes (i.e scaling with the order of the finite difference scheme required) and hence take significantly longer. As evidenced by Figures \ref{ch3_discotex_wave_conv_test} and \ref{ch3_discotex_hyperboloidal_wave_conv_test} $N=\{34,45\}$ nodes with $J= \{15/15\}$ jumps are required for an highly accurate implementation of \texttt{DiscoTEX} for numerical solutions in the coordinate charts of $(t,x)$ and $(\tau,\sigma)$ respectively. Furthermore, we decided to use a timestep of $\Delta t \approx 0.0029 $ and $\Delta \tau \approx 0.0067$ supported by Figures (\ref{ch3_discotex_wavetr_stepsize}, \ref{ch3_discotex_hyperboloidal_stepsize}) where we can see the numerical weak-form solutions via \texttt{DiscoTEX} scale at fourth-(or second-) order as expected for a fourth-(or second-) order numerical \texttt{IMTEX} Hermite time-integration scheme. To complement these results and facilitate future comparison work with new algorithms we include the amplitudes of the field and its first two spatial derivatives evaluated at the particle solution in Table \ref{tab_interpolator_wave_results}. We can hence conclude that highly accurate computations of \texttt{[REQ 1,2]} are possible with \textit{the correct considerations and implementations} of \texttt{DiscoTEX}. 

\begin{table}
\begin{small}
\begin{tabular}{ ||c|| c|| }
\hline
\textrm{$\Psi(t,x)$ }& 
\textrm{Accuracy, $\eta$} \\ \hline \hline
\rowcolor{retroyellow} $-0.131448548684 + 0.128643183774 \iu$ & $2.9\times 10^{-11}$  \\ \hline \hline 
\textrm{$\Pi(t,x)$ }& 
 \\ \hline \hline
\rowcolor{retroyellow} $-0.2572863675538771 - 0.31781337994184816\ \iu $& $3.7 \times 10^{-15}$  \\ \hline \hline 
\textrm{$\partial_{x}\Psi^{+}(t,x)$ }& 
 \\ \hline \hline
\rowcolor{retroyellow} $-0.3859295513282838 + 0.084127071161650865 \iu$ & $1.3\times 10^{-15}$   \\ \hline 
\textrm{$\partial_{x}\Psi^{-}(t,x)$ }& 
 \\ \hline \hline
\rowcolor{retroyellow} $0.6432159188872233 + 0.23368630878134367\ \iu$ & $1.3 \times 10^{-15}$  \\ \hline 
\textrm{$\partial_{x}^{2}\Psi^{+}(t,x)$ }& 
\\ \hline \hline 
\rowcolor{retroyellow} $0.08412707116495977 + 0.246994912838705\ \iu$ & $4.7 \times 10^{-14}$  \\ \hline 
\textrm{$\partial_{x}^{2}\Psi^{-}(t,x)$ }& 
 \\ \hline \hline
\rowcolor{retroyellow} $0.2336863087846508 - 1.1434949669191923\ \iu$ & $4.7 \times 10^{-14}$  \\ \hline \hline  \end{tabular}
\caption{Comparison of the amplitudes obtained via the \texttt{generic discontinuous interpolator} described in equations (\eqref{ch34_psi_discotex_interpol}-\eqref{ch34_psi_discotex_interpol_gen_deriva}) and those obtained from the exact solution as in equation \eqref{ch34_distr_wavequation_delta1_plus_delta0_sol}. The results were interpolated at the particle position (and its right and left limits) computed as $r_{p}(t) =  v t$ for a velocity $v=1/4$ and a numerical spatial domain of $x \in [-4,4]$ and $t = [-1.52, 6.60]$. One can conclude \texttt{DiscoTEX} is capable of meeting both \texttt{[REQ 1,2]}.} \label{tab_interpolator_wave_results}
\end{small}
\end{table}
Finally, and for completion of the algorithm we complement these results and discussions with \ref{app_discoimp} where the numerical weak-form solution to equations of Type II and Type III distributionally sourced wave-equations are computed, and compared, using both the \texttt{DiscoTEX} algorithm and the discontinuous collocation machinery with the purely implicit \texttt{IM/(P)} Hermite fourth-order time-integrator with both radiation boundary conditions in the $(t,x)$ chart and the $(\tau,\sigma)$ coordinate chart. Ultimately, the algorithm has been proved in its entirety and computation of \texttt{[REQ 1,2]} is not just feasible as a weak-form numerical solution, it can be done with very high accuracies. 
\begin{figure*}
\includegraphics[width=99mm]{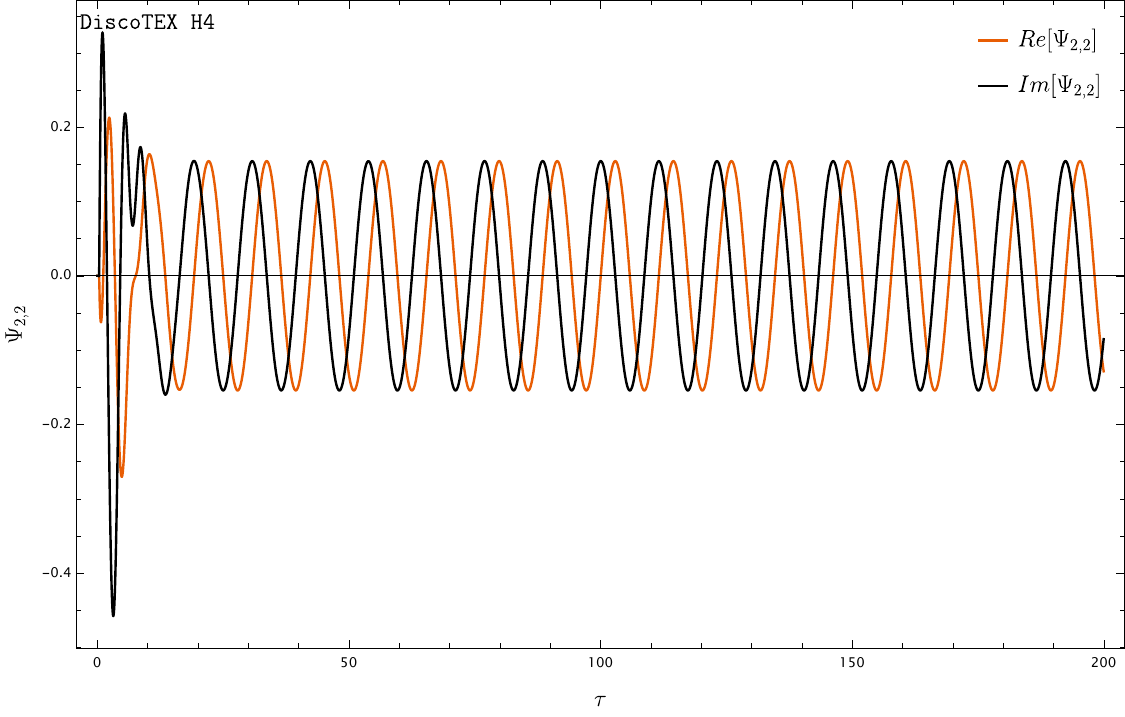}
\quad
\includegraphics[width=79mm]{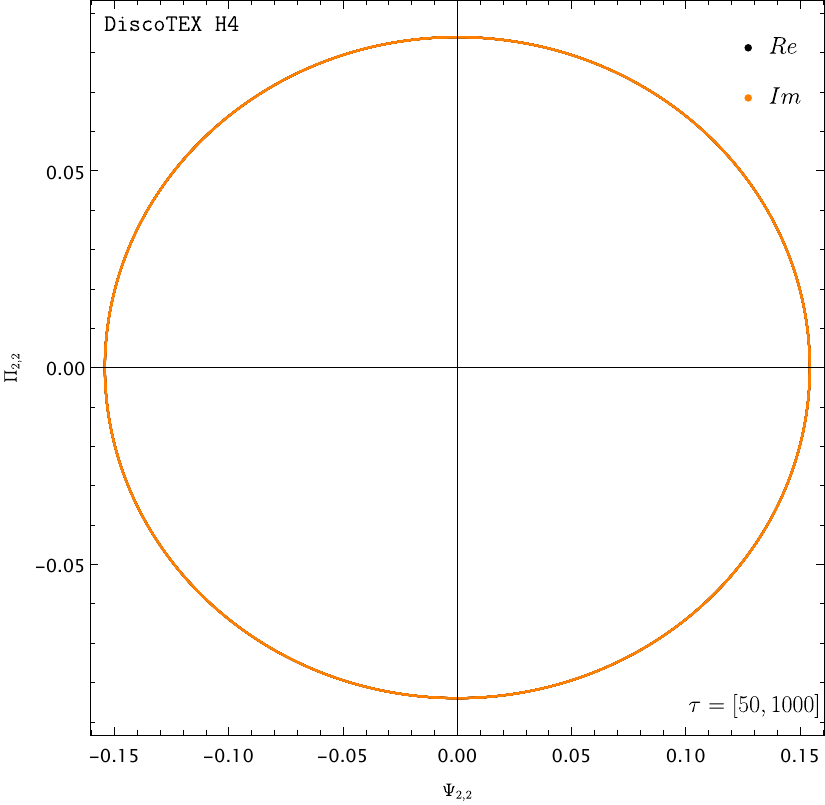}
\caption{\textbf{Left:} Waveform at $\mathscr{I}^{+}$ of a scalar point-charge on a circular geodesic in a SMBH Schwarzschild background at $\sigma_{p} = 2M/6M$ for the $(l,m) = (2,2)$ mode. \textbf{Right:} Here one shows the phase portrait of the numerical evolution for the field $\Psi_{2,2}(\tau,\sigma)$ at $\mathscr{I}^{+}$.  One determines a reasonable time to assume all junk radiation to dissipate away and our simulation to be in steady-state to be at least $ \tau > 50 $, corresponding to ignoring the first 1.5 orbits as per the plot on the left.\label{ch3_sclar_plots}}
\end{figure*} 
\section{Applications to numerical black hole perturbation theory}\label{Sec4}

To demonstrate further the main application of the \texttt{DiscoTEX} algorithm, particularly with a focus on numerical black hole perturbation theory, we will now briefly demonstrate how the algorithm can be used to calculate the amplitude of the scalar and gravitational self-forces for a point-particle on a circular geodesic in a Schwarzschild black hole. The work here does not investigate the numerical optimisations, this will be subject to upcoming collaborative work \cite{paper2, paper3} and it merely intends to demonstrate the ability of the algorithm to give highly accurate amplitude results through numerical weak-form solutions via \texttt{DiscoTEX}. Here, in the first section, we will briefly demonstrate this for the computation of scalar perturbations and in the next section how weak-form solutions can be recovered for metric perturbation amplitudes and gravitational-self force calculations will be demonstrated. To get a better idea of how the numerical optimisation control factors are determined in the context of gravitational self-force applications work we refer the reader to our previous work \cite{da2023hyperboloidal} and first paper of the series \cite{paper2,paper3}. 

\subsection{Modelling perturbations in Schwarzschild via \texttt{DiscoTEX} }

Here we will show how key quantities for (X)/(E)MRI modelling can be computed via the \texttt{DiscoTEX} algorithm for when a scalar point-charge, $q$ or a point-particle $\mu$ prescribes a circular geodesic around an SMBH of mass $M$. The SMBH is described by a four-dimensional manifold $\mathcal{M}$, the Schwarzschild spacetime, which in Schwarzschild coordinates, $x^{\mu} = (t,r,\theta,\phi)$ is given as,
\begin{equation}
    ds^{2} = - f(r) dt^{2} + f(r)^{-1} dr^{2} + r^{2} d\omega^{2} \ \ \ \ f(r) = \bigg(1 - \frac{2M}{r}\bigg), 
    \label{ch4_schwarzsc_metric}
\end{equation}
with $d\omega^{2} =  d\theta^{2} + \sin^{2}\theta \ d\theta d\phi$. In numerical BHPT it is standard to split the spacetime into two sub-manifolds $\mathcal{M} = \mathcal{M}^{2} \times \mathcal{S}^{2}$, where $\mathcal{M}^{2} $ is a Lorentzian sub-manifold with coordinates $x^{a}= (t,r)$ and $\mathcal{S}^{2}$ is a the spherical space described by $\mathcal{S}^{2}$ described by coordinates $\theta^{A} = (\theta, \phi)$. 

The point-particle $\mu$, has the worldline $x^{\mu} = \big(t_{p}(s), r_{p}(s),$
$\theta_{p}(s), \phi_{p}(s) \big)$, where $s$ denotes the proper time on the circular geodesic given as $x^{\mu}_{p} = (t_{p}, r_{p}, \theta = \frac{\pi}{2}, \phi_{p})$, with constants of motion
\begin{equation}
    \mathcal{E} = \frac{f}{\sqrt{1 - 3M/r}}, \  \mathcal{L} = \frac{\sqrt{r M}}{1 - 3M/r}, \ \Omega_{\phi} = \frac{M}{r^{3}}.
    \label{ch4_circ_orb_parameters}
\end{equation}

The four-velocity $u^{\mu} = dx^{\mu}/ds$ has the components
\begin{equation}
    u^{t} = \frac{\mathcal{E}}{f}, \ (u^{r})^{2} = \mathcal{E}^{2} - \mathcal{U}^{2},\ u^{\theta}=0,\ u^{\phi} = \frac{\mathcal{L}}{r^{2}}, 
    \label{ch4_4velocity}
\end{equation}
where $\mathcal{U}^{2} = f(r) + (1 + \mathcal{L}/r^{2})$. 
\begin{figure*}
\includegraphics[width=90mm]{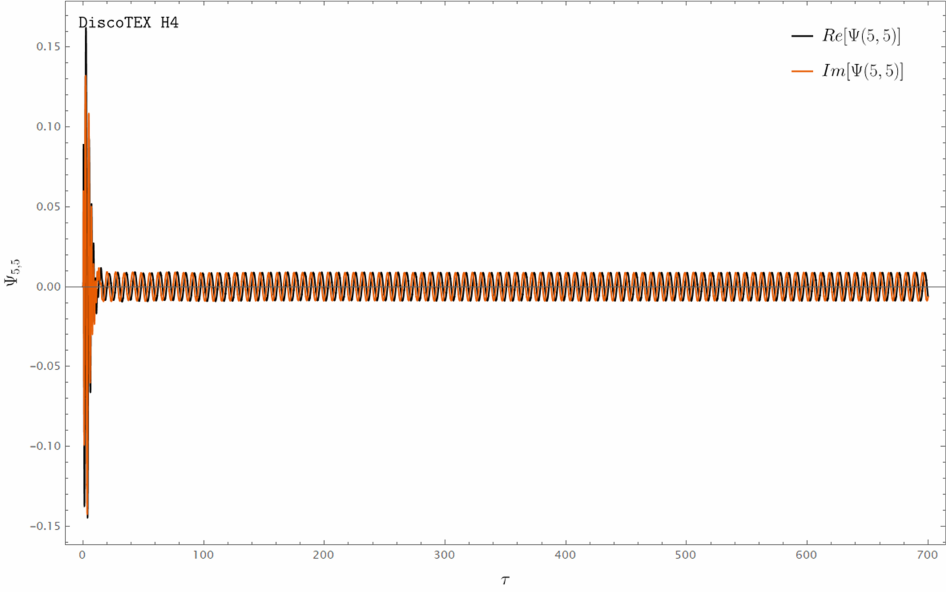}
\quad
\includegraphics[width=90mm]{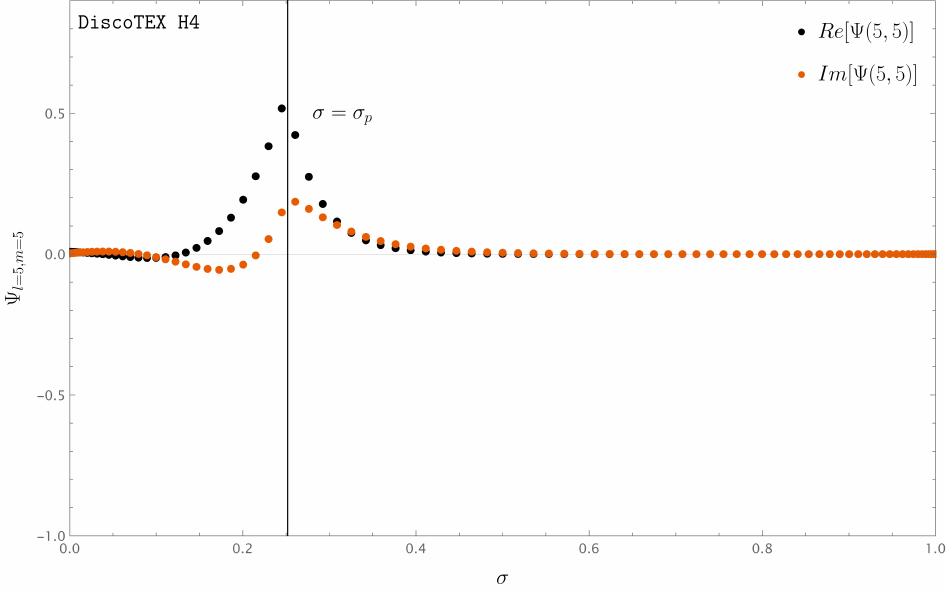}
\caption{\textbf{Left:} Waveform at $\mathscr{I}^{+}$ of a point-particle on a circular geodesic in a SMBH Schwarzschild background at $\sigma_{p} = 2M/7.9456M$ for the $(l,m) = (5,5)$ mode. \textbf{Right:} Here we show the retarded hyperboloidal field $\Psi(\tau\,\sigma)$ at $\sigma_{p} = 2M/7.9456M$ and $\tau=500$ extending from the future null infinity $\mathscr{I}^{+}$ at $\sigma=0$ and to the event horizon $\sigma=1$. \label{ch4_gsfFields_plots}}
\end{figure*} 
\subsection{Modelling scalar perturbations in Schwarzschild via \texttt{DiscoTEX}}
We start by demonstrating the usefulness of the algorithms by modelling the behaviour of a scalar point-particle charge $q$ around the SMBHs described by, 
\begin{equation}
    \square_{g} \Psi = \frac{1}{\sqrt{-\textbf{g}}} \partial_{\mu} [ \sqrt{-\textbf{g}} \ g^{\mu\nu}\partial_{\nu} ] \Psi = S_{\Psi}, 
    \label{ch4_scalarwave}
\end{equation}
with the determinant of the background SMBH given as $\textbf{g} = -r^{4}\sin^{2}\theta$. The source term associated with the scalar point charge is given as
\begin{eqnarray}
    S_{\Psi} &=& -4 \pi \rho = -4\pi \int \frac{\delta^{4}(x^{\mu} - x^{\mu}_{p}(s))}{\sqrt{-\textbf{g}}} \ ds. 
    \label{ch4_sem_scalar}
\end{eqnarray}
Taking advantage of the spherical symmetry of the SMBH background metric we apply spherical harmonic decomposition expanding the field as
\begin{equation}
    \Psi(t,r,\theta,\phi) = \frac{1}{r} \sum^{\infty}_{l,m} \psi^{lm}(t,r) Y^{lm}(\theta, \phi),  
    \label{ch4_scalar_hd}
\end{equation}
and $S_{\Psi}$ as
\begin{equation}
    S_{\Psi} = -\frac{4 \pi q f}{E} \delta(r-r_{p}) Y^{lm}(\theta, \phi) Y^{*lm}(\theta_{p}, \phi_{p}), 
    \label{}
\end{equation}
with the subscript $p$ denoting it is with respect to the particle's position and where the following Dirac delta propriety to collapse the spherical components, 
\begin{equation}
    \delta(\theta - \theta_{p}) \delta(\phi - \phi_{p}) = \sin \theta \sum_{l,m} Y^{lm}(\theta,\phi)Y^{*lm}(\theta_{p}, \phi_{p})
    \label{ch4_dirac_delta_spherical_propriety},  
\end{equation}
has been used. 

Applying these expansions into equation \eqref{ch4_scalarwave}, we get the \texttt{1+1D} distributional sourced wave-like equation
\begin{equation}
    \square \psi = G^{lm}(t,r) \delta(r-r_{p}),  
    \label{ch4_scalar_wave_withsource}
\end{equation}
where the source term is given by
\begin{equation}
    G^{lm}(t,r) = -\frac{4 \pi f^{2}}{Er}Y^{*lm}(\theta_{p}, \phi_{p}) 
    \label{ch4_sourcegterm}
\end{equation}
and with the d'Alembert-like wave operator
\begin{equation}
    \square = - \frac{\partial^{2}}{\partial t^{2}} + \frac{\partial^{2}}{\partial x^{2}} - V_{l}(r), 
    \label{ch4_dalembert_scalar}
\end{equation}
with the potential associated with the scalar perturbation given by
\begin{equation}
    V_{l}(r) =  \frac{f}{r^{2}} \bigg[ l(l+1) + \frac{2M}{r}\bigg]. 
    \label{ch4_ScalarPotential}
\end{equation}
For the scalar case, the jumps are given as, 
\begin{eqnarray}
    [[\psi]]= 0 , \hspace{0.4cm} {}[[\psi_{r}]]= \frac{G^{lm}(t,r)}{f^{2}}.\label{ch4_scalar_jumps}
\end{eqnarray}

To compute the scalar self-force (SSF) the field $\Psi$ shall be regularised as demonstrated in \cite{macedo2022hyperboloidal}. The first step is to decompose the field into, 
\begin{equation}
    \Psi^{\text{ret} } = \Psi^{R} +  \Psi^{S},  
    \label{ch4_ssf_fielddec}
\end{equation}
where $\Psi^{S}$ is the singular part of the field equation and $\Psi^{\rm{ret}}, \Psi^{R}$ are the retarded and regular fields respectively. The SSF is then given as
\begin{equation}
    F_{\alpha} := q\nabla_{\alpha} \Psi_{R} = q (\nabla_{\alpha} \Psi^{ret}  - \nabla_{\alpha}\Psi^{S}). 
    \label{ch4_ssf_general}
\end{equation}

Furthermore, the computations presented here follow the mode-sum computational strategy as put forward by \cite{barack2000mode}. The retarded field $\Psi^{\text{ret}}$ is decomposed as
\begin{equation}
    \Psi^{\text{ret,l}}_{\alpha} = \nabla_{\alpha} \sum^{l}_{m=-l} \psi^{lm}(t,r) Y^{*lm}(\theta_{p}, \phi_{p}). 
    \label{ch4_retarded}
\end{equation}
In this work the interest lies in demonstrating the algorithm's main application so one will only look at the $r-$component. To compute the self-force, $F_{r}$, we need to correct for the singular terms contributions as demonstrated by \cite{Heffernan:2012su}, 
\begin{equation}
  \underset{x^{\mu} \rightarrow x^{\mu}_{p}}{\lim} = q \bigg[
    A_{\alpha}(l(l+1)) + B_{\alpha} + \frac{C_{\alpha}}{l+ \frac{1}{2}} +  ... \bigg]. 
    \label{ch4_ssf_r_singular_cont}
\end{equation}
Finally, with substituting equation \eqref{ch4_ssf_r_singular_cont} into (\eqref{ch4_ssf_general}, \eqref{ch4_ssf_r_ret}) we get
\begin{equation}
    F_{lr} = F^{\textrm{ret} \pm}_{lr} - A_{r}(l(l+1)) - B_{r} -  \sum^{3}_{n=1} F^{l}_{r}[2n], 
    \label{ch4_ssf_r_ret_full}
\end{equation}
where
\begin{equation}
     F^{\textrm{ret} \pm}_{lr} = \sum^{l}_{m=-l} \partial_{r} \bigg[ \frac{1}{r} \psi^{l,m}(t,r) \bigg]^{\pm} Y^{*lm}(\theta_{p}, \phi_{p}). 
    \label{ch4_ssf_r_ret}
\end{equation}
As emphasised, the goal here is to demonstrate the accuracy and the flexibility of the \texttt{DiscoTEX} (and relatives) algorithm, to that effect we choose not to include the numerical recipe description in detail as it closely follows the previous section but where now the minimal gauge implementation is exactly as described in our previous work in Section III of \cite{da2023hyperboloidal}. The distinction here is in the jump implementation which in the scalar case greatly simplifies and can be trivially computed via chain-rules from equations \eqref{ch4_scalar_jumps}. With a preliminary numerical optimisation study\footnote{We note these results are part of upcoming work \cite{paper2}. The inclusion and presentation here is merely to demonstrate \texttt{DiscoTEX}'s applicability. The author thanks Rodrigo Panosso Macedo and Benjamin Leather for providing their frequency-domain data of \cite{macedo2022hyperboloidal} for comparison along with sharing their regularisation conventions. In this work, we used acceleration regularisation parameters up to order $n=3$ as defined by in equation \eqref{ch4_ssf_r_ret_full} and following the definitions of \cite{Heffernan:2012su}.} The the following optimisation parameters were determined: $N=60$ nodes, $J=12$ jumps, $\Delta \tau = 0.05$, and $\tau_{\text{extr}} = 10,000$. In Figure \ref{ch3_sclar_plots} on the left plot the scalar waveforms for the point-particle on a circular geodesic at $r_{p}=6M$ for the $(l,m) = \{2,2\}$ mode are shown.  Unlike in the previous section, where the exact solutions were used as initial data for the evolution, here, as is usual in BHPT simulations, we have given it trivial initial data \cite{da2023hyperboloidal}. This causes the simulation to be contaminated at earlier times by junk radiation as evidenced on this figure at least during $\tau = [0,10]$, as customary, we wait until the simulation reaches steady-state evolution and introduce a fourth numerical optimisation control factor i.e. 
\begin{enumerate}
    \item \textcolor{gray}{\texttt{[CTRL F\ref{controlfactor1}]} - Number of \texttt{N} nodes;}
    \item \textcolor{gray}{\texttt{[CTRL F\ref{controlfactor2}]} - Number of \texttt{J} jumps;} 
    \item \textcolor{gray}{\texttt{[CTRL F\ref{controlfactor3}]} - $\Delta \tau$, time step-size;} 
    \item \texttt{[CTRL F4]} - Minimal time for steady state evolution. \label{controlfactor4}
\end{enumerate}
Aided by the phase portraits given by the plot on the right of Figure \ref{ch3_sclar_plots} we estimated the minimal time for steady-state evolution to be at least $\tau > 50$, at earlier times symplectic structure was not preserved corresponding to about 1.5 orbits. We decided to extract the quantities at a much longer time, $\tau = 10, 000$ after quick convergence tests showing that the quantities evaluated at the particle limit needed longer times to converge. This value is nevertheless an overestimate, as explained this will be given in detail \cite{paper2}. 
\begin{figure*}
\includegraphics[width=90mm]{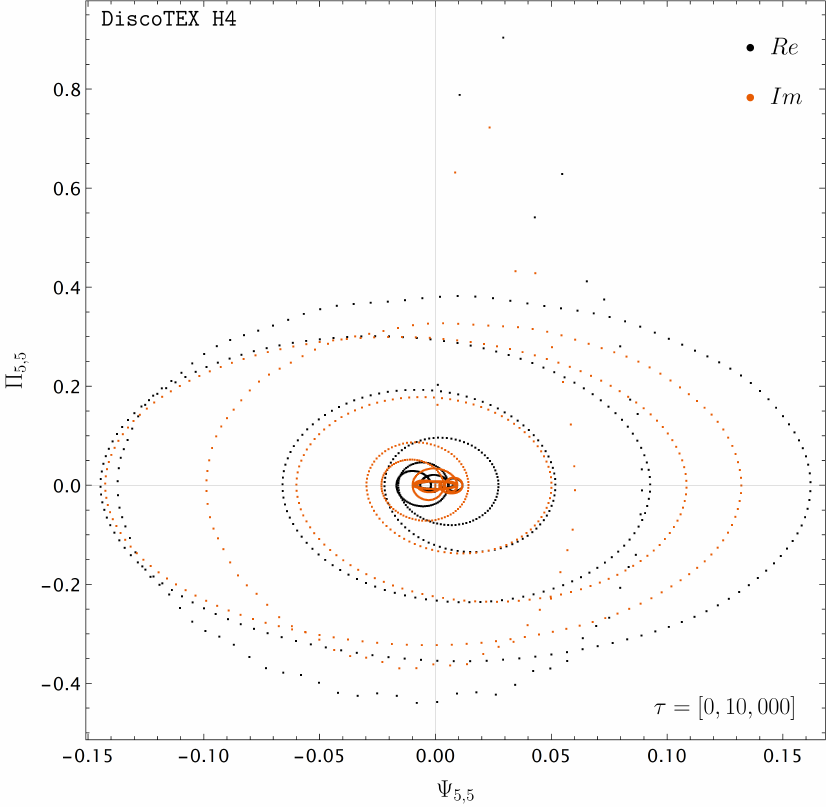}
\quad
\includegraphics[width=93mm]{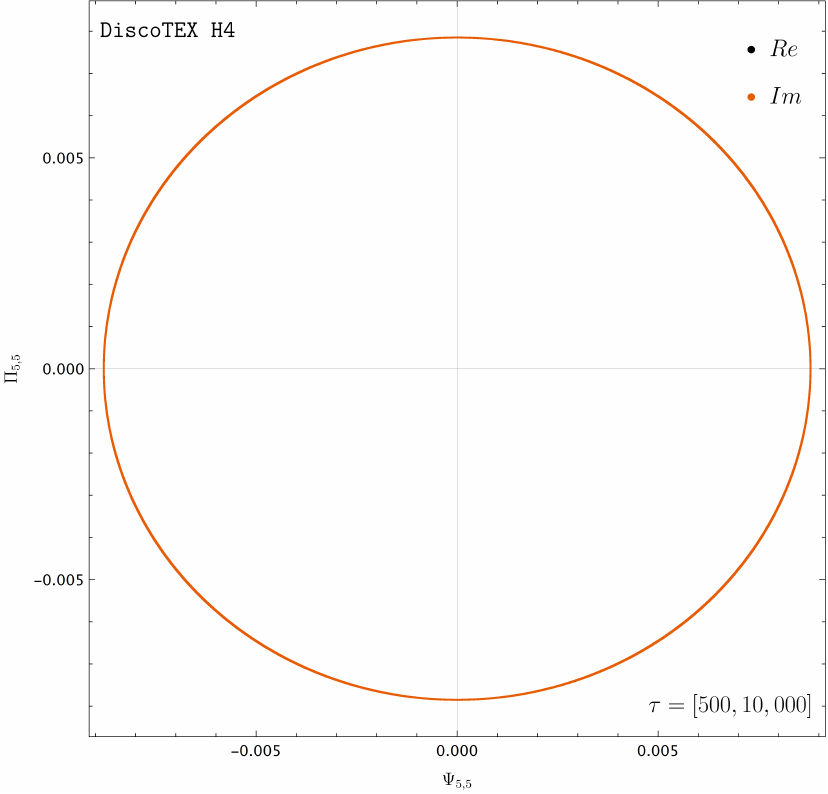}
\caption{Here we show the phase portrait of our numerical evolution for the field $\Psi^{5,5}(\tau,\sigma)$ at infinity $\mathscr{I}^{+}$. \textbf{Left:} Phase portrait for the entire simulation $\tau = [0,10,000]$. \textbf{Right:} Phase portrait after we waited a sufficiently long time, $\tau > 500$, allowing all junk radiation to dissipate away and our simulation to reach steady-state.\label{ch3_gsfFields_PP_plots}}
\end{figure*} 
To calculate the $r-$component of the SSF as given by equation \eqref{ch4_ssf_r_ret} all is needed is to take the numerical solution obtained by solving equation \eqref{ch4_scalar_wave_withsource} and interpolate at the particle limit via \texttt{DiscoTEX}. We note equation \eqref{ch4_ssf_r_ret} expands as
\begin{equation}
    F^{\textrm{ret} \pm}_{lr} = \sum^{l}_{m=-l} \bigg[ \frac{1}{r} \partial_{r}\psi^{l,m}(t,r) - \frac{1}{r^{2}} \psi^{l,m}(t,r)  \bigg]^{\pm} Y^{*lm}(\theta_{p}, \phi_{p}). 
    \label{ch4_SSF_r_expanded}
\end{equation}
This means we need, not only to interpolate the field at the particle limit but also its $r$-derivative. These interpolations are done on the hyperboloidal field $\Psi(\tau,\sigma)$ spanning the spatial domain $\sigma=[0,1]$, with the aid of chain rules between the minimal gauge chart to the Schwzraschild chart i.e $(t \to t(\tau,\sigma), \ r \to r(\sigma))$, which will naturally introduce extra-derivatives. For the SSF, we, using the \texttt{DiscoTEX} algorithm, interpolate directly at the hyperboloidal field, $\Psi(\tau,\sigma)$
\begin{equation}
    \Psi(\tau,\sigma) \bigg|_{\sigma = \sigma_{p}}\approx \sum^{N}_{j=0} \bigg[ \Psi_{j}(\tau) + \Delta_{\Psi}(\sigma_{j} - \sigma_{p}(\tau); \sigma - \sigma_{p}(\tau))  \bigg] \pi_{j}(\sigma)\bigg|_{\sigma = \sigma_{p}}.
    \label{genericinterpol}
\end{equation}
The $r-$derivative by the chain-rule is given as
\begin{equation}
    \partial_{r} = -\frac{\sigma^{2}}{2M} \bigg[\partial_{\sigma} + \partial_{\sigma} H(\sigma) \partial_{\tau}  \bigg], 
    \label{ch4_dr_chainrule}
\end{equation}
and thus we interpolate at $\Pi(\tau,\sigma)$, which, itself, is simply obtained by  
\begin{equation}
    \Pi(\tau,\sigma) \bigg|_{\sigma = \sigma_{p}}\approx \sum^{N}_{j=0} \bigg[ \Pi_{j}(\tau) + \Delta_{\Pi}(\sigma_{j} - \sigma_{p}(\tau); \sigma - \sigma_{p}(\tau))  \bigg] \pi_{j}(\sigma)\bigg|_{\sigma = \sigma_{p}}. 
    \label{genericinterpol_pi}
\end{equation}
Finally the $\sigma$- derivative of the field, $\partial_{\sigma} \Psi(\tau,\sigma) $ is simply given by
\begin{align}
       &\partial_{\sigma} \Psi(\tau,\sigma) \bigg|_{\sigma = \sigma_{p}}\approx \sum^{N}_{j=0} \bigg[ \Psi_{j}(\tau) \pi_{j}'(\sigma) \nonumber \\ 
       &+ \Delta_{\Psi}(\sigma_{j} - \sigma_{p}(\tau); \sigma - \sigma_{p}(\tau)) \pi'_{j}(\sigma) \bigg] \bigg|_{\sigma = \sigma_{p}}.
    \label{genericinterpol_dsigma} 
\end{align}
To check the accuracy of \texttt{DiscoTEX} the first 5 modes for the $r-$component of the self-force were computed to get $F_{r} = 4.152070\times 10^{-3}$ with a numerical error of $6.0\times 10^{-9}$. We note the numerical error was calculated against that of \cite{macedo2022hyperboloidal} and the value is only for the $l=1$ to $l=5$ modes. In upcoming work, we will give the full results \cite{paper2}. 

\subsection{Modelling gravitational perturbations in Schwarzschild via \texttt{DiscoTEX}}
To demonstrate the main application of the \texttt{DiscoTEX} algorithm we will use some of the results in \cite{da2023hyperboloidal} (the full results and discussion will be given in \cite{paper2,paper3}) and compute the gravitational perturbations as described by the Regge-Wheeler-Zerilli formalism. Specifically, unlike in \cite{da2023hyperboloidal}, where we solved for the Regge-Wheeler and Zerilli master functions \cite{thompson2017gauge, thompson2019gravitational} as the goal is to compute the self-force as regularised by \cite{thompson2019gravitational}, we will now, for the first time and for completion, solve for the Cunningham-Price-Moncrief (CPM), $\Psi^{a}_{lm}(t,r)$, \cite{cunningham1978radiation, cunningham1979radiation} and Zerilli-Moncrief (ZM) master functions $\Psi^{p}_{lm}(t,r)$, \cite{moncrief1974gravitational} as given in \cite{field2009discontinuous, hopper2010gravitational,martel2005gravitational}\footnote{An incorrect attempt at solving these equations was given in Table 6.1 of \cite{o2022timeQMUL, o2022time}. It is not clear what the mistake in their implementation is. The author has tried to reproduce their results but obtained a different answer, also incorrect. These references do not provide convergence tests.}. 

\subsubsection{Phase I - Numerical weak-form solution to CPM/ZM master functions via  \texttt{DiscoTEX}}\label{phase1}
The CPM and the ZM satisfy the following wave equation 
\begin{equation}
    \bigg[ -\frac{\partial^{2}}{\partial t ^{2}} + \frac{\partial^{2}}{\partial x ^{2}} - V^{a/p}_{l}(r) \bigg] \Psi^{a/p}_{lm}(t,r) = S_{lm}^{a/p}(t,r),
    \label{ch4_CPM_ZM_wavequation}
\end{equation}
where $x$ remains as the tortoise coordinate. The axial and polar potentials are given as
\begin{eqnarray}
    \label{ch4_axial_potential}
    V^{a}_{l}(r) =  \frac{1}{r^{2}} \bigg[ l(l+1) - \frac{6M}{r}\bigg], \\
    V^{p}_{l}(r) = \frac{1}{\tilde{\Lambda}^{2}} \bigg[ \mu^{2} \bigg( \frac{\mu +2}{r^{2}} + \frac{6M}{r^{3}} \bigg) + \frac{36 M^{2}}{r^{4}} \bigg( \mu + \frac{2M}{r}\bigg) \bigg],
    \label{ch4_polar_potential}
\end{eqnarray}
where $\tilde{\Lambda}:= (l-1)(l+2)+ 6M/r$ and $\mu:= (l-1)(l+2)$. 
The source $S_{lm}^{a/p}(t,r)$ is given as, 
\begin{eqnarray}
   \nonumber S_{lm}^{a/p}(t,r) &=& \bar{G}^{a/p}_{lm}(t)\delta\left(r-r_{p}(t)\right) 
    + \bar{F}^{a/p}_{lm}(t)\delta'(r-r_{p}(t)). 
    \label{cha4_sourceterm}
\end{eqnarray} 
For simplicity, and following \cite{da2023hyperboloidal, hopper2010gravitational} we define the source-term functions as,
\begin{eqnarray}
       \bar{F}_{lm}^{a/p}(t) &=& F^{a/p}_{lm}(t,r_{p}(t)), \\
       \bar{G}_{lm}^{a/p}(t) &=& \bigg[ G^{a/p}_{lm}(t,r) - \partial_{r} F^{a/p}_{lm}(t,r) \bigg]\bigg|_{r=r_{p}}. 
       \label{cha4_sourceterm_wrt_partworldline}
\end{eqnarray}
The jumps of the retarded field at the particle's trajectory have been given generically in \cite{da2023hyperboloidal} to facilitate implementation across different versions of the RWZ equations and are given as
\begin{align}
\label{ch3_j0} 
   &[[\Psi]](t) =  \frac{\mathcal{E}^{2}}{f^{2}_{p} U^{2}_{p}} \bar{F}^{a,p}_{lm}(t),\\  
  &{}[[\Psi_{r}]](t) =  \frac{E^{2}}{f_{p}^{2} U^{2}_{p}} \bigg[ \bar{G}^{a,p}_{lm}(t)\nonumber \\
    & + \frac{1}{U^{2}_{p} r^{2}_{p}} \bigg( 3M - \frac{\mathcal{L}^{2}}{r_{p}}+ \frac{5 M \mathcal{L}^{2}}{r^{2}_{p}}  \bigg) \bar{F}^{a,p}_{lm}(t)  - 2 \dot{r}_{p} \frac{d}{dt} \big( [[\Psi]](t) \big) \bigg],\label{ch3_j1}\\
  {}\label{ch3_jt}&[[\Psi_{t}]](t)  \partial_{t}[[\Psi]](t) - \dot{r}_{p}[[\Psi_{r}]](t),\\
     {}\label{ch3_jx}&[[ \Psi_{x}]](t) = f_{p} [[ \Psi_{r}]](t). 
\end{align}
The reader should now refer to \ref{app1_bhpt} in this manuscript to get the correct $\bar{G}(t), \bar{F}(t)$ functions. To confirm the accuracy of the solution as done in \cite{da2023hyperboloidal}, we evaluate the energy fluxes at infinity $\mathscr{I}^{+}$ and the horizon $\mathcal{H}$, as given by the formulas
\begin{eqnarray}
    \label{ch4_energyFlux}
    \dot{E}^{\pm}_{lm} = \frac{1}{64\pi}\frac{(l+2)!}{(l-2)!}\bigg|\dot{\Psi}^{\pm}_{lm}(t,r)\bigg|^{2}, \\
    \dot{L}^{\pm}_{lm} = \frac{i m}{64\pi}\frac{(l+2)!}{(l-2)!}\bigg|\dot{\Psi}^{\pm}_{lm}(t,r)\Psi^{\pm*}_{lm}(t,r)\bigg|. 
    \label{ch4_angularmomentumFlux}
\end{eqnarray}
The numerical error associated with the \texttt{DiscoTEX} algorithm is computed by comparing against the frequency domain results as given by the BHPT toolkit \cite{BHPToolkit}. The relative error difference is thus defined as, 
\begin{equation}
    \eta = \bigg|1 - \frac{\dot{E}_{\textrm{total}}}{\dot{E}^{\textrm{ref}}_{\textrm{total}}}\bigg|.
    \label{ch4_relative_error}
\end{equation}

\begin{table*}
\begin{center}
\begin{tabular}{||l|| c ||c ||c ||c ||c|| }
\hline \hline
\textrm{$(l,m)$}& 
\textrm{$\dot{E}^{\infty}_{lm}$}&
\textrm{$\dot{E}^{\infty, GDS}_{lm}$} \cite{da2023hyperboloidal}&
\textrm{$\dot{L}^{\infty}_{lm}$}&
\textrm{$\dot{L}^{\infty, GDS}_{lm}$} \cite{da2023hyperboloidal} \\ \hline \hline 
(2,2) & $1.70621955\times 10^{-4}$ & $1.70621954 \times 10^{-4}$ & $3.8214216\times 10^{-3}$  &  $3.8214217\times 10^{-3}$  \ \  \\ 
(2,1) & $8.16304\times 10^{-7}$ & $8.163040 \times 10^{-7}$ & $1.82828\times 10^{-5}$  & $1.8282769 \times 10^{-5}$ \ \  \\
(3,3)  & $2.547061\times 10^{-5}$ & $2.547061\times 10^{-4}$ & $5.704656\times 10^{-3}$  &  $5.704656\times 10^{-4}$  \ \  \\ 
(3,2)  & $2.51984\times 10^{-7}$ & $2.519844\times 10^{-7}$ & $5.643699\times 10^{-6}$  &  $5.643699\times 10^{-6}$  \ \  \\ 
(3,1) &  $2.17303\times 10^{-9}$ & $2.1730\times 10^{-9}$ &  $4.86693\times 10^{-8}$ &  $4.86693\times 10^{-9}$  \ \  \\ 
(4,4) & $4.7254\times 10^{-6}$ & $4.72538\times 10^{-6}$ & $1.058344\times 10^{-4}$  &  $1.058345\times 10^{-4}$  \ \  \\ 
(4,3)  & $5.77490\times 10^{-7}$ & $5.7749\times 10^{-9}$ & $1.293404\times 10^{-6}$  &  $1.29340\times 10^{-9}$  \ \  \\ 
(4,2) & $2.5098\times 10^{-9}$ & $2.5090\times 10^{-9}$ & $5.6194\times 10^{-8}$  &  $5.619\times 10^{-9}$  \ \  \\ 
(4,1)  & $8.395\times 10^{-13}$ & $8.40\times 10^{-13}$ & $1.8803\times 10^{-11}$  &  $1.880\times 10^{-9}$  \ \  \\ 
(5,5) & $9.4559\times 10^{-7}$ & $9.4560\times 10^{-7}$ & $2.11785\times 10^{-5}$  &  $2.11785\times 10^{-5}$  \ \  \\ 
(5,4)  & $1.23238\times 10^{-7}$ & $1.232\times 10^{-9}$ & $2.7602\times 10^{-7}$  &  $2.760\times 10^{-9}$  \ \  \\ 
(5,3) & $1.0932\times 10^{-9}$ & $1.093\times 10^{-9}$ & $2.4485\times 10^{-3}$  &  $2.448 ×\times 10^{-8}$  \ \  \\
(5,2)  & $2.79\times 10^{-12}$ & $2.78\times 10^{-12}$ & $6.25\times 10^{-11}$  &  $6.2\times 10^{-11}$  \ \  \\ 
(5,1)  & $1.3\times 10^{-15}$ & $1.0\times 10^{-15}$ & $2.8\times 10^{-14}$  &  $3.0\times 10^{-12}$  \ \  \\  \hline \hline 
\rowcolor{retroorange} Total & $2.0290770\times 10^{-4}$ & $2.0290770 \times 10^{-4}$  & $4.5445256\times 10^{-3}$  & $4.5445256 \times 10^{-3}$ \ \  \\  \hline \hline 
\rowcolor{retroyellow} $\eta$ & $1.4\times 10^{-8}$ & $3.8 \times 10^{-8}$  & $1.2\times10^{-8}$  & $2.0 \times 10^{-8}$  \ \  \\
\hline \hline
\end{tabular}
\caption{Comparison of energy fluxes at $\mathscr{I}^{+}$, in units of $(M / \mu)^{2}$ against reference FD values comparing to our previous time-domain work \cite{da2023hyperboloidal} which was solving a different set of RWZ equations as described in \cite{thompson2017gauge, thompson2019gravitational}. These results verify our numerical implementation, the gauge invariance associated of energy and angular momentum fluxes and the equivalency between the different RWZ equations formalisms.} \label{tab_energyfluxes}
\end{center}
\end{table*}

\begin{table*}
\begin{center}
\begin{tabular}{||l|| c || c ||c|| c ||c||  }
\hline \hline 
\textrm{$(l,m)$}& 
\textrm{$\dot{E}^{\infty}_{lm}$}&
$\eta$&
\textrm{$\dot{L}^{\infty}_{lm}$} &
$\eta$\\ \hline \hline 
(2,1) & $7.81752\times 10^{-5}$& $7.5\times 10^{-7}$ & $1.76890 \times 10^{-5}$ & $7.5\times 10^{-7}$   \ \  \\ 
(2,2) & $1.6481030\times 10^{-4}$& $4.8\times 10^{-8}$ & $3.729231 \times 10^{-3}$ & $4.9\times 10^{-8}$   \ \  \\ 
(3,1)& $2.0883\times 10^{-9}$& $1.3\times 10^{-5}$ & $4.725 \times 10^{-8}$ & $1.3\times 10^{-5}$  \ \  \\ 
(3,2)& $2.3974\times 10^{-7}$& $1.1\times 10^{-6}$ & $5.4247 \times 10^{-6}$ & $1.1\times 10^{-6}$  \ \  \\ 
(3,3) & $2.44440\times 10^{-5}$& $4.1\times 10^{-6}$ & $5.5301 \times 10^{-4}$ & $4.1\times 10^{-6}$  \ \  \\
\hline \hline 
\end{tabular}
\caption{Comparison of the energy and angular momentum fluxes at $\mathscr{I}^{+}$ for when the point-particle is at $r_{p} = 8M$. The results presented here were obtained with the numerical factors of \cite{o2022timeQMUL, o2022time} for comparison and to avoid erroneous implementations of the \texttt{DiscoTEX} algorithm. One too uses, $J=10$ jumps, $N=45$ Chebyshev nodes and a timestep of $\Delta \tau = 0.01$. The numerical error was calculated against the frequency-domain results of \cite{BHPToolkit}. See footnote 12.}\label{tab_energyfluxes_8m}
\end{center}
\end{table*}
For consistency and to fairly compare against our work in \cite{da2023hyperboloidal} we have decided to use the same optimisation factors: $N=89$ Chebyshev collocation nodes, $J=11$ jumps, a time-step of $\Delta \tau = 0.02$ with the physical quantities extracted at around $\tau = 10,000$. It should be stressed implementation of \texttt{DiscoTEX} should always be accompanied by thorough numerical studies studying the optimal control factors, given the difference in the equations and source terms, with this version and the version in \cite{da2023hyperboloidal} there could be other optimal choices. On the left plot of Figure \ref{ch4_gsfFields_plots} we see the gravitational waveform at $(l,m)=(5,5)$ with the field being highly irregular at earlier times due to the junk radiation. As explained in \cite{da2023hyperboloidal} after waiting a minimal time, \textit{steady-state} is reached and we observe the expected periodic pattern for a point-particle on a circular geodesic with an angular frequency $m\Omega_{\phi}$, where $\Omega_{\phi}$ is as defined in equation \eqref{ch4_circ_orb_parameters}. In Table \ref{tab_energyfluxes} the energy fluxes are computed for the first 5 modes against previous frequency-domain work \cite{thompson2019gravitational}. Furthermore, in Table \ref{tab_energyfluxes_8m}, we give results for $r_{p} =8M$ with numerical optimisation factors $N=45$ Chebyshev nodes, $J=10$ and $\Delta \tau = 0.01$ as the solutions to these equations have previously been attempted, albeit erroneously, in the literature by \cite{o2022timeQMUL, o2022time} with those optimisation factors (see footnote 12). 
\begin{small}
\begin{table*}
\begin{tabular}{||l||  c c ||c|| ||c c ||c||}
\hline \hline 
\textrm{$r_{p}$}& 
\textrm{$\dot{\mathcal{E}}_{\text{total}}$}&
\textrm{$\mathcal{F}_{t}^{lm}$}&
\textrm{$\eta$}&
\textrm{$\dot{\mathcal{L}}_{\text{total}}$}&
\textrm{$\mathcal{F}_{\phi}^{lm}$}&
\textrm{$\eta$}\\ \hline \hline 
$7.9456 M$ & $2.03294497*10^{-4}$ & $3.20775924*10^{-4}$ & $1.3*10^{-9}$ & $4.5531889267*10^{-3}$ & $-7.1844216179*10^{-2}$& $2.9*10^{-11}$ \\ \hline \hline 
\end{tabular}
\caption{ Comparison of the gravitational self-force $t-$ component, $\mathcal{F}_{t}$ from our time domain algorithm against state-of-the-art frequency domain results of \cite{thompson2019gravitational} for the first $l_{max} =20$ modes.}
\label{ch3g3_table_Flt_fluxbalancelaw}
\end{table*}
\end{small}
\subsubsection{Phase 2 and Phase 3 - Numerical weak-form solutions for metric perturbation reconstruction  and gravitational self-force computations via \texttt{DiscoTEX} interpolations}

In this section, we show how the metric perturbations prescribed by a point-particle on a circular geodesic can be reconstructed and used to compute the conservative gravitational self-force (GSF) through the \texttt{DiscoTEX} algorithm. The results of \cite{da2023hyperboloidal} will be used and, as in the previous section, the intention is merely to demonstrate the algorithm's main application. The complete results are to be included in upcoming work with collaborators \cite{paper2, paper3}. \\

\textbf{GSF computations form dissipative laws}\\

As briefly mentioned in Section \ref{Sec1_intro} it is possible to compute the $t$-component of the self-force, $\mathcal{F}_{t}$ by computing the total energy flux radiated into/out of the black hole \cite{barack2007gravitational, detweiler2008consequence, phdthesis-jonathan} using the numerical results from Phase I from the previous section for the master functions of RW/Z,  
\begin{equation}
   \mathcal{F}_{t}  = - \frac{1}{2u^{t}} u^{\alpha}u^{\beta}\partial_{t}h_{\alpha\beta} = 2 \ \  \frac{1}{u^{t}}\frac{d\mathcal{E}}{dt}. 
   \label{self_force_tcomp_fluxbalance}
\end{equation}
Furthermore $\phi$-component of the gravitational self-force can be further computed, due to the helical symmetry present in circular geodesic motion, 
\begin{align}
      \label{energy_ang_total_Helicalsym}
    &\dot{\mathcal{E}}_{\text{total}} + \Omega \ \dot{\mathcal{J}}_{\text{total}} = 0, \\
    &\mathcal{F}_{t} + \Omega \ \mathcal{F}_{\phi} = 0. 
    \label{ft_fphi__total_Helicalsym}
\end{align}

On Table \ref{tab_energyfluxes}  the energy fluxes for the first $l$ modes of the self-force components are computed, as $\mathcal{F}_{t}$ and $\mathcal{F}_{\phi}$ and with an unforeseen accuracy of $10^{-9}$/$10^{-11}$ through time-domain methods from the direct numerical solution of a distributionally sourced PDE such as equation \eqref{ch4_CPM_ZM_wavequation} showing the power of the \texttt{DiscoTEX} numerical algorithm directly addressing all the three main difficulties highlighted in the introduction. \\
\begin{figure}
\includegraphics[width=92mm]{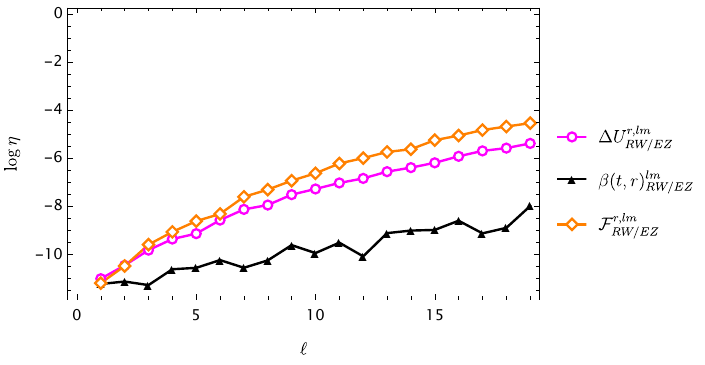}
\caption{\label{modesum_odd} Numerical convergence result for all the un-regularised physically relevant quantities for Phase II and Phase III for the computation of the gravitational self-force in the time domain in the Regge-Wheeler gauge. All $l_{\text{max}} = 20 $ modes against reference frequency domain values of \cite{thompson2019gravitational} at $r_{p} = 7.9456M$ are given. } 
\end{figure}
\textbf{Conservative GSF computations}\\

As a preview into our upcoming work \cite{paper2,paper3} and for the sake of completion of the numerical algorithm highlighted in Section \ref{Sec3}, we will also show how the numerical machinery here can be used to reconstruct the metric and compute the \textit{un-regularised} \footnote{Regularised self-force terms require no numerical computations and are obtained analytically as explained in \cite{Heffernan:2012su, thompson2017gauge, thompson2019gravitational}} gravitational self-force. To compute this, the fields are evaluated at the vicinity of the particle i.e. at the particle limit from the right and the left as the radiation propagates towards infinity and the horizon of the black hole. For simplicity, we choose to demonstrate \texttt{DiscoTEX} for the odd-component given the equations are not singular as shown in \cite{thompson2019gravitational}. The results here are computed against the state-of-the-art frequency domain results of \cite{thompson2019gravitational} transformed to the time-domain case. We do not explicitly give the magnitudes as this is part of upcoming collaborative work \cite{paper2, paper3}. The \textit{un-regularised} odd-parity self-force given for Regge-Wheeler and Eazy gauges radiative modes is the same and one has
\begin{eqnarray}
    \mathcal{F}^{r,lm}_{RW/EZ}(r_{p}) =  \frac{r_{p} f_{p} \Omega}{r_{p}-3M} \bigg[\beta + r_{p} \partial_{r}\beta \bigg] \partial_{\theta}Y_{lm}\bigg( \frac{\pi}{2}, \phi_{p}(t) \bigg). 
    \label{gsf_odd_RW_EZ}
\end{eqnarray}
The only required metric perturbation is given as \cite{thompson2019gravitational, hopper2010gravitational}
\begin{equation}
    \beta(t,r) = -\frac{1}{\lambda r} \bigg[f \Psi^{a}(t,r) + r f \partial_{r}\Psi^{a}(t,r) - r^{3} E_{C} \bigg].
    \label{oddmetric_beta}
\end{equation}
where $\lambda$, $E_{C}$ are constant quantities defined in Appendix B.2 of \cite{da2023hyperboloidal}. Even though this quantity contains a singular term $E_{C}$ one can show by writing its weak-form solution that all components are regular and should $vanish$ at the particle, see for example \cite{hopper2010gravitational}. The weak form is as given in the Introduction section of this manuscript by equation \eqref{sec1_pert_metric}. Similarly, we must also calculate the red-shifts as in \cite{thompson2019gravitational}
\begin{equation}
    \Delta U^{r,lm}_{RW/EZ}(r_{p}) = \frac{2 r_{p}^{2} \Omega }{r_{p} - 3M} \beta \ \partial_{\theta}Y_{lm}\bigg( \frac{\pi}{2}, \phi_{p}(t) \bigg). 
    \label{uredshift_odd}
\end{equation}
For the numerical weak-form solutions here, we choose to work with Chebyshev collocation nodes due to the significantly faster results and comparable accuracy at the particles. The numerical choices are $N=89$ Chebyshev nodes, $J=11$ jumps, and it was found a time of $\tau=400$ with a $\Delta \tau =0.02$ to be sufficient for accurate computations. As demonstrated in the previous sections, we use the numerical solutions obtained via \texttt{DiscoTEX} with the generic discontinuous interpolator inequation \eqref{ch3_sb2_WFsolution} and evaluate at the particle position of $\sigma_{p}$, 
\begin{equation}
    \Psi(\tau,\sigma) \bigg|_{\sigma = \sigma_{p}}\approx \sum^{N}_{j=0} \bigg[ \Psi_{j}(\tau) + \Delta_{\Psi}(\sigma_{j} - \sigma_{p}(\tau); \sigma - \sigma_{p}(\tau))  \bigg] \pi_{j}(\sigma)\bigg|_{\sigma = \sigma_{p}}, 
    \label{genericinterpol}
\end{equation}
Interpolation for the field derivative, necessary for equations (\eqref{gsf_odd_RW_EZ} -  \eqref{uredshift_odd}) computations, follows trivially from this formula, we note though care must be taken when transforming back to the physical quantities from the hyperboloidal ones. \footnote{This will be explained with more care in upcoming work \cite{paper2, paper3}, given the simplicity of these equations careful application of the chain rule suffices, see equation \eqref{ch4_dr_chainrule} for example. }

In Figure \ref{modesum_odd} we demonstrate the numerical results for the first $l=20$ modes for the metric perturbation amplitudes gravitational self-force and redshift amplitudes given as numerical weak-form solutions via \texttt{DiscoTEX}. For all these results the cumulative error for each mode was calculated, and we highlight that, as seen in results of \cite{da2023hyperboloidal} in Section IV, the physical quantities magnitudes and hereby contributions to final results of equations (\eqref{gsf_odd_RW_EZ}- \eqref{uredshift_odd}) decrease as the number of $l$ modes increases, and hereby we expect the final error contributions to become less and less significant as we will show in the third paper \cite{paper3} of the series starting with \cite{da2023hyperboloidal}. The main observation is that the \texttt{DiscoTEX} numerical algorithm is capable of highly accurate computations required for the gravitational self-force programme \cite{pound2022black, afshordi2023waveform}.  \\

\subsubsection{Numerical computation of Phase I via \texttt{DiscoTEX}'s direct relatives: \texttt{DiscoIMP} and \texttt{DiscoREX}}\label{sec_discotex_family}

For completion and to complement the section that follows, where we will review the previous numerical methods used when modelling (X)/(E)MRIs, we will further compute numerical solutions to Phase I as done in Section \ref{phase1}. Specifically, the even solution of the RWZ equations will be computed and compared using three different numerical integration methods: the purely implicit algorithm, i.e. \texttt{DiscoIMP} and the purely explicit numerical algorithm i.e. \texttt{DiscoREX}. 

The parent algorithm i.e. \texttt{DiscoIMP} at second- (\texttt{H2}) and fourth-order (\texttt{H4}) is given as, respectively, 
\begin{align}
    \Pi_{n+1} &= \texttt{LIMPH2} \cdot \bigg[\texttt{RIMPH2} \cdot \Pi_{n} + (\Delta \tau \ \textbf{L}_{1}) \cdot \Psi_{n} \nonumber  \\ 
    &+ \frac{\Delta \tau}{2} \bigg( \textbf{s}_{\Psi, n} + \textbf{s}_{\Psi, n+1} + \textbf{s}_{\Pi, n} + \textbf{s}_{\Pi, n+1} \bigg)  \bigg], \nonumber \\ 
    \Psi_{n+1} &= \Psi_{n} + \frac{\Delta \tau}{2}\bigg(\Pi_{n} + \Pi_{n+1} \bigg), 
    \label{emris_hyper_discoimph2}
\end{align}

where here,  
\begin{align}
    &\texttt{LIMPH2} = \bigg[ \textbf{I} - \frac{\Delta \tau \ \textbf{L}_{2} }{2} + \frac{\Delta \tau^{2} \ \textbf{L}_{1} }{4}  \bigg]^{-1}, \nonumber \\
    &\texttt{RIMPH2} = \bigg[ \textbf{I} + \frac{\Delta \tau \ \textbf{L}_{2} }{2} + \frac{\Delta \tau^{2} \ \textbf{L}_{1} }{4}  \bigg].\nonumber
\end{align}

\begin{align}
    \Pi_{n+1} &= \texttt{LIMPH4} \cdot \bigg[\texttt{RIMPH4} \cdot \Pi_{n} + \frac{\Delta \tau^{2}}{12} \bigg( \dot{\textbf{s}}_{\Psi, n} - \dot{\textbf{s}}_{\Psi, n+1} + \dot{\textbf{s}}_{\Pi, n} - \dot{\textbf{s}}_{\Pi, n+1} \bigg) \nonumber \\
    &+ \texttt{SIMPH4}\cdot \bigg(\textbf{s}_{\Psi, n}  + \textbf{s}_{\Psi, n+1} \bigg)+ \bigg(\frac{\Delta \tau \ \textbf{L}_{1}}{2} + \frac{\Delta \tau^{2} \ \textbf{L}_{2} \cdot \textbf{L}_{1}}{12} \nonumber \\
    &+ \bigg( \frac{\Delta \tau \ \textbf{L}_{1}}{2} - \frac{\Delta \tau^{2} \ \textbf{L}_{2} \cdot \textbf{L}_{1}}{12} \bigg) \bigg) \cdot \Psi_{n} 
    + \texttt{SNIMPH4}\cdot \bigg(\textbf{s}_{\Pi, n} - \textbf{s}_{\Pi, n+1} \bigg) \bigg], \nonumber \\ 
    \Psi_{n+1} &= \Psi_{n} +  \texttt{LLIMPH4}\cdot \bigg[ \frac{\Delta \tau}{2}\bigg(\Pi_{n} + \Pi_{n+1} \bigg) + \frac{\Delta \tau^{2} \ \textbf{L}_{2}}{2}\bigg(\Pi_{n} - \Pi_{n+1} \bigg) \nonumber \\ 
    &+ \frac{\Delta \tau^{2}}{12} \bigg(\textbf{s}_{\Psi, n} - \textbf{s}_{\Psi, n+1} + \textbf{s}_{\Pi, n} - \textbf{s}_{\Pi, n+1} \bigg) \bigg].
    \label{emris_hyper_discoimph4}
\end{align}
Here, 
\begin{align}
    &\texttt{LIMPH4} = \bigg[ \textbf{I} - \frac{\Delta \tau \ \textbf{L}_{2} }{2} + \frac{\Delta \tau^{2} \ (\textbf{L}_{1} + \textbf{L}_{2}\cdot \textbf{L}_{2}) }{12}\nonumber \\
    &- \bigg(\frac{\Delta \tau \ \textbf{L}_{1} }{2} - \frac{\Delta \tau^{2} \ \textbf{L}_{2} \cdot \textbf{L}_{1}}{12} \bigg) \cdot \texttt{LLIMPH4} \cdot\bigg( \frac{\Delta \tau \ \textbf{I}}{2} - \frac{\Delta \tau^{2} \ \textbf{L}_{2}}{12} \bigg) \bigg], \nonumber \\
    &\texttt{RIMPH4} = \bigg[ \textbf{I} + \frac{\Delta \tau \ \textbf{L}_{2} }{2} + \frac{\Delta \tau^{2} \ (\textbf{L}_{1} + \textbf{L}_{2}\cdot \textbf{L}_{2}) }{12} \nonumber \\ 
    &+ \bigg(\frac{\Delta \tau \ \textbf{L}_{1} }{2} -\frac{\Delta \tau^{2} \ \textbf{L}_{2} \cdot \textbf{L}_{1}}{12} \bigg) \cdot \texttt{LLIMPH4} \cdot\bigg( \frac{\Delta \tau \ \textbf{I}}{2} + \frac{\Delta \tau^{2} \ \textbf{L}_{2}}{12} \bigg) \bigg], \nonumber \\
    &\texttt{LLIMPH4} = \bigg[ \textbf{I} + \frac{\Delta \tau^{2} \ \textbf{L}_{1} }{12}\bigg]^{-1},\nonumber\\
    &\texttt{SIMPH4} = \bigg[ \frac{\Delta \tau}{2} \textbf{I} + \frac{\Delta \tau^{2} \ \textbf{L}_{2} }{12} + \frac{\Delta \tau^{2}}{12} \nonumber \\ 
    &\times \bigg(  \frac{\Delta \tau \ \textbf{L}_{1} }{2} -  \frac{\Delta \tau^{2} \ \textbf{L}_{2} \cdot \textbf{L}_{1} }{12} \bigg) \cdot \texttt{LLIMPH4}\bigg],\nonumber\\
    &\texttt{SNIMPH4} = \bigg[ \frac{\Delta \tau}{2} \textbf{I} - \frac{\Delta \tau^{2} \ \textbf{L}_{2} }{12} - \frac{\Delta \tau^{2}}{12} \nonumber \\&\times \bigg(  \frac{\Delta \tau \ \textbf{L}_{1} }{2} -  \frac{\Delta \tau^{2} \ \textbf{L}_{2} \cdot \textbf{L}_{1} }{12} \bigg) \cdot \texttt{LLIMPH4}\bigg].\nonumber
\end{align}

To further justify the choice of \texttt{DiscoTEX} at \texttt{4th-} order as the optimal numerical scheme for modelling (X)/(E)MRIs we compare its results to solutions computed at lower (\texttt{2th}) and higher-orders (\texttt{6th-12th}). The details of these schemes can be found in \cite{discotexII}, where here the following nullifications $\textbf{J}_{\texttt{H2-H12}} (\Delta \tau_{\times}, \Delta \tau) \rightarrow 0 $ and $\Upsilon \rightarrow 0 $ apply. 

Finally, we further compare our results with a purely-explicit scheme \texttt{DiscoREX} at order-2 given as, 

\begin{align}
    &\textbf{U}_{n+1} = \textbf{U}_{n} + (\textbf{HFRK2} - \textbf{I}) \cdot \textbf{U}_{n} + \frac{\Delta \tau}{2} \textbf{s}_{n+1} + \frac{\Delta \tau}{2} \bigg( \textbf{I} + \textbf{A} \bigg) \cdot \textbf{s}_{n} , 
    \label{emris_hyper_discoexrk2}
\end{align}

with the term \textbf{HFRK2} defined in equations (\eqref{ch3_hornerform_rk2}-\eqref{ch3_hornerformer_rk2mat}) in \ref{app_ex_rk}. Unlike all the aforementioned evolution algorithms \texttt{DiscoTEX} and \texttt{DiscoImp}, \texttt{DiscoEX} as a purely explicit \texttt{Runge-Kutta} method will need a fifth user-specifiable numerical optimisation control factor, i.e. we need to determine the Courant limit given as, 
\begin{align}
    \textnormal{CFL} = \frac{\Delta t}{\Delta x} \leq \textnormal{CFL}_{\rm{max}}.
    \label{cfl}
\end{align}
Thus, for \texttt{DiscoREX} we need to investigate five control factors, i.e. 
\begin{enumerate}
    \item \textcolor{gray}{\texttt{[CTRL F\ref{controlfactor1}]} - Number of \texttt{N} nodes;}
    \item \textcolor{gray}{\texttt{[CTRL F\ref{controlfactor2}]} - Number of \texttt{J} jumps;} 
    \item \textcolor{gray}{\texttt{[CTRL F\ref{controlfactor3}]} - $\Delta \tau$, time step-size;} 
    \item \textcolor{gray}{\texttt{[CTRL F\ref{controlfactor4}]} - Minimal time for steady state evolution;}
    \item \texttt{[CTRL F5]} - Assess CFL condition, as given in equation \eqref{cfl} is obeyed by, for example, calculating the eigenvalues. \label{controlfactor5_rex}
\end{enumerate}

To find out a reasonable CFL, we calibrate our choice by computing the eigenvalues, through the Mathematica command, 
\begin{align}
\lambda = \texttt{Max[Abs[Eigenvalues[\textbf{HFRK2}]]]}.
\label{chapter7_rk2_eigenvalues}
\end{align}

The full details, for completion, will be given in an upcoming publication \cite{discoREX} up to eight orders. \\

\begin{table*}
\centering
\begin{small}
\begin{tabular}{||l|| c  || c ||c || c|| }
\hline \hline 
\texttt{Algorithm}&
\textrm{$\dot{E}^{\infty}_{2,2}$ }&
$\eta$ &
\texttt{Wall-clock times, s}&
\texttt{[\# CTRL Factors]}\\ \hline \hline 
\rowcolor{Hred} \texttt{DiscoEX RK2}&$1.70\times10^{-4} $ & $2.2\times10^{-5}$& 143.1869502& 5 \\ \hline \hline
\rowcolor{retroorange} \texttt{DiscoIMP H2}& $1.70\times10^{-4}$ & $2.8\times10^{-5}$ &84.3186792& 4 \\ 
\rowcolor{retroorange} \texttt{DiscoIMP H4}& $1.7062\times10^{-6}$  &$4.7\times10^{-9}$ & 211.5136043& 4\\ \hline \hline 
\rowcolor{mygreen}\texttt{DiscoTEX H2}& $1.70\times10^{-4}$ & $2.8\times10^{-5}$ &83.0924737& 4 \\
\rowcolor{mygreen}\texttt{DiscoTEX H4}&$1.7062\times10^{-4}$&$4.7\times10^{-9}$& 107.1625516& 4\\ 
\rowcolor{mygreen}\texttt{DiscoTEX H6}&$1.7062\times10^{-4}$ & $4.7\times10^{-9}$ &172.3661989& 4 \\
\rowcolor{mygreen}\texttt{DiscoTEX H8}&$1.7062\times10^{-4}$&$4.7\times10^{-9}$&  311.9818151& 4\\ 
\rowcolor{mygreen}\texttt{DiscoTEX H10}& $1.7062\times10^{-4}$ & $4.7\times10^{-9}$ & 585.3051247& 4 \\
\rowcolor{mygreen}\texttt{DiscoTEX H12}&$1.7062\times10^{-4}$&$4.7\times10^{-9}$&  1128.9236532& 4\\ \hline \hline 
\end{tabular}
\end{small}
\caption{Comparison between \texttt{DiscoTEX H2-H12} with its direct relatives i.e. parent \texttt{DiscoIMP H2-H4} and cousin \texttt{DiscoEX RK2}.  Remarkably, \texttt{DiscoTEX} and \texttt{DiscoIMP} retain the same accuracy, with \texttt{DiscoTEX H4} offering the best compromise with accuracy and minimal simulation running time. \texttt{DiscoEX RK2}, unlike all others, requires 5 user-specifiable numerical optimisation control factors, is conditionally stable and, of all order-2 methods takes the longest time, even more time than \texttt{DiscoTEX H4}. With this, it is recommended the use of \texttt{DiscoTEX H4}.} 
\label{table_energies_timedomain_scalar_allalgorithms} 
\end{table*}

For our numerical simulations, the following numerical optimisation control factors were chosen: an intermediary number of Chebyshev collocation nodes was chosen as \texttt{[CTRL F \ref{controlfactor1}- F\ref{controlfactor2}]} $N=60$ and jumps, $J=11$, were chosen and for \texttt{[CTRL F\ref{controlfactor3}]} we choose a time-step of $\Delta \tau = 0.02$, with the exception of the implementation of \texttt{DiscoREX RK2} which, due to being a conditionally stable scheme required the timestep to be lowered to $\Delta \tau_{\texttt{[CTRL F\ref{controlfactor3}]}} = 0.002$, yielded a $\texttt{CFL}_{\texttt{[CTRL F\ref{controlfactor5_rex}]}} \approx 6.6$. For comparison we compute the relative error of the energy flux, for example for the $(l,m) = (2,2)$ mode at $\mathscr{I}^{+}$. From Table \ref{table_energies_timedomain_scalar_allalgorithms} it is clear that \texttt{DiscoTEX H4} has a significant advantage over all other algorithms including its parent \texttt{DiscoIMP H4}, which even though registers, remarkably, the exact same numerical accuracy it takes a significantly longer time which, in addition to its more complex implementation, as evidenced by equation \eqref{emris_hyper_discoimph4}, places it at a significant disadvantage when compared to its child, \texttt{DiscoTEX H4}. The accuracy between all the algorithms tested, at the same order, registers the same accuracy, with the exception, albeit only marginally, of \texttt{DiscoREX RK2}. Furthermore and for completion we observe in Figures \ref{grav_pps_discoEX_RK2} and \ref{grav_pps_parent_higher}, that all algorithms for this set of numerical control factors \texttt{[CTRL F\ref{controlfactor1}-F\ref{controlfactor5_rex}]}, conserve symplectic structure numerically. From Figure \ref{grav_pps_discoEX_RK2}, in \ref{app_discotex_family} it is clear \texttt{DiscoREX} takes slightly longer to achieve such a feat. 

\subsection{State of numerical algorithms in the time-domain for numerical black hole-perturbation theory applications}
\begin{table*}
\begin{small}
\begin{tabular}{||c ||c|| c|| c||  c||  c||  }
\hline
\textrm{Numerical Scheme}& 
\textrm{Difficulty 1}& 
\textrm{Difficulty 2}& 
\textrm{Difficulty 3}& 
\textrm{Accuracy }&
\textrm{Applications} \\ \hline \hline 
\rowcolor{Hred} Lousto et al & Finite-difference & RBCs. & \texttt{EX}  Finite-difference & Accurate &\texttt{[REQ 1, 2]} \\ 
\rowcolor{Hred}  \cite{lousto1997head}  &  & &  c. stable, constrained & \texttt{2n} order &\cite{martel2002one, martel2004gravitational,barack2017time, long2021time} \\ \hline 
\rowcolor{Hred} Lousto et al & Finite-difference &  RBCs. & \texttt{EX}  Finite-difference & Accurate &\texttt{[REQ 1,2]} \\ 
\rowcolor{Hred} \cite{lousto2005time}  &  & &  c. stable, constrained & \texttt{4th} order &  \cite{haas2007scalar, barack2010gravitational} \\ \hline \hline  
\rowcolor{myorange} Sopuerta et al & Finite element & RBCs.& \texttt{IMEX} & Accurate & \texttt{[REQ 1]} \ \  \\ 
\rowcolor{myorange} \cite{sopuerta2005toy, sopuerta2006finite} & method & & See \cite{sopuerta2006finite} & \texttt{4th} order  &  \cite{sopuerta2006finite} \\ \hline
\rowcolor{myorange} Sopuerta et al & Multi-domain & RBCs. & \texttt{EX} \texttt{RK4} &Accurate & \texttt{[REQ 1,2]} \cite{canizares2009efficient, canizares2010efficient} \ \  \\ 
\rowcolor{myorange} \cite{canizares2009efficient} & pseudospectral  &  &c. stable, constrained  & \texttt{4th} order  & \cite{canizares2010modelling, canizares2010pseudospectral, canizares2011tuning, canizares2011time}\\ \hline  \hline 
\rowcolor{hyellow}Hughes et al& Finite impulse  & \cite{krivan1997dynamics} BCs & \texttt{EX} Lax-Wendroff &Accurate   & \texttt{[REQ 1]} \cite{barausse2012modeling}  \\ 
\rowcolor{hyellow} \texttt{$\delta$-code} \cite{sundararajan2007towards, sundararajan2008towards} &representation &  & c.stable, constrained&  \texttt{2nd} order &  Sev. \cite{ taracchini2014small,apte2019exciting, lim2019exciting}  \\ \hline  \hline 
\rowcolor{lred} Dolan, Barack  & Finite-difference &  RBCs. & \texttt{EX} \texttt{RK4}& Accurate &\texttt{[REQ 1,2]} \cite{dolan2011self, dolan2011self2} \\ 
\rowcolor{lred} \cite{barack2007m, barack2007scalar} & Puncture  & &  c. stable, constrained & \texttt{4th} order & \cite{dolan2023metric} \\ \hline
\rowcolor{lred} Dolan & Finite-difference &  RBCs. & \texttt{EX} \texttt{RK4} & Accurate &\texttt{[REQ 1]}  \\
\rowcolor{lred} \cite{dolan2013superradiant} &  & &  c. stable, constrained & \texttt{4th} order & \cite{dolan2013superradiant} \\ \hline \hline 
\rowcolor{mygreen}Khana et al & Gaussian approximation &  \cite{krivan1997dynamics} BCs & \texttt{EX} Lax-Wendroff & Accurate  & \texttt{[REQ 1]}  \\ 
\rowcolor{mygreen}\cite{krivan1997dynamics, lopez2003perturbative} & the $\delta$ distribution  &   &c.stable, constrained & \texttt{2nd} order &  \cite{khanna2004teukolsky, burko2007accurate} \\ \hline  \hline 
\rowcolor{jgreen}Thornburg et al& Finite difference & Pure Outflow BCs & \texttt{EX} \texttt{RK4} &Accurate   & \texttt{[REQ 1,2]} \\ 
\rowcolor{jgreen} \cite{thornburg2017scalar} & with & with FD & c.stable, constrained&  \texttt{4th} order &   \cite{thornburg2017scalar}  \\  
\rowcolor{jgreen} &Berger-Oliger & ghost-zones & \texttt{IMEX RK} considered & \texttt{3rd} order  &   \\ 
\rowcolor{jgreen} &mesh refinement  &  & c.stable, \cite{boscarino2007error, boscarino2009accurate, boscarino2009class} & & See App.B11 \cite{thornburg2017scalar} \\ \hline  \hline
\rowcolor{mybblue}Nagar et al & Gaussian approximation & RBCs &  \texttt{EX} Lax-Wendroff & Accurate & \texttt{[REQ 1]}  \\ 
\rowcolor{mybblue} \cite{nagar2007binary}  &the $\delta$ distribution  & &c.stable, constrained&  \texttt{2nd} order  &  \cite{damour2007final, damour2007faithful, damour2008binary}  \\  \hline
\rowcolor{mybblue}Bernuzzi et al & Gaussian approximation & RBCs &  \texttt{EX} \texttt{RK4}  & Accurate & \texttt{[REQ 1]}  \\ 
\rowcolor{mybblue} \cite{bernuzzi2010binary}  &the $\delta$ distribution  & &c.stable, constrained& \texttt{4th} order  &    \\  \hline
\rowcolor{mybblue} \texttt{Teukode} & Gaussian approximation&hyperboloidal layer&  \texttt{EX} \texttt{RK4}  & Accurate & \texttt{[REQ 1]} \ \  \\ 
\rowcolor{mybblue}\cite{bernuzzi2011binary, harms2013numerical, harms2014new}  & the $\delta$ distribution &  & c.stable, constrained& \texttt{4th}order&  Sev. \cite{albanesi2023faithful, harms2016asymptotic, harms2016spinning} \\ \hline  \hline 
\rowcolor{mydblue} Diener et al& \texttt{8th} order & Penalty & \texttt{EX} \texttt{RK4} & Accurate & \cite{schnetter2006multi, vega2013scalar}\\ 
\rowcolor{mydblue}  \cite{vega2013scalar}  &Finite-difference& BCs \cite{diener2005new} &c.stable, constrained&\texttt{4th} order   &  \texttt{[REQ 1,2]}[$1^{st}$ \texttt{SCE}]\  \\ \hline
\rowcolor{mydblue} Diener et al & Discontinuous & hyperboloidal layer &\texttt{EX} \texttt{RK4}& Highly acc. & \texttt{[REQ 1]} \\ 
\rowcolor{mydblue} \cite{samuelcuppC25, phdthesis-cupp}  &Garlerkin  & & c.stable, constrained& \texttt{4th} order &  \cite{samuelcuppC25, phdthesis-cupp}  \\ \hline  \hline 
\rowcolor{mypurple} Field et al & Discontinuous  & RBCs & \texttt{EX} \texttt{RK4} & Accurate &  \texttt{[REQ 1]} \\ 
\rowcolor{mypurple} \cite{field2009discontinuous, field2010persistent}  &Garlerkin  & &c.stable, constrained&\texttt{4th} order &  \cite{field2009discontinuous, field2010persistent} \\  \hline
\rowcolor{mypurple} Field et al & Discontinuous & hyperboloidal layer & \texttt{EX} \texttt{RK4}& Highly acc.  &  \texttt{[REQ 1]}\\ 
\rowcolor{mypurple} \cite{vishal2023towards}  &Garlerkin  & & c.stable, constrained& \texttt{4th} order & \cite{vishal2023towards} \\ \hline  \hline 
\rowcolor{mypinkt} \texttt{DiscoREX}  & Discontinuous  &hyperboloidal slice& \texttt{EX} \texttt{RK2, RK4}& Highly acc. &  \texttt{[REQ 1,2]} \cite{phdthesis-lidia, discoREX} \\ 
\rowcolor{mypinkt} \cite{reviews-lidia, phdthesis-lidia, discoREX} & collocation LIP &minimal gauge  & c.stable, constrained  &\texttt{2nd/4th} order&  \  \\ \hline   
\rowcolor{mypinkt} \texttt{DiscoIMP}  &Discontinuous &RBCs& \texttt{IM} Hermite 2,4 & -- & \texttt{[REQ 1,2]} \cite{phdthesis-lidia} \\ 
\rowcolor{mypinkt}  \cite{2014arXiv1406.4865M,reviews-lidia, phdthesis-lidia} &collocation LIP  &--  &unconditionally stable &  &  \  \\ \hline
\rowcolor{mypinkt} \texttt{DiscoIMP}  & Discontinuous &hyperboloidal slice& -- & -- & \texttt{[REQ 1,2]} \cite{phdthesis-lidia} \\ 
\rowcolor{mypinkt}  \cite{reviews-lidia, phdthesis-lidia} &collocation LIP &minimal gauge  &  &  &   \cite{paper2, paper3} \\   \hline
\rowcolor{mypinkt}  \texttt{DiscoTEX}  & Discontinuous & RBCs & \texttt{IMTEX} Hermite 2-12& -- &  \texttt{[REQ 1,2]} \cite{phdthesis-lidia, discotexII}\\ 
\rowcolor{mypinkt}  \cite{reviews-lidia, phdthesis-lidia, discotexII} &collocation LIP  &  &  &  &  \  \\   \hline
\rowcolor{mypinkt}  \texttt{DiscoTEX}  & Discontinuous & hyperboloidal slice& -- & -- & \texttt{[REQ 1,2]}\\ 
\rowcolor{mypinkt}  \cite{reviews-lidia, phdthesis-lidia, discotexII} &collocation LIP&minimal gauge  &  &  &   \cite{da2023hyperboloidal, paper2, paper3}  \\   \hline  \hline 
\rowcolor{raspberry} Fully-spectral,& Multi-domain  & hyperboloidal slice& Spectral \texttt{IM} \texttt{SDIRK} & Highly acc. &  \texttt{[REQ 1,2]} \cite{paper2, paper3}\\ 
\rowcolor{raspberry} P.Macedo et al \cite{macedo2014axisymmetric}  & spectral decomposition &minimal gauge  & unconditionally stable &  \texttt{3rd} order &  \  \\   \hline  \hline 
\end{tabular}
\caption{Overview of numerical methods used by the radiation-reaction community for the modelling of EMRIs in the time domain. Here the short notation, c.stable, refers to conditionally stable. } \label{tab_emris_timedomain}    
\end{small}
\end{table*}

One of the key goals for the Laser Interferometer Space Antenna (LISA) scheduled for 2034 is the detection of EMRIs \cite{afshordi2023waveform}, and potentially, XMRIs (in Sgr A*) \cite{amaro2019extremely, gourgoulhon2019gravitational, vazquez2022revised, vazquez2023sgr, barsanti2022detecting, huang2024primordial}. These systems are more likely formed from the scattering capture mechanism where the small BH (or compact object) spirals radiatively into the massive BH with high initial eccentricities and a small periastron passage emitting gravitational-wave bursts as they radiatively inspiral until their orbit shrinks and circularises with a very sudden transition from inspiral to plunge. Before it plunges, it will though spend a very long time in inspiral in the strong-field regime, potentially taking years, making it an exceptionally promising probe for strong-gravity due to the very high signal-to-noise ratio of gravitational wave emission. One of the most promising tests will be the ability to give high precision measurements, that will allow to test the \enquote{no-hair} conjecture, as LISA is expected to be able to measure mass and spin with a precision of $10^{[-3,-6]}, 10^{[-4,-6]}$ respectively. Furthermore, it can help test alternative theories of general relativity by potentially probing extra polarisation modes of gravitational waves and they can also act as dark sirens providing independent measurements for the expansion rate of the Universe \cite{pound2022black, afshordi2023waveform}. \\

Most of the progress \cite{pound2020second, warburton2021gravitational, wardell2023gravitational, afshordi2023waveform} in the modelling of these systems comes from the radiation-reaction programme \cite{pound2022black, afshordi2023waveform} using the two-timescale expansion in the Fourier domain and hence all current waveform templates are available and generated through this framework. A fully alternative approach is to perform a self-consistent evolution (SCE) in the time domain where we could simply couple the field equations and the equations of motion evolving it as a system self-consistently sourcing the orbit by computing the GSF at each time step. So far only \cite{diener2012self} has performed this for a toy model describing a scalar charge on a non-rotating background. The SCE under the gravitational self-force, however, has remained a challenge. To understand why, it is important to emphasise the key proprieties of these physical systems:

\begin{enumerate}[leftmargin=0.75in]
    \item[ \hspace{0.1cm} \texttt{[F1]}]: Extreme disparity in length scales: mass and initial separation between the two bodies; \label{f1} 

    \item [ \hspace{0.1cm} \texttt{[F2]}]: Several thousands of orbital cycles, very long-term evolutions: days to years (EMRIs), or millions of years (XMRIs); \label{f2}

    \item[\hspace{0.1cm} \texttt{[F3]}]: Weak amplitude GW emission but strong SNR due to \texttt{[F2]}; \label{f3}
    
    \item[ \hspace{0.1cm} \texttt{[F4]}]: Complex intricate orbits - Expected to have high initial eccentricities, be spinning and inclined - if the formation scenario is \textit{scattering gravitational capture}; \label{f4}
 
    \item[\hspace{10cm} \texttt{[Alt F4]}]: Circular orbits - if the formation scenario is via \textit{tidal binary disruption}; \label{alt_f4}
    
    \item[ \hspace{0.1cm} \texttt{[Alt F4]}]: Equatorial circular orbits - if the formation of stellar remnants via accretion of the circumstellar disk surrounding the SBMH occurs; \label{altf4_accs}
    
    \item[ \hspace{0.1cm} \texttt{[F5]}]: Complex IMR with a sudden transition from inspiral to plunge; \label{f5}

    \item[\hspace{0.1cm} \texttt{[F6]}]: Possibility of GW emission as EMRBs from scattering/unbound hyperbolic systems before the EMRI have the chance to form during its slow inspiral.\label{f6}  
\end{enumerate} 

These proprieties, along with the aforementioned difficulties $1)-4)$ which must address \texttt{[REQ 1,2]} place clear \textbf{key} \textbf{constraints} on the range of suitable numerical methods. One of the longstanding struggles in modelling (X)/(E)MRIs in the Fourier domain, pertains to feature\texttt{[F4]}, i.e. handling eccentricity as it requires a high number of modes for convergence which in the time-domain translates to requiring a very small step-size \cite{barton2008computational, afshordi2023waveform}. Another concern comes from feature \texttt{[F3]}, as it will be a very long-term evolution and hence numerical methods which have their time step-size constrained are likely to fall short. Immediately, these two factors place a clear constraint in the ideal numerical evolution time-integrators used. Most of the progress in the time domain within the radiation-reaction self-force community, see Table \ref{tab_emris_timedomain}, has come from the application of time-step size constrained explicit numerical integrators mostly at \texttt{2nd}- and \texttt{4th}- order convergence. As such, time-domain results have usually been significantly less accurate and more computationally expensive than frequency-domain strategies. With the time-steppers, Hermite implicit \texttt{IM} and implicit-turned-explicit \texttt{IMTEX} implemented here with \texttt{DiscoIMP/DiscoTEX}, respectively, it was shown that these issues can be avoided altogether if unconditionally/long-term stable schemes are used. We have shown that highly accurate GSF computations are possible in the time domain and with competitive accuracy to frequency domain methods. It is also worth mentioning, that the results in \cite{da2023hyperboloidal} and here were meant to show agreement with the frequency domain approach and hence why we have chosen to compute as much as $l=20$ modes. However, careful convergence studies and further gauge of \texttt{DiscoTEX}'s $3$ numerical optimisation control factors, \texttt{CTRL F\ref{controlfactor1}-F\ref{controlfactor3}}, i.e number of jumps, nodes, times-step size can result in faster convergence and reduce the required number of modes, as expected in the time-domain, showing exceptional promise for modelling eccentric systems. Furthermore, in future work, we will show the implementation of a fully independent algorithm to \texttt{DiscoIMP/DiscoTEX} \footnote{Not fully independent as both algorithms \textbf{are hyperboloidal} addressing difficulty $2$ in the same manner, however, this just means we solve the same equation in the same coordinate chart $(\tau,\sigma)$. }, which is fully-spectral and also implements an advanced fully-spectral \texttt{SDIRK} unconditionally stable $3^{th}$ order time-stepper \cite{macedo2014axisymmetric} which has registered significant highly accurate results \cite{paper2, paper3}, conserves symplectic structure (albeit numerically) and it is faster than \texttt{DiscoTEX}. \\ 

Altogether, the simplicity of \texttt{DiscoTEX} as evidenced by the evolution algorithms in equations (\eqref{ch3_discoTEX4_Wave_tr}, \eqref{ch3_discoTEX4_Wave_hyper}), and its remarkable stability and coordinate preserving proprieties make it a well-suited algorithm for the modelling of (X)/(E)MRI systems, and further motivates our current work in modelling the most realistic system where the background black hole is a rotating Kerr BH on eccentric and spinning orbits. It is also noteworthy, the times-steppers here mentioned have the potential to improve any of the other algorithms on Table \ref{tab_emris_timedomain}, and it is strongly recommended for more realistic and accurate implementations. The implementation of more adequate time-stepper schemes would also allow for a more fair comparison between the rich choices that were made to address feature \texttt{[F1]} translating to difficulty $1)$ as evidenced by Table $\ref{tab_emris_timedomain}$, the impact here would be multi-disciplinary. Implementation of more suitable strategies to accurately handle boundary conditions has single-handedly already resulted in the improvement of the algorithms used by Bernuzzi and Field as evidenced in Table \ref{tab_emris_timedomain}, and hence here is recommended that hyperboloidal slices automatically enforcing boundary conditions be used when solving these equations \cite{macedo2014axisymmetric, macedo2018hyperboloidal, macedo2020hyperboloidal, macedo2022hyperboloidal}. 

\section{Conclusion}

In the body of this work, we have presented a numerical algorithm, \texttt{DiscoTEX}, which uses discontinuous collocation methods with implicit-turned-explicit \texttt{IMTEX} time-integrator. We have proved that implicit \texttt{IM} and implicit-turned-explicit \texttt{IMTEX} Hermite integrators are capable of preserving the symplectic structure, exactly and numerically, while retaining high accuracies and long-term stability for both homogeneous and distributional sourced wave and wave-like equations. We have further demonstrated that \texttt{IM \& IMTEX} time integrators, when solving the same problem, are more accurate than traditionally explicit schemes such as \texttt{Runge-Kutta}, (even when writing in optimised Horner-Form), which requires an order-7 scheme to achieve the same accuracy attained with an \texttt{IMTEX} order-2 scheme (lowest order). When applied to the same distributionally sourced problem it was also demonstrated that \texttt{DiscoTEX} achieves the same accuracies as \texttt{DiscoIMP} while being faster, retaining stability and coordinate-preserving proprieties and less cumbersome to implement. \\

Furthermore, this work gives and compares implementations of the \texttt{IMTEX} Hermite time-integration scheme and the discontinuous time-integration rule up to twelve orders and justifies our decision to use an order-4 scheme in \cite{da2023hyperboloidal} and upcoming work \cite{phdthesis-lidia, paper2, paper3}. In a companion paper the full details of the computation of \texttt{DiscoTEX} at higher orders is provided \cite{discotexII}. The stagnation in accuracy results from the increase in round-off error associated with the increase of the evolution matrix complexity and increase in the number of operations. These results could potentially be improved with the use of compensation-summation for example to reduce round-off error. An increase in the precision used in the operations prior to the computation of the evolution algorithm, for example, given in equation \eqref{ch3_discoTEX4_Wave_tr}, also results in a significant improvement of these results at the expense of extra computational time. Future work will improve on this and explore this avenue further. \\

Here we presented the work in \cite{reviews-lidia, phdthesis-lidia} and previously discussed in \cite{24thCapraTalk, 25thCapraTalk} for simpler toy models and give the numerical studies that led to the final form of the algorithm we have implemented in \cite{da2023hyperboloidal, paper2, paper3}. Furthermore this work solves three issues: $i)$ It fully verifies the recurrence relation, \textit{for the first time} given in equations (\eqref{ch34_wavetx_REC_j0}- \eqref{ch34_wavetx_REC_jmp2}), initially given in \cite{2014arXiv1406.4865M}, proving the algorithm numerically and exactly with comparisons against existing exact solutions to distributionally sourced wave-equations \cite{field2009discontinuous, field2010persistent, field2022discontinuous} and implementations in either coordinate charts $(t,x)$ and $(\tau,\sigma)$; $ii)$ It demonstrates and proves, that highly accurate computations are possible both at the particle and it's right and left-limits, by using the numerical solution provided by the \texttt{DiscoTEX} solver through generic interpolation; $iii)$ It demonstrates there are no losses in accuracy when implementing a hyperboloidal slice nor when using an Implicit-turned-explicit \texttt{IMTEX} instead of an implicit scheme; $iv)$ We further demonstrated by applying \texttt{DiscoTEX} to both distributionally sourced wave-like equations governing the scalar and gravitational perturbations describing an (X)/(E)MRI problem that highly accurate computations are possible, with ease, in the time-domain. An overview of the requirements for successful modelling of EMRIs through a time-evolution is also provided and complemented with an overview of all previous numerical methods utilised within the radiation-reaction community which highly influenced the results at the time on Table \ref{tab_emris_timedomain}. \\

Finally, we have shown it is possible to compute the required quantities for modelling of (X)/(E)MRIs with high accuracies that comfortably surpass the requirements for detection by LISA\cite{afshordi2023waveform}. In this work, the focus was on demonstrating how the \texttt{DiscoTEX} (and \texttt{DiscoIMP}) numerical machinery is applied, however in previous work \cite{da2023hyperboloidal} and future work we will demonstrate the full computations for the scalar and gravitational perturbations of point-particle on a Schwarzschild background BH \cite{paper2, paper3}. Additionally in future work, we will include an alternative solution with a fully independent numerical scheme which is fully spectral, developed by Panosso Macedo \cite{macedo2014axisymmetric} which not only implements a state-of-the-art \cite{macedo2014axisymmetric} implicit integrator and hence retains the necessary unconditionally stability proprieties for successful evolutions of EMRIs, it has the benefit of not requiring higher order jumps. A back of the envelop calculation, estimates \texttt{DiscoTEX}/ \texttt{DiscoIMP}/ \texttt{DiscoREX} \cite{discoREX} \cite{reviews-lidia, phdthesis-lidia} to require a high number of jumps: say 10 $(l,m)$ modes are necessary, by applying the formula $N_{\rm{evolution}} = ((l_{\rm{max}}+1 )(l_{\rm{\max}} +2))/2$, we estimate a total number of $J_{\rm{total}} = J_{max}*N_{\rm{evolution}} = 726 $ jumps to be necessary, where $J_{\rm{max}} =11$ jumps as in this work and \cite{da2023hyperboloidal}. This further estimates the computational cost of our simulations, it is likely this method will always be slower than our fully-spectral implementation, however, effective use of high-performance computing can significantly attenuate the cost as the evolution algorithm is relatively simple to implement and reduce to, mostly, vector-matrix and matrix-matrix multiplications. In future work, we will explore implementations across different architectures. Particularly, we will show how both the hyperboloidal \texttt{DiscoTEX}, (with extra comparisons via \texttt{DiscoIMP/DiscoEX}) and the fully-spectral \textbf{implicit} scheme can be used as tools to achieve this with the required accuracies for detection by LISA \cite{afshordi2023waveform} for both non-rotating Schwarzschild \cite{paper2,paper3} and rotating Kerr black holes. This work \cite{reviews-lidia, phdthesis-lidia, discotexII, discoREX}, our results in \cite{da2023hyperboloidal, macedo2014axisymmetric} and our collaborative incoming work \cite{paper2, paper3} in the time-domain offers promising new prospects for the modelling of (X)/(E)MRIS. Particularly, it provides a clear avenue for the first full alternative generation waveform model to current strategies relying on the two-time scale expansion in the Fourier domain through a self-consistent evolution. \\

\section*{Acknowledgements}
This work makes use of the Black Hole Perturbation Toolkit \cite{BHPToolkit}. I would like to thank all the co-authors of \cite{da2023hyperboloidal} for discussions that greatly shaped the body of this work as we prepared that manuscript and our anonymous referee. I thank my PhD supervisors Juan Valiente Kroon and Pau Figueras for their guidance and encouragement. Furthermore, I would also like to thank Scott E. Field and his collaborators for outstanding previous work \cite{field2009discontinuous}, without it, this work and its derivatives would have taken significantly longer. I also thank the welcoming Capra community, particularly everyone who has worked in the time domain and is mentioned in Table 8. Thoroughly researching and analysing previous numerical methods in the time domain is what ultimately motivated the Capra meetings archive restoration, made possible by previous efforts of Soichiro Isoyama \cite{CapraArchive}. Moreso, I thank Nelson Eiro for his contributions and discussions to \cite{o2022conservative} pertaining to \texttt{IMTEX} geometric integrators during group meetings with Charalampos Markakis, Michael O'Boyle, myself and collaborators. Additionally, I thank him for his work on applying these methods to the $1+1$D Teukolsky solver during the first years of his PhD. Finally, it is a pleasure to thank Subrahmanyan Chandrasekhar, whose intelligence and work ethic have been a source of motivation and (inspiration) in the last years of my PhD \cite{chandra, phdthesis-lidia}. 

\appendix

\section{Complementary work to Section \ref{sec_3_2_discojumps} }\label{app2}
Note that the following section is from \cite{da2023hyperboloidal} and it has been included here for completeness. This is this author's own PhD original work \cite{reviews-lidia} and has appeared in \cite{phdthesis-lidia}. 

\subsection{Jump condition derivation through the Frobenius Method}\label{appA_frobs}
As explained in \cite{da2023hyperboloidal} we reconsider equation \eqref{ch1_bhpt_wave} by transforming the d'Alembert operator $\square \Psi$ back to its pre-tortoise coordinate transformation $(t,r)$ form, reverting eq.~\eqref{ch3_general_pdes} into a PDE of the form, 
\begin{equation}
    -\frac{\partial^{2} \Psi}{\partial t^{2}} +  \eta(r) \frac{\partial^{2} \Psi}{\partial r^{2}} +  \varrho(r) \frac{\partial \Psi}{\partial r} - V_{l}(r)\Psi = S_{lm}(t,r),  \ \ \ \ 
    \label{aa_ugly_pde}
\end{equation}
where $\eta(r) = f^{2}$, $\rho(r) = ff'$. \newline
At the end of deriving the recurrence relation for $J_{m+2}(t)$ we will then revert back to its tortoise coordinate form, which is slightly different as it has been previously done in the literature \cite{sopuerta2006finite, field2009discontinuous, hopper2010gravitational}. The numerical algorithm here can be seen as an application of the Frobenius method \cite{weisstein2002frobenius, frobenius1873} for the case where we have a function in \textit{weak-form}. We will therefore revisit it briefly here. The Frobenius method states if $x_{0}$ is an ordinary point of an ODE describing a smooth function $\Phi(x)$ we can expand it in a Taylor series about $x_{0}$. A Maclaurin series is then attained by taking the expansion point as $x_{0} = 0$, resulting in
\begin{equation}
    \Phi =\sum_{n=0}^{\infty} a_{n}x^{n}.
    \label{maclaurinseries}
\end{equation}
Plugging $\Phi$ back in the ODE and grouping the coefficients by power we obtain a recurrence relation for the $nth$ term and write the series expansion in terms of the $a_{n}$ coefficients. It's relevant to the calculations that follow to highlight the first two derivatives, 
\begin{eqnarray}
 \Phi &&=\sum_{n=0}^{\infty} a_{n}x^{n}. \label{frobenius_0}\\
 \Phi' &&= \partial_{x}\bigg( \sum_{n=0}^{\infty} a_{n}x^{n}\bigg),\\
  &&=\sum_{n=1}^{\infty} n a_{n}x^{n-1},\\
 &&=\sum_{n=0}^{\infty} (n+1) a_{n+1}x^{n}.\\
  \Phi'' &&= \partial_{x}^{2} \bigg(\sum_{n=0}^{\infty} a_{n}x^{n}\bigg),\\
   &&=\sum_{n=2}^{\infty} n(n-1) a_{n}x^{n-2},\\
 &&=\sum_{n=0}^{\infty}(n+2)(n+1) a_{n+2}x^{n}. \label{frobenius_der2} 
\end{eqnarray}

\subsubsection{Unit Jump relations}\label{appA1a}
As given in Section \ref{sec2}, the wave and wave-like distributional sourced equations studied here assume a \textit{weak-form} as described by eq.~\eqref{ch1_weakform_master}. We thus write the ansatz: 
\begin{equation}
    \Psi_{lm}(t,r) = \Psi^{S}_{lm}(t,r) + \Psi^{J}_{lm}(t,r),
    \label{AppA_WFansatz}
\end{equation}
where $\Psi^{S}_{lm}(t,r)$, like $\Psi^{+/-}(t,r)$ functions in eq.~\eqref{ch1_weakform_master}, are smooth satisfying everywhere the homogeneous PDE, 
\begin{equation}
     -\frac{\partial^{2} \Psi^{S}}{\partial t^{2}} +  \eta(r) \frac{\partial^{2} \Psi^{S}}{\partial r^{2}} +  \varrho(r) \frac{\partial \Psi^{S}}{\partial r} - V_{l}(r) \Psi^{S} = 0 ,
    \label{AppA_SmoothPart}
\end{equation}
and $\Psi^{J}_{lm}(t,r)$ describe the inhomogeneous part of the master function in an infinite series as with the Frobenius method. We then write,
\begin{equation}
    \label{wave_eq_jumps1}
    \Psi^{J}(r,t;r_{p}(t)) = \sum^{\infty}_{n=0} J_{n}(t) \Psi_{n}(r;r_{p}(t)).
\end{equation}
where \(\Psi_n\) are piecewise monomials centred at the discontinuity $r = r_{p}(t)$:
\begin{small}
\begin{align}
       &\Psi_{n}\big(r,t;r_{p}(t)\big) =  \Psi_{n}\big(r - r_{p}(t)\big) = \nonumber \\
       &=\Psi_{n}, = \frac{1}{2} \text{sgn} \big(r-r_{p}(t)\big)\frac{\big(r-r_{p}(t)\big)^{n}}{n!}.
       \label{aa_sigmum_piecewisemonos} 
\end{align}
\end{small}
The expansion in eq.~\eqref{wave_eq_jumps1} allows the computation of the discontinuities 
\begin{small}
\begin{eqnarray}
    \label{disco1}
\hspace{-3em}\frac{\partial^{m} \Psi}{\partial{r}^{m}} \bigg\vert_{r =r_{p}^{+}} - \frac{\partial^{m} \Psi}{\partial{r}^{m}} \bigg\vert_{r = r_{p}^{-}}  &=& \frac{\partial^{m} \Psi^{J}}{\partial{r}^{m}} \bigg\vert_{r= r_{p}^{+}}  - \frac{\partial^{m} \Psi^{J}}{\partial{r}^{m}} \bigg\vert_{r= r_{p}^{-}} = J_{m}(t).   
\end{eqnarray}
\end{small}
These functions have the property that all spatial derivatives have matching left and right limits at $r= r_{p}(t)$, 
\begin{eqnarray}
    \label{disco2}
\frac{\partial^{m} \Psi_{n}}{\partial{r}^{m}} \bigg\vert_{r = r_{p}^{+}} - \frac{\partial^{m} \Psi_{n}}{\partial{r}^{m}} \bigg\vert_{r = r_p^{-}} = \delta_{m,n},   
\end{eqnarray}
except at the $nth$ derivative, where the jump is one.  
\begin{equation}
    \label{disco3}
\frac{\partial^{n} \Psi_{n}}{\partial{r}^{n}} \bigg\vert_{r = r_{p}^{+}} - \frac{\partial^{n} \Psi_{n}}{\partial{r}^{n}} \bigg\vert_{r = r_{p}^{-}} = 1.
\end{equation}
By differentiating the monomial we get,  
\begin{equation}\frac{\partial \Psi_n}{\partial r} = \left\{ 
\begin{array}{ccc}
\Psi_{n-1} &\text{for}& n\geq 1 \\
\delta(w) &\text{for}& n=0 
\end{array}
\right. .
\label{mondelta}
\end{equation}

\begin{equation}\frac{\partial^{2} \Psi_n}{\partial r^{2}} = \left\{ 
\begin{array}{ccc}
\Psi_{n-2} &\text{for}& n\geq 2 \\
\delta(w) &\text{for}& n=1 \\
\delta'(w) &\text{for}& n=0 
\end{array}
\right.
\label{mondelta2}
\end{equation}
For simplicity here the notation $w = (r - r_{p}(t))$ was introduced. The idea to use unit jump functions to compute higher-order jumps was put forward in \cite{2014arXiv1406.4865M}. This work further included an attempt at understanding and using it for the computation of the numerical solution to the simplest case of the distributionally sourced wave-equation, i.e. Type II. 

\subsubsection{Spatial Jumps}\label{appA1b}
The inhomogeneous part of the solution is given by equation \eqref{wave_eq_jumps1}. To determine the jumps in eq.~\eqref{aa_ugly_pde}, it is useful to find the spatial and temporal derivatives of the unit jump functions. These derivations are an application of the Frobenius method. More explicitly, the first order derivative of eq.~\eqref{wave_eq_jumps1} is:
\begin{align}
 &\partial_{r}\Psi^{J}(w) =  \partial_{r} \bigg[\sum^{\infty}_{n = 0}\bigg(J_{n}(t) \Psi_{n}(w)\bigg) \bigg],\\
 &= \sum^{\infty}_{n=1}\bigg(J_{n}(t) \Psi_{n-1}(w)\bigg), \nonumber \\
 &= \sum^{\infty}_{n=0}\bigg(J_{n+1}(t) \Psi_{n}(w) \bigg) + J_{0}(t)\delta(w).  \ \ \ \  \ \ \label{spatial_o1_fb} \ \ 
\end{align}
where the selection property of Dirac-$\delta$, as given below by equation \eqref{appendix_a_dirac_selectionI}, was used to ensure the problem is fully with respect to the particle's worldline, $r_{p}(t)$. For second-order it extends to,  
\begin{eqnarray}
    \partial_{r}^{2} \Psi(w) &=& \partial_{r}^{2} \bigg[\sum^{\infty}_{n = 0}\bigg(J_{n}(t) \Psi_{n}(w)\bigg) \bigg], \nonumber\\
    &=& \sum^{\infty}_{n=0} \bigg[J_{n+2}(t) \Psi_{n}(w)  \bigg]
    + J_{1}(t)\delta(w) + J_{0}(t)\delta'(w).  \nonumber \\
    \label{spatial_o2_fb}
\end{eqnarray}
Given the coefficients present in the differential operators in eq.~\eqref{aa_ugly_pde} and the potential term, it is important to revisit eq.~\eqref{aa_sigmum_piecewisemonos}. One of its key properties is 
\begin{eqnarray}
    \text{sgn} = \frac{x}{|x|} = \frac{|x|}{x}.
    \label{sigmum_property}
\end{eqnarray}
Using this, eq.~\eqref{aa_sigmum_piecewisemonos} is rewritten as 
\begin{eqnarray}
\Psi_{n} &=& \frac{1}{2} \frac{|w|}{w} \frac{w^{n}}{n!},\label{wave_eq_jumps3} \\ 
  & = &\frac{1}{2n!} (w)^{n-1} |w|,\label{wave_eqs_jumps4} \\
 w \; \Psi_{n} & = &\frac{1}{2n!} |w| w^{n}. \label{wave_eqs_jumps5}
\end{eqnarray}
We then generalise for $\Psi_{n + m}$: 
\begin{eqnarray}
\Psi_{n+m}(w) &=& \frac{1}{(n+m)!} \big|w\big| \big(w\big)^{n + m - 1}, \label{wave_eqs_jumps6}\\
\Psi_{n+m}(w)(n+m)! &=&\frac{\big|w\big| \big(w\big)^{n+m-1}}{2}.   \ \ \ \label{wave_eqs_jumps7}
\end{eqnarray}
Substituting for the term $|w| \ w^{n}/2$ in the RHS of this equation by the LHS of equation \eqref{wave_eqs_jumps5} multiplied by $n!$ we get, 
\begin{eqnarray}
    n! \ (w) \Psi_{n} &=& (w)^{1-m} (n+m)! \Psi_{n+m},  \nonumber \\ 
    (w)(w)^{m-1} \Psi_{n} &=& \frac{(m+n)!}{n!} \Psi_{n+m}, \nonumber \\ 
    (w)^{m} \Psi_{n} &=& \frac{(m+n)!}{n!} \Psi_{n+m}.
    \label{piecewise_mon_generic}
\end{eqnarray}
We want to be able to expand any coefficient in $r$ acting on the master functions $\Psi^{a/p}_{lm}(t,r)$ and its differential operators as given by equation \eqref{aa_ugly_pde}. As an example, let's consider the potential term $V_{l}(r)$ acting on the master function, we can expand it around $r=r_{p}$ 
\begin{equation}
    V_{l}(r) = \sum^{\infty}_{k=0} V^{(k)}_{l}(r_{p}) \frac{(r-r_{p})^{k}}{k!}, 
\end{equation}
where $k \leq n$ and compute the product $V(r) \Psi^{J}(r;r_{p})$ following the ansatz of Eq.~\eqref{wave_eq_jumps1}, 

\begin{align}
   &V_{l}(r) \Psi^{J}(r;r_{p})= \bigg[ \sum^{\infty}_{k=0} V^{(k)}_{l} (r_{p}) \frac{(r-r_{p})^{k}}{k!} \bigg] \times\bigg[ \sum^{\infty}_{n=0}J_{n}(t) \Psi_{n}(r;r_{p}) \bigg]. \ \ 
    \label{proof_coefficients_1}
\end{align}
Furthermore, given $k \leq n $ we can define $n = m-k$ where at $n=0$, $m=k$, and the preceding equation simplifies to, 
\begin{align}
&V_{l}(r) \Psi^{J}(r;r_{p}) =  \sum^{\infty}_{k=0} \sum^{\infty}_{m=k} V_{l}^{(k)}(r_{p}) \nonumber \\
&\times \frac{(r-r_{p})^{k}}{k!} J_{m-k}(t) \Psi_{m-k}(r;r_{p}). 
    \label{proof_coefficients_2}
\end{align}
From eq.~(\ref{piecewise_mon_generic}) with $m=k$ and $n=m-k$ we get, 
\begin{eqnarray}
    (r-r_{p})^{k} \Psi_{m-k} = \frac{(m-k+k)!}{(m-k)!} \Psi_{m-k+k}, \nonumber \\
    (r-r_{p})^{k} \Psi_{m-k} =  \frac{m!}{(m-k)!} \Psi_{m}.
    \label{monos_generilisationtok}
\end{eqnarray}
Eq.~\eqref{proof_coefficients_2} then simplifies to, 
\begin{small}
\begin{align}
  &V_{l}(r) \Psi^{J}(r;r_{p}) = \sum^{\infty}_{m=0} \sum^{m}_{k=0} V_{l}^{(k)}(r_{p}) \frac{m!}{(m-k)!k!} J_{m-k}(t) \Psi_{m}(r;r_{p}) \nonumber \\
  &= \sum^{\infty}_{m=0} \bigg[ \sum^{m}_{k=0}  {m\choose k}  V_{l}^{(k)}(r_{p}) J_{m-k}(t) \bigg] \Psi_{m}(r;r_{p}), 
    \label{final_proof_coefficents_handling}
\end{align}
\end{small}
where to go from the first to the second line factorial notation is adapted.  
Furthermore, we extend this to the coefficients of the first-order derivative of the master functions, as given by equation \eqref{spatial_o1_fb}, 
\begin{align}
    &\varrho(r) \partial_{r} \Psi(w) = \sum^{\infty}_{m=0} \bigg[\sum^{m}_{k=0}  \varrho^{(k)}(r_{p}) {m\choose k} J_{m+1-k}(t)\Psi_{m}(w) \bigg] \nonumber \\ 
    &+ \varrho(r_{p}) J_{0}(t)\delta(w). 
    \label{corrected_firstorderdiff}
\end{align}
The second-order jump, as given in eq.~\eqref{spatial_o2_fb}, is similarly now given by 
\begin{align}
   &\eta(r) \partial_{r}^{2} \Psi(w) = \sum^{\infty}_{m=0} \bigg[\sum^{m}_{k=0} \eta^{(k)}(r_{p}) {m\choose k} J_{m+2-k}(t)\Psi_{m}(w) \bigg]  \nonumber \\
   & + \eta(r) J_{1}(t)\delta(w) + \eta(r) J_{0}(t)\delta'(w), \nonumber \\
   &= \sum^{\infty}_{m=0} \bigg[\sum^{m}_{k=0} \eta^{(k)}(r_{p}) {m\choose k} J_{m+2-k}(t)\Psi_{m}(w) \bigg] \nonumber \\
   &+  \eta(r_{p}) J_{1}(t)\delta(w) + \eta(r_{p}) J_{0}(t)\delta'(w) \nonumber \\
   &-2 \varrho(r_{p})J_{0}(t)\delta(w), 
   \label{corrected_secondorderdiff}
\end{align}
For both derivations selection properties as given by equations (\eqref{appendix_a_dirac_selectionI}, ~\eqref{appendix_a_dirac_selectionII}) were applied on the coefficient of the Dirac delta distribution and its $r$-derivative respectively, to go from the second to the third line. 

\subsubsection{Temporal Jumps}\label{appA1b}
The first order temporal derivative and associated jumps of the function $\Psi^{J}(r,t;r_{p})$ are: 
\begin{align}
     &\partial_{t}\Psi^{J}(w) = \partial_{t} \sum^{\infty}_{m = 0}\bigg[J_{m}(t) \Psi_{m}(w)\bigg], \nonumber \\
     &= \sum^{\infty}_{m=0} \bigg[ \dot{J}_{m}(t) - \dot{r}_{p}  J_{m+1}(t) \bigg]\Psi_{m}(w) \nonumber \\
     &- \dot{r_{p}} J_{0}(t)\delta(r-r_{p}). 
     \label{app_temp_approx}
\end{align}
The second-order derivative can be obtained by calculating: 
\begin{align}
    \label{temporal_o2_ansa}
 &\partial_{t}^{2} \Psi^{J}(w) =\partial_{t} \bigg[\sum^{\infty}_{m=0} \dot{J}_{m}(t) \Psi_{m}(w) \bigg] - \nonumber \\ 
 &\partial_{t} \bigg[ \sum^{\infty}_{m=0} \dot{r_{p}} J_{m+1}(t) \Psi(w)\bigg] - \partial_{t}\bigg[\dot{r_{p}} J_{0}\delta(w)\bigg]. 
\end{align}
For simplicity and to clarify we will be showing the  explicit calculation of this derivative as $D1,D2,D3$ given respectively as, 
\begin{small}
\begin{align}
   &D1 =\partial_{t} \bigg[ \sum^{\infty}_{m=0} \dot{J}_{m}(t) \Psi_{m} \bigg] =  \sum^{\infty}_{m=0}\bigg[ \ddot{J}_{m}(t)  \nonumber \\
   &- \dot{r}_{p}\dot{J}_{m+1} \bigg] \Psi_{m}(w)  - \dot{r_{p}} \dot{J}_{0} \delta (w),  
   \label{temporal_D1}
\end{align}
\end{small}
followed by, 
\begin{align}
       &D2 = - \partial_{t} \bigg[ \sum^{\infty}_{m=0} J_{m+1}(t)\dot{r}_{p} \Psi(w) \bigg] \nonumber \\
       &= \sum^{\infty}_{m=0} \bigg( -\dot{r}_{p} \dot{J}_{m+1}(t) - \ddot{r}_{p} J_{m+1}(t) + 
      \dot{r}_{p}^{2} J_{m+2}(t) \bigg)\Psi_{m}(w) \nonumber \\
      &+ \dot{r}_{p}^{2} J_{1}(t) \delta(w), 
     \label{temporal_D2}
\end{align}
and finally by, 
\begin{align}
    &D3  =- \bigg[ \bigg(\dot{r}_{p} \dot{J}_{0} + \ddot{r}_{p} J_{0}(t)\bigg) \delta(w) - \dot{r}_{p}^{2} J_{0} \delta'(w) \bigg] \nonumber \\ 
    &= r_{p}^{2} J_{0}(t)\delta'(w) - \dot{r}_{p}\dot{J}_{0}(t) \delta(w) \nonumber \\
    &- \ddot{J}_{0}(t) \delta(w).
     \label{temporal_D3}
\end{align}
Plugging this results into equation \eqref{temporal_o2_ansa}, we get:
\begin{align}
  &\partial^{2}_{t}\Psi^{J}(w) =\sum^{\infty}_{m=0} \bigg[ \ddot{J}_{m}(t) - 2\dot{r}_{p} \dot{J}_{m+1}(t) - \ddot{r}_{p} J_{m+1}(t) + \dot{r}_{p}^{2} J_{m+2}(t) \bigg]\psi_{m}(w) \nonumber \\
    &+  \dot{r}_{p}^{2} J_{1}(t) \delta(w)  - 2\dot{r}_{p}\dot{J}_{0}(t) \delta(w) - \ddot{J}_{0}(t) \delta(w) +  r_{p}^{2} J_{0}(t)\delta'(w) 
   \label{temporal_fb_final}    
\end{align}
Matching coefficients in $\delta'(w)$ with respect to the particle worldline gives the first initialising jump, 
\begin{equation}
    J_{0}(t) = \frac{F^{a/p}_{lm}(t,r_{p})}{\big(f^{2}_{p} - \dot{r}_{p}^{2}\big) }.
\end{equation}
Similarly, for the second and final initialising jump, we match the coefficients of the Dirac $\delta(w)$, 
\begin{align}
&J_{1}(t) = - 2 \dot{r}_{p} \partial_{t}J_{0}(t) - \big(\ddot{r}_{p} - f_{p}f'_{p} \big) J_{0}(t) \nonumber \\
&+ \big[ G^{a/p}_{lm}(t,r_{p}) - \partial_{r_{p}}F^{a/p}_{lm}(t,r_{p}) \big] / \big(f^{2}_{p} - \dot{r}_{p}^{2}\big). 
\label{lit_j1}
\end{align}
These jumps match those stated in the literature \cite{field2009discontinuous, hopper2010gravitational} without the $f$ factor. The equations in (\eqref{ch3_j0}, \eqref{ch3_j1}) match and follow the approach of \cite{hopper2010gravitational}. To calculate the higher-order jumps we could then collect all remainder terms by solving for $J_{m+2}(t)$ as given in equation \eqref{cha3_rec_relation_mjumps_PhysicalChart_RWZ}. 

\subsection{Dirac delta distribution proprieties}\label{appA2_DDDsProps}
In this work, the following selection proprieties are used
\begin{eqnarray}
\label{appendix_a_dirac_selectionI}
f(a)\delta(a - b) =  f(b)\delta(b-a), \\
f(a)\delta'(a - b) = f(b)\delta'(a-b) - f(b)\delta(b-a). 
\label{appendix_a_dirac_selectionII}
\end{eqnarray}
Furthermore, the following composition rules were also needed: 
\begin{eqnarray}
\label{appendix_a_dirac_compI}
\delta(f(a)) = \frac{1}{|f'(b)|} \delta(a - b), \ \ \  \\
\delta'(f(a)) = \frac{f'(b)}{|f'(b)|^{3}} \delta'(a - b) + \frac{f''(b)}{|f'(b)|^{3}} \delta(a - b). \ \ \ 
\label{appendix_a_dirac_compositionII}
\end{eqnarray}
where the last distribution property was firstly given in Appendix F of \cite{mathews2022self}. 

\section{Complementary work on time-integrators such as Hermite, Hermite \texttt{IMTEX}, \texttt{HF IMTEX} and \texttt{EX HF RK}  and discontinuous time-integrators as presented in Section \ref{sec_discotex_justime}}\label{appb}
\subsection{Higher-order \texttt{IMTEX} time integration}\label{app_higherorder_imtex}
Here we give the explicit equations for all Hermite Horner-Form implicit-turned-explicit \texttt{HF IMTEX} time-integrators derived as explained in \ref{sec_imtex} through equations (\eqref{ch33_reduction_odes}, \eqref{ch3_hfh6}) from second- to twelve-order: \\

Hermite order-2, \texttt{IMTEX HFH2}, 
\begin{small}
\begin{align}
    \label{ch3_hornerfor_sh2}
     &\textbf{U}_{n+1} = \textbf{U}_{n}+ (\Delta t \ \textbf{L}) \cdot \textbf{HFH2} \cdot \textbf{U}_{n}, \\
     &\textbf{HFH2} = \bigg( \textbf{I} - \frac{\Delta t}{2} \textbf{L}\bigg)^{-1}.
    \label{ch3_hfh2}
\end{align}   
\end{small}

Hermite order-4, \texttt{IMTEX HFH4}, 
\begin{small}
\begin{align}
    \label{ch3_hornerfor_sh4}
    &\textbf{U}_{n+1} = \textbf{U}_{n}+ \textbf{A} \cdot \textbf{HFH4} \cdot \textbf{U}_{n}, \\
    &\textbf{HFH4} = \bigg( \textbf{I} - \frac{\textbf{A}}{2} \cdot \bigg( \textbf{I} - \frac{\textbf{A}}{6}  \bigg)\bigg)^{-1}, 
    \label{ch3_hfh4}
\end{align}  
\end{small}
where here $(\Delta t \ \textbf{L}) = \textbf{A}$, $(\Delta t \ \textbf{L}) \cdot (\Delta t \ \textbf{L})  = \textbf{A}^{2} = \textbf{A} \cdot \textbf{A} $. \\

Hermite order-8 \texttt{IMTEX HFH8}, 
\begin{small}
\begin{align}
    \label{ch3_hornerfor_sh8}
     &\textbf{U}_{n+1} = \textbf{U}_{n}+ \textbf{A} \cdot \bigg(\textbf{I} + \frac{1}{42} \textbf{A}^{2}\bigg) \cdot \textbf{HFH8} \cdot \textbf{U}_{n}, \\
    &\textbf{HFH8} = \bigg( \textbf{I} + \textbf{A}\cdot \bigg( -\frac{\Delta t}{2} + \textbf{A}\cdot \bigg( \frac{3}{28} \textbf{I} + 
    \bigg( -\frac{1}{84}\textbf{I} + \frac{\textbf{A}}{1680}\bigg) \cdot \textbf{A} \bigg)\bigg)\bigg)^{-1}.
    \label{ch3_hfh8}
\end{align}   
\end{small}

Hermite order-10 \texttt{IMTEX HFH10}, 
\begin{small}
\begin{align}
\label{ch3_hfh10}
&\textbf{U}_{n+1} = \textbf{U}_{n}+ \textbf{A} \cdot \bigg( \textbf{I} + \textbf{A}^{2} \cdot \bigg(  
\textbf{A}\cdot \bigg( \frac{1}{36} \textbf{I} + \frac{\textbf{A}}{15120}\bigg) \bigg)\bigg) \cdot   \textbf{HFH10} \cdot \textbf{U}_{n}, \\
&\textbf{HFH10} = \bigg( \textbf{I} +\textbf{A} \cdot \bigg( 
-\frac{1}{2}\textbf{I} + \textbf{A} \cdot \bigg( \frac{1}{9}\textbf{I} + \nonumber \\ 
&\textbf{A} \cdot \bigg( -\frac{1}{72} \textbf{I} + \bigg( \frac{1}{1008} \textbf{I} - \frac{\textbf{A}}{30240}\bigg)\cdot \textbf{A}\bigg)\bigg)\bigg)\bigg)\bigg)^{-1}. 
\label{ch3_hornerfor_sh10}
\end{align} 
\end{small}

Hermite order-12 \texttt{IMTEX HFH12}, 
\begin{small}
\begin{align}
\label{ch3_hornerfor_sh12}
&\textbf{U}_{n+1} = \textbf{U}_{n} + \textbf{A}\cdot \bigg( \textbf{A}^{2} \cdot \bigg( 
\frac{1}{33}\textbf{I} + \frac{\textbf{A}^{2}}{7920}  \bigg) \bigg)\cdot \textbf{HFH12} \cdot \textbf{U}_{n}, \\
& \textbf{HFH12} = \bigg( \textbf{I} + \textbf{A}\cdot \bigg( -\frac{1}{2}\textbf{I} + \textbf{A}\cdot\bigg( \frac{5}{44} \textbf{I} + \nonumber \\
&\textbf{A}\cdot \bigg( \textbf{A}\cdot \bigg(-\frac{1}{66} \textbf{I} + \textbf{A}\cdot\bigg(
\frac{1}{792}\textbf{I} + \bigg( -\frac{1}{15840} \textbf{I} + \frac{(\textbf{A})}{665280}
\bigg) \cdot \textbf{A} \bigg) \bigg) \bigg) \bigg) \bigg)\bigg)^{-1}. 
\label{ch3_hfh12}
\end{align}
\end{small}
\subsection{Explicit Horner-Form Runge-Kutta \texttt{EX RK} time integration schemes}\label{app_ex_rk}

To complement the results showed in Section \ref{sec_imtex} we give the time-integration Explicit Runge-Kutta numerical schemes in Horner-form \texttt{EX HF RK} from second- to sixth-order: 

Runge-Kutta order-2, \texttt{EX HF RK2}:
\begin{small}
 \begin{align}
    \label{ch3_hornerform_rk2}
    &\textbf{U}_{n+1} = \textbf{U}_{n} + (\textbf{HFRK2}-\textbf{I})\cdot \textbf{U}_{n}, \\
    &\textbf{HFRK2} = \textbf{I} + \bigg(\textbf{I} + \frac{\textbf{A}}{2} \bigg) \cdot \textbf{A}. 
    \label{ch3_hornerformer_rk2mat}
\end{align}
\end{small}

Runge-Kutta order-3, \texttt{EX HF RK3}:
\begin{small}
\begin{align}
    \label{ch3_hornerform_rk3}
    &\textbf{U}_{n+1} = \textbf{U}_{n} + (\textbf{HFRK3}-\textbf{I})\cdot \textbf{U}_{n}, \\
    &\textbf{HFRK3} = \textbf{I} + \textbf{A} \cdot \bigg( \textbf{I} + \bigg(\frac{1}{2}\textbf{I} + \frac{1}{6}\textbf{A}\bigg) \cdot \textbf{A}\bigg).
    \label{ch3_hornerformer_rk3mat}
\end{align}
\end{small}

Runge-Kutta order-4, \texttt{EX HF RK4}:
\begin{small}
\begin{align}
\label{ch3_hornerformer_rk4mat}
&\textbf{U}_{n+1} = \textbf{U}_{n} + (\textbf{HFRK4}-\textbf{I})\cdot \textbf{U}_{n}, \\
&\textbf{HFRK4} = \textbf{I} + \textbf{A} \cdot \bigg(\textbf{I} + \textbf{A} \cdot \bigg( \frac{1}{2}\textbf{I} + \bigg( \frac{1}{6}\textbf{I} + \frac{1}{4}\textbf{A} \bigg) \cdot \textbf{A} \bigg).
\label{ch3_hornerform_rk4}
\end{align}
\end{small}

Runge-Kutta order-5, \texttt{EX HF RK5}:
\begin{small}
\begin{align}
\label{ch3_hornerform_rk5}
&\textbf{U}_{n+1} = \textbf{U}_{n} + (\textbf{HFRK5}-\textbf{I})\cdot \textbf{U}_{n}, \\
&\textbf{HFRK5} = \textbf{I} +  \textbf{A} \cdot \bigg(\textbf{I} + \textbf{A} \cdot \bigg(\frac{1}{6}\textbf{I} + \bigg( \frac{1}{24}\textbf{I} + \frac{1}{120}\textbf{A}\bigg) \cdot \textbf{A} \bigg).
\label{ch3_hornerformer_rk5mat}
\end{align}
\end{small}

Runge-Kutta order-6, \texttt{EX HF RK6}:
\begin{small}
\begin{align}
\label{ch3_hornerform_rk6}
&\textbf{U}_{n+1} = \textbf{U}_{n} + (\textbf{HFRK6}-\textbf{I})\cdot \textbf{U}_{n}, \\
&\textbf{HFRK6} = \textbf{I} + \textbf{A} \cdot \bigg( \textbf{I} +  \textbf{A} \cdot +  \nonumber \\ 
&\bigg( \frac{1}{2}\textbf{I} + \textbf{A} \cdot \bigg(\frac{1}{6} \textbf{I} + \textbf{A}\cdot\bigg( \frac{1}{24}\textbf{I} + \bigg( \frac{1}{120}\textbf{I} + \frac{1}{720}\textbf{A}\bigg)\cdot\textbf{A}\bigg) \bigg)  \bigg)\bigg).
\label{ch3_hornerformer_rk6mat} 
\end{align}
\end{small}

\subsection{\texttt{IMTEX} time-integrator proprieties}\label{app_imtex_props}
In this section, the aim is to demonstrate that the Hermite \texttt{IMTEX} algorithm discussed in this manuscript and in previous collaborative work \cite{o2022conservative, da2023hyperboloidal} preserves energy and symplectic structure both exactly and numerically, along with its Horner-form counterpart i.e Hermite \texttt{HF IMTEX}. For simplicity we will demonstrate this by using the simple Hamiltonian as an example, however, the proofs here hold for any quadratic Hamiltonian. Generically we have the following canonical equations, 
\begin{eqnarray}
 \begin{cases}
      \displaystyle \dot{q} =  \frac{dq}{dt} = \frac{\partial H}{\partial p} = \nabla_{p}H,\\
     \displaystyle \dot{p} =  \frac{dp}{dt} = - \frac{\partial H}{\partial q}= - \nabla_{q}H, 
      \end{cases}
\label{app_hamiltoniansys_general}
\end{eqnarray} 
compactly given as
\begin{align}
    \frac{d}{dt} \begin{bmatrix}q \\ p \end{bmatrix} = J^{-1}\nabla H(q,p),
    \label{ham_sys_simplified}
\end{align}
where $J$ is the Jacobian matrix,
\begin{align}
    J = \begin{pmatrix} 0& I\\ -I & 0 \end{pmatrix},
    \label{hamil_jacobian}
\end{align}
with $I$ being the $d \times d$ identity matrix. For simplicity we let $u = (q,p)$ be a point in the phase space in $\mathbb{R}^{2d}$, with equation ~\eqref{ham_sys_simplified} being alternatively written as, 
\begin{equation}
    u = J^{-1} \nabla H(u). 
    \label{ham_simp_u}
\end{equation}
As the later is a canonical equation, the following definition holds,

\textbf{Definition \cite{hairer2010geometric} - Flow of the system:} The phase flow, $\varphi_{t}$ of an Hamiltonian system $H(q,p)$ is a one-parameter family of mappings $\varphi_{t}: \mathbb{R}^{2d} \rightarrow \mathbb{R}^{2d}$, that for each time $t$ 
\begin{align}
    \varphi_{t}: (q(0), p(0)) \rightarrow (q(t),p(t)). 
    \label{phase_flow_ham}
\end{align}
In other words, the phase flow $\varphi$ of the Hamiltonian system is the mapping that advances the solution by time $t$, i.e. $\varphi_{t_{0}}(q_{0},p_{0}) = (q(t,q_{0},p_{0}), p(t,q_{0},p_{0}))$ where $q(t,q_{0},p_{0}), p(t,q_{0},p_{0})$ is the solution of the Hamiltonian system corresponding to the initial values $q(0) = q_{0}, \; p(0) = p_{0}$. We can thus conclude that the phase flow is a differentiable mapping.\\

\textbf{I. Energy-conserving time-integrators}

In most cases, and in the body of this work, the Hamiltonian corresponds, physically, to the total energy of a system. \footnote{This isn't, however always true, see for example \cite{su2016application, hu2021energy}.} Discretising these derivatives from an \texttt{nth} to \texttt{(n+1)th} step as, 
\begin{eqnarray}
    \label{ch3_discrete_sho_1}
    \frac{q_{n+1} - q_{n}}{\Delta t} = \frac{H(q_{n},p_{n+1}) - H(q_{n},p_{n})}{p_{n+1} - p_{n}},\\
    \frac{p_{n+1} - p_{n}}{\Delta t} = \frac{H(q_{n},p_{n+1}) - H(q_{n+1},p_{n+1})}{q_{n+1} - q_{n}}, 
    \label{ch3_discrete_sho_2}
\end{eqnarray}
and solving them as a system of equations we have, 
\begin{equation}
     H(q_{n+1},p_{n+1}) - H(q_{n},p_{n}) = 0, 
     \label{ch3_sho_conservation}
\end{equation}
demonstrating equations (\eqref{ch3_discrete_sho_1}, \eqref{ch3_discrete_sho_2}) conserve the Hamiltonian \textit{exactly}. Thus we define, that a time integrator is said to conserve energy \textbf{\textit{exactly}} if 
\begin{equation}
     H(q_{n+1},p_{n+1}) = H(q_{n},p_{n}), 
     \label{ch3_sho_conservation}
\end{equation}
for every smooth Hamiltonian $H(q,p)$. \\

\textbf{II. Symplectic time-integrators \cite{weyl1939classical,hairer2010geometric}}
We define a numerical time-integrator advancing the solution from $t_{n}$ to $t_{n+1}$, as $u_{n+1} = U_{h}(u_{n})$, where $U_{h}$ is the governing evolution matrix depending on the type of time-stepper under study. This numerical integrator will be symplectic if for every smooth Hamiltonian such as $H(q,p)$ and every step-size \texttt{h} the mapping \cite{hairer2010geometric}, 
\begin{equation}
    u_{h}:\mathbb{R}^{2d} \rightarrow \mathbb{R}^{2d}, \hspace{0.4 cm} u_{h}: (q_{n}, p_{n}) \rightarrow (q_{n+1}, p_{n+1}), 
\end{equation}
is symplectic, meaning its Jacobian matrix is, 
\begin{equation}
   U_{h} := \bigg[\frac{\partial u_{h}(q_{n}, p_{n})}{\partial q_{n}} \frac{\partial u_{h}(q_{n},p_{n})}{\partial p_{n}} \bigg], 
    \label{app_def_simp_jac}
\end{equation}
is sympletic for all \texttt{h} i.e., $U^{T}_{h} J U_{h}  = J$, where $J$ is as defined in equation \eqref{ham_simp_u}. 

\textbf{III. Symmetric time-integrators}
A numerical time-stepper, as introduced above, i.e. $u_{n+1} = U_{h}(u_{n})$, is \textit{symmetric} or \textit{time-reversible} if \cite{hairer2010geometric}, 
\begin{equation}
    U_{h} \circ U_{-h} = \texttt{id} \quad \textnormal{or equivalently,} \quad U_{h} = U^{-1}_{-h}, 
\end{equation}
where \texttt{id} is the identity map. 
Using the adjoint method definition, $U^{*}_{h} = U^{-1}_{-h} $, the requirement for symmetry/time-reversibility is  $U_{h} = U^{*}_{h} $. Thus, a time-integrator is \textit{symmetric} if under the exchange $u_{n} \leftrightarrow u_{n+1}$ and $h \leftrightarrow -h$ it remains unaltered. Furthermore, it follows, that a symmetric time-integrator is $\rho -$ reversible if, 
\begin{equation}
    \rho \circ U_{h} = U_{-h} \circ \rho, 
\end{equation}
then the discrete flow $U_{h}$ is a $\rho-$reversible map, if and only if, $U_{h}$ is a \textit{symmetric/time-reversible} time-integrator. 

We will now prove that Hermite \texttt{IMTEX} schemes conserve both energy and symplectic structure exactly while being symmetric. For simplicity, we consider the simple harmonic oscillator given as,
\begin{equation}
H(p,q) = \frac{1}{2}(p^{2} + q^{2}). 
    \label{ch3_harmonic_oscillator}
\end{equation}
Rewriting eq.~(\eqref{app_hamiltoniansys_general}), following eq.~(\eqref{ch3_general_pdes}), as a reduced system of ODEs in matrix form one has, 
\begin{equation}
    \frac{d \textbf{U}}{dt} = \textbf{L} \ \textbf{U}, 
    \label{ch3_reduced_systemODEs}
\end{equation}
where, 
\begin{eqnarray}
    \textbf{U} = 
    \begin{pmatrix}
        p \\ q 
    \end{pmatrix}, \qquad 
  \textbf{L} = 
            \begin{pmatrix}
                0 & -1\\
                1&0
                \label{ch3_ho_operator}
            \end{pmatrix}. 
    \label{ch3_reduced_ode}
\end{eqnarray}

For all the proofs, for simplicity, one considers \texttt{IMTEX} Hermite schemes of solely order-2 given as, 
\begin{align}
&\bigg[ \textbf{I} - \frac{\Delta t}{2}\textbf{L} \bigg] \cdot  \textbf{U}_{n+1}=\bigg[ \textbf{I} + \frac{\Delta t}{2}\textbf{L} \bigg] \cdot \textbf{U}_{n}, \nonumber \\
&\textbf{U}_{n+1}=\bigg[ \textbf{I} - \frac{\Delta t}{2}\textbf{L} \bigg]^{-1} \cdot \bigg[ \textbf{I} + \frac{\Delta t}{2}\textbf{L} \bigg] \cdot \textbf{U}_{n},  \ \ \ 
\label{ch3_imtex_order2}
\end{align}
and in its \texttt{HF IMTEX} form, 
\begin{align}
    \label{ch3_hornerfor_sh2}
    &\textbf{U}_{n+1} = \textbf{U}_{n}+ (\Delta t \ \textbf{L}) \cdot \textbf{HFH2} \cdot \textbf{U}_{n},  \\
    &\textbf{HFH2} = \bigg( \textbf{I} - \frac{\Delta t}{2} \textbf{L}\bigg)^{-1}. 
    \label{ch3_hfh2}
\end{align}
All that follows, nevertheless, trivially extends to higher orders. 

\subsubsection{Hermite \texttt{IMTEX} and \texttt{HF IMTEX} \newline as energy-preserving schemes}

Expanding eq.~\eqref{ch3_imtex_order2} with the terms as defined for the harmonic oscillator in eq.~\eqref{ch3_ho_operator} we have, 
\begin{align}
    \begin{pmatrix}
        q_{n+1} \\ p_{n+1} 
    \end{pmatrix}=  
    \begin{bmatrix}
      -1 + \frac{8}{4+\Delta t^{2}} & \frac{4\Delta t}{4 + \Delta t^{2}}\\
     -\frac{4\Delta t}{4 + \Delta t^{2}}&-1 + \frac{8}{4+\Delta t^{2}}
    \end{bmatrix}\cdot \begin{pmatrix}
        q_{n} \\ p_{n} 
    \end{pmatrix}, 
    \label{proof_imtex_ep_1}
\end{align}
subtracting both sides by $q_{n}$ and $p_{n}$ and dividing the equations by $\Delta t$ we get, 
\begin{align}
&\frac{q_{n+1} - q_{n}}{\Delta t} = \frac{1}{\Delta t} \bigg[-2 + \frac{8}{4+\Delta t^{2}} \bigg] q_{n} +  \frac{4\Delta t}{4 + \Delta t^{2}} p_{n}\\
&\frac{p_{n+1}-p_{n}}{\Delta t} = \frac{1}{\Delta t} \bigg[-2 + \frac{8}{4+\Delta t^{2}} \bigg] p_{n} - \frac{4\Delta t}{4 + \Delta t^{2}} q_{n}. 
\label{proof_imtex_ep_2}
\end{align}
As evidenced by equations (\eqref{ch3_discrete_sho_1}, \eqref{ch3_discrete_sho_2}) we can establish the equality, 
\begin{align}
&\frac{1}{\Delta t} \bigg[-2 + \frac{8}{4+\Delta t^{2}} \bigg] q_{n} +  \frac{4\Delta t}{4 + \Delta t^{2}} p_{n} =    \frac{H(p_{n+1}, q_{n}) - H(p_{n},q_{n})}{p_{n+1} - p_{n}}, \\
&\frac{1}{\Delta t} \bigg[-2 + \frac{8}{4+\Delta t^{2}} \bigg] p_{n} - \frac{4\Delta t}{4 + \Delta t^{2}} q_{n} =  \frac{H(p_{n+1}, q_{n}) - H(p_{n+1},q_{n+1})}{q_{n+1} - q_{n}}. 
\label{proof_imtex_ep_3}
\end{align}
Defining for simplicity $c_{1} = (-2 + 8/ 4+\Delta t^{2})$ and $c_{2} = 4 /4 + \Delta t^{2}$, we have, 
\begin{small}
\begin{align}
\label{proof_imtex_ep_4_1}
&\frac{1}{\Delta t} c_{1}^{2}p_{n}q_{n} - c_{2}c_{1}q_{n}^{2} + c_{2}c_{1}p_{n}^{2} - \Delta t c_{2}^{2}p_{n}q_{n} =    H(p_{n+1}, q_{n}) - H(p_{n},q_{n}), \\
&\frac{1}{\Delta t} c_{1}^{2}p_{n}q_{n} - c_{2}c_{1}q_{n}^{2} + c_{2}c_{1}p_{n}^{2} - \Delta t c_{2}^{2}p_{n}q_{n} =  H(p_{n+1}, q_{n}) - H(p_{n+1},q_{n+1}). 
\label{proof_imtex_ep_4_2}
\end{align}
\end{small}
Subtracting equations (\eqref{proof_imtex_ep_4_1} - \eqref{proof_imtex_ep_4_2}) we directly verify eq.~\eqref{ch3_sho_conservation} and thus Hermite \texttt{IMTEX} schemes \textbf{\textit{conserve energy exactly}}. 

Now to assess \texttt{HF IMTEX} when expanded with the terms as defined for the harmonic oscillator in equation \eqref{ch3_ho_operator} also exactly conserves energy we compute
\begin{align}
    \begin{pmatrix}
        q_{n+1} \\ p_{n+1} 
    \end{pmatrix} = \begin{pmatrix}
        q_{n} \\ p_{n} 
    \end{pmatrix} +  
    \begin{bmatrix}
      \frac{-\Delta t^{2}}{2(1+\frac{\Delta t^{2}}{4})} & \frac{\Delta t}{1+\frac{\Delta t^{2}}{4}}\\
     -\frac{\Delta t}{1+\frac{\Delta t^{2}}{4}}&  \frac{-\Delta t^{2}}{2(1+\frac{\Delta t^{2}}{4})} 
    \end{bmatrix}\cdot \begin{pmatrix}
        q_{n} \\ p_{n} 
    \end{pmatrix}, 
    \label{proof_imtex_hf_ep_1}
\end{align}
yielding, 
\begin{align}
    \begin{pmatrix}
        q_{n+1} \\ p_{n+1} 
    \end{pmatrix}=  
    \begin{bmatrix}
      -1 + \frac{8}{4+\Delta t^{2}} & \frac{4\Delta t}{4 + \Delta t^{2}}\\
     -\frac{4\Delta t}{4 + \Delta t^{2}}&-1 + \frac{8}{4+\Delta t^{2}}
    \end{bmatrix}\cdot \begin{pmatrix}
        q_{n} \\ p_{n} 
    \end{pmatrix}, 
    \label{proof_imtex_hf_ep_2}
\end{align}
which is precisely equation \eqref{proof_imtex_ep_1} and hereby we conclude \texttt{HF IMTEX} exactly conserve energy too. 

\subsubsection{Hermite \texttt{IMTEX} and \texttt{HF IMTEX} as sympletic schemes}
Calculating the determinant of equation \eqref{proof_imtex_ep_1} pertaining to the Hermite \texttt{IMTEX} scheme we get, 

\begin{equation}
    \textrm{det} \ \textbf{U}_{\texttt{IMTEX H2}} =\bigg(-1 + \frac{8}{4+\Delta t^{2}} \bigg)^{2} + \bigg(\frac{4\Delta t}{4 + \Delta t^{2}}\bigg)^{2} = 1.  
    \label{det_imtex_h2}
\end{equation}
Thus, the required condition for symplecticity holds. Further inspecting that $\textbf{U}_{\texttt{IMTEX H2}}$ is symplectic we test,  
\begin{align}
    &U^{T}_{h} J  U_{h} = \begin{pmatrix}
       -1 + \frac{8}{4+\Delta t^{2}} &\frac{\Delta t}{1+\frac{\Delta t^{2}}{4}}\\
     -\frac{\Delta t}{1+\frac{\Delta t^{2}}{4}}& -1 + \frac{8}{4+\Delta t^{2}}
    \end{pmatrix} \cdot J \cdot \begin{pmatrix}
       -1 + \frac{8}{4+\Delta t^{2}} &-\frac{\Delta t}{1+\frac{\Delta t^{2}}{4}}\\
     \frac{\Delta t}{1+\frac{\Delta t^{2}}{4}}& -1 + \frac{8}{4+\Delta t^{2}}
    \end{pmatrix}  \nonumber \\
    &= J.
\end{align}
Finally, by equation \eqref{proof_imtex_hf_ep_2} Hermite \texttt{HF IMTEX} also \textbf{conserves symplectic structure exactly}. 

\section{Complementary work to Section \ref{Sec34_DiscoTEX_for_wave} describing \texttt{DiscoTEX}}

In this section, we include the corrections necessary to build the correct discontinuous time-integration scheme used. We then explicitly give the first jump terms to facilitate comparisons with potentially alternative strategies and, naturally, for completion.

\subsection{Hermite discontinuous time-integration schemes}\label{app_disco_time_h4}

We give the first two orders discontinuous Hermite time-integration schemes. We note in \cite{discotexII} that the schemes are given and applied to the computation of the numerical solution to the distributionally sourced wave-equation up to \texttt{12th}-order. 

\begin{align}
  &f(t)_{\texttt{DH2}} = \frac{\Delta t }{2} \bigg(f(t_{n}) + f(t_{n+1} )  \bigg) + \textbf{J}_{\texttt{H2}}(\Delta  t_{\times}, \Delta t), 
\label{app3_disco_time_h2}
\end{align}

\begin{align}
    &f(t)_{\texttt{DH4}} =\frac{\Delta t }{2} \bigg(f(t_{n}) + f(t_{n+1} )  \bigg) + \frac{\Delta t^{2} }{12}  \bigg(\dot{f}(t_{n}) - \dot{f}(t_{n+1} )  \bigg) \nonumber \\
    &+ \textbf{J}_{\texttt{H4}}(\Delta  t_{\times}, \Delta t), 
    \label{app3_disco_time_h4}
\end{align}

where $\textbf{J}_{\texttt{H2-H4}}(\Delta  t_{\times}, \Delta t)$ are given respectively as,  
\begin{align}
    &\textbf{J}_{\texttt{H2}}(\Delta  t_{\times}, \Delta t) 
    = \frac{1}{2}(\Delta t - 2 \Delta  t_{\times}) \textbf{J}_{0} 
    + \frac{\Delta  t_{\times}}{2}(\Delta  t_{\times} - \Delta t) \textbf{J}_{1}, 
    \label{app3_disco_time_Jh2}
\end{align}

\begin{align}
    &\textbf{J}_{\texttt{H4}}(\Delta  t_{\times}, \Delta t) = \frac{1}{2}(\Delta t - 2 \Delta t_{\times} )  \textbf{J}_{0}  + \frac{\Delta  t_{\times}^{2}}{12}(\Delta t^{2} - 6 \Delta t \Delta  t_{\times} + 6 \Delta t_{\times} ^{2})  \textbf{J}_{1} \nonumber \\
    &- \frac{1}{12}\Delta  t_{\times} (\Delta t^{2} - 3 \Delta t \Delta  t_{\times} + 2 \Delta  t_{\times} ^{2}) \textbf{J}_{2} + \frac{1}{24}\Delta  t_{\times}^{2}, (\Delta t-\Delta  t_{\times}) \textbf{J}_{3}. \ \ \ \ \ \ \ \ 
    \label{app3_disco_time_Jh4}
\end{align}

\subsection{First jumps in the $(t,x)$ coordinate chart}\label{app_space_jumps}
The first jumps pertaining to equations (\eqref{ch34_wavetx_REC_j0}, \eqref{ch34_wavetx_REC_j1},\eqref{ch34_wavetx_REC_jmp2}) are,  
\begin{align}
    \label{wave_tx_j0}
    &J_{0}(t) = -\frac{\iu \cos{t}}{v^{2}-1}, \\
    \label{wave_tx_j1}
    &J_{1}(t) = \frac{\cos{t} -v^{2} \cos{t} + 2\iu v \sin{t}}{(v^{2}-1)^{2}}, \\
    \label{wave_tx_j2}
    &J_{2}(t) =  \frac{\iu (1+ 3v^{2})\cos{t} + 2 v (v^{2}-1) \sin{t}}{(v^{2}-1)^{3}}, \\
    \label{wave_tx_j3}
    &J_{3}(t) = \frac{(-1 -2v^{2} +3v^{4}) \cos{t} - 4 \iu v (1+v^{2}) \sin{t}}{(v^{2}-1)^{4}}, \\
    \label{wave_tx_j4}
    &J_{4}(t) =  \frac{-\iu(1+5v^{2}(2 + v^{2}))\cos{t} - 4 v (-1+v^{4}) \sin{t}}{(v^{2}-1)^{5}}, \\
    &J_{19}(t) = \frac{1}{(v^{2}-1)^{20}} \bigg[(-1 + v) (1 + v) \bigg( 1 + 19v^{2} \big( 9 + 204 v^{2} + 1428v^{4} \nonumber \\
    &  +3978v^{6} + 4862v^{8} + 2652v^{10} + 612v^{12} +51v^{14} + v^{16} \big)\bigg)\cos{t} \nonumber  \\
    & -4\iu v(1+v^{2})(5+10v^{2} +v^{4})(1+44v^{2} +166v^{4}+44v^{6} +v^{8})\nonumber \\
    &(1+5v^{2}(2+v^{2}))\sin{t} \bigg]. 
    \label{wave_tx_j20}
\end{align}

\subsection{Time jumps for the implementation of \texttt{DiscoTEX} in the $(t,x)$ coordinate chart}\label{app_timejumps}
The first time jumps associated with the implementation of the term $\textbf{J}_{\texttt{H4}}(\Delta t_{\times}, \Delta t)$ and $\textbf{J}(t)$ effective vector in equations (\eqref{ch3_discoTEX4_Wave_tr}-\eqref{ch34_wavetr_timejumps}) are, 
\begin{align}
    \label{wave_jb_0}
    &\mathbb{J}_{0} = -\frac{v(v^{2}-1) \cos{t} - \iu (v^{2}-1) \sin{t}}{(v^{2}-1)^{2}}, \\
    \label{wave_Kb_0}
    &\mathbb{K}_{0} = \frac{ \iu (1+ 3v^{2}) \cos{t} + 2v(v^{2} -1)\sin{t}}{(v^{2}-1)^{3}}, \\
    \label{wave_Lb_0}
    &\mathbb{L}_{0} = \frac{v(-3 + 2 v^{2} +v^{4}) \cos{t} - \iu (1+ 6 v^{2} +v^{4})\sin{t}}{(v^{2}-1)^{4}}, \\
    \label{wave_Mb_0}
    &\mathbb{M}_{0} = \frac{ \iu (1+ 5v^{2}(2+v^{2}))\cos{t} + 4v(v^{4} -1)\sin{t}}{(v^{2}-1)^{5}}, 
\end{align}
and, 
\begin{align}
    \label{wave_jc_0}
    &\mathcal{J}_{0} = -\frac{i(1+3v^{2}) \cos{t} + 2v(v^{2}-1)\sin{t} }{(v^{2}-1)^{3}}, \\
    \label{wave_Kc_0}
    &\mathcal{K}_{0} = -\frac{v(-3 + 2v^{2} + v^{4}) \cos{t} - \iu (1+6v^{2}+v^{4})\sin{t} }{(v^{2}-1)^{4}}, \\
    \label{wave_Lc_0}
    &\mathcal{L}_{0} = \frac{\iu (1+5v^{2}(2+v^{2}))\cos{t} + 4 v (v^{4}-1)\sin{t}}{(v^{2}-1)^{5}}, \\
    \label{wave_Mc_0}
    &\mathcal{M}_{0} = \frac{-v(-5 -5v^{2}+9v^{4} +v^{6})\cos{t}}{(v^{2}-1)^{6}} \nonumber \\
    & +\frac{\iu 4 (1+v^{2}) (1+ 14v^{2}+v^{4})\sin{t}}{(v^{2}-1)^{6}}. 
\end{align}
Explicitly, 
\begin{align}
    &\textbf{J}_{0} = \begin{pmatrix} \mathbb{J}_{0} \\ \mathcal{J}_{0}\end{pmatrix}, \textbf{J}_{1} = \begin{pmatrix} \mathbb{K}_{0} \\ \mathcal{K}_{0}\end{pmatrix},\textbf{J}_{2} = \begin{pmatrix} \mathbb{L}_{0} \\ \mathcal{L}_{0}\end{pmatrix},\textbf{J}_{3} = \begin{pmatrix} \mathbb{M}_{0} \\ \mathcal{M}_{0}\end{pmatrix}. 
    \label{jtime_tx}
\end{align}

It is important to emphasise that the time jumps $\textbf{J}_{j}$ are taken in the limit of $t_{i} \rightarrow x_{i}/v$. For the results discussed in Section \ref{ch34_waveTR_l_op_boundaryCondos} we have $v = \dot{r}_{p}(t) = 1/4$ and $x_{i} \in [-4,4]$ where \texttt{i = 0, $\cdots$, N}, with \texttt{N=34}. Concretely, the time jumps are fully evaluated on the numerical simulation domain and hence can be stored prior to the numerical evolution outputting the numerical weak-form solution, as given in equation \eqref{ch3_discoTEX4_Wave_tr}, thus, reducing the total simulation running time.

\subsection{Hyperboloidal recurrence relation derivation and initialising jumps}\label{app_hyper_rec_hyper_jumps}
\subsubsection{Hyperboloidal recurrence relation}
For completion, the corrected hyperboloidal differential operators are included here as described in detail \ref{app2}. The procedure follows directly and all the remainder terms give the $J_{m+2}(\tau)$ recurrence relation given in Eq.~\eqref{cha3_rec_relation_mjumps_PhysicalChart_RWZ}. 
The function $\Psi^{J}(\sigma,\tau;\sigma_{p})$ first-order derivative has the form,  
\begin{align}
    &\upvarrho(\sigma) \partial_{\tau}\Psi(\zeta) = \sum^{\infty}_{m=0} \bigg[ \sum^{m}_{k=0} \upvarrho(\xi_{p})^{(k)} {m\choose k} \bigg( \dot{J}_{m-k}(\tau) - \dot{\xi}_{p} J_{m+1-k}(\tau) \bigg)  \bigg] \Psi(\zeta) \nonumber \\ 
    & -\upvarrho(\xi_{p})\dot{\xi}_{p} J_{0}(\tau)\delta(\zeta).   
    \label{hyper_partialtau_rho}
\end{align}
For the second-order temporal form, we attain, 
\begin{align}
&\Gamma(\sigma) \partial_{\tau}^{2}\Psi(\zeta) = \sum^{\infty}_{m=0} \bigg[ \sum^{m}_{k=0} \Gamma(\xi_{p})^{(k)} {m\choose k}  \bigg( \ddot{J}_{m-k}(\tau) - 2\dot{\xi}_{p} \dot{J}_{m+1-k}(\tau) \nonumber \\
    &- \ddot{\xi}_{p} J_{m+1-k}(\tau)  +  \dot{\xi}_{p}^{2}J_{m+2-k} \bigg) \bigg]\Psi(\zeta) + \Gamma(\xi_{p}) \dot{\xi}_{p}^{2}J_{1}(\tau)\delta(\zeta) \nonumber \\
    &- 2 \Gamma(\xi_{p}) \dot{\xi}_{p} \dot{J}_{0}(\tau) \delta(\zeta)  -\Gamma(\xi_{p}) \ddot{\xi}_{p} J_{0}(\tau) \delta(\zeta) - \Gamma(\xi_{p}) \dot{\xi}_{p}^{2}J_{0}(\tau)\delta'(\zeta). 
   \label{hyper_partiald2tau_Gamma}   
\end{align}
The spatial form of $\Psi^{J}(\sigma,\tau;\xi_{p})$ is given at first order as, 
\begin{align}
    &\iota(\sigma)\partial_{\sigma}\Psi(\zeta) = \sum^{\infty}_{m=0} \bigg[ \sum^{m}_{k=0} \iota^{(k)}(\xi_{p})  {m\choose k} \bigg( J_{m+1-k}(\tau) \bigg) \bigg]\Psi(\zeta) \nonumber \\
    &+\iota(\xi_{p})J_{0}(\tau)\delta(\zeta)
    \label{hyper_partialsigma_iota}
\end{align}
and second-order,
\begin{align}
     &\upchi(\sigma)\partial_{\sigma}^{2}\Psi(\zeta) = \sum^{\infty}_{m=0} \bigg[ \sum^{m}_{k=0}  \upchi(\xi_{p})^{(k)}  {m\choose k} \bigg( J_{m+2-k}(\tau))\bigg) \bigg]\Psi(\zeta) \nonumber \\
     &+  \upchi(\xi_{p}) J_{1}(\tau)\delta(\zeta) +  \upchi(\xi_{p})J_{0}(\tau)\delta'(\zeta). 
    \label{hyper_partiald2sigma_chi}
\end{align}
Finally the first order crossed derivative in $(\tau,\sigma)$ is given as, 
\begin{align}
    &\upvarepsilon({\sigma})\partial_{\sigma}\partial_{\tau}\Psi(\zeta) = \sum^{\infty}_{m=0} \bigg[ \sum^{m}_{k=0} \upvarepsilon(\xi_{p})^{(k)}{m\choose k} \nonumber \\ 
    &\times \bigg( \dot{J}_{m+1-k}(\tau) - \dot{\xi}_{p}J_{m+2-k}(\tau) \bigg) \bigg] \Psi(\zeta) \nonumber \\
    &+ \upvarepsilon(\xi_{p}) \big(\dot{J}_{0}(\tau) - \dot{\xi}_{p}J_{1}(\tau)\big)\delta(\zeta) - \upvarepsilon(\xi_{p})\dot{\xi}_{p} J_{0}(\tau)\delta'(\zeta).  
    \label{hyper_partialrhopartialtau_epsilon} 
\end{align}

Collecting all higher-order terms we obtain the following recurrence relation, 
\begin{align}
    &J_{m+2}(\tau) =  -\bar{\gamma}^{2}   \bigg[ \sum^{m}_{k=0} {m \choose k} \bigg(\upvarepsilon^{(k)}(\xi_{p}) \dot{J}_{m+1-k} + \iota^{(k)}(\xi_{p}) J_{m+1-k}  \nonumber \\
    &- V^{(k)}(\xi_{p}) J_{m-k} + \upvarrho^{(k)} (\xi_{p}) (\dot{J}_{m-k} - \dot{\xi}_{p} J_{m+1-k}) \nonumber \\ 
    &+ \Gamma^{(k)}(\ddot{J}_{m-k} - 2 \dot{\xi}_{p} \dot{J}_{m+1-k}- \ddot{\xi_{p}}J_{m+1-k}  ) \bigg)   \nonumber \\
    &+  \sum^{m}_{k=1} {m \choose k} \bigg(\dot{\xi}_{p}^{2} \Gamma^{(k)}(\xi_{p}) + \dot{\xi}_{p} \upvarepsilon^{(k)}(\xi_{p}) + \upchi^{(k)}(\xi_{p}) \bigg) J_{m+2-k} \bigg].
\label{cha34_rec_relation_njumps_RWZ_hyper}
\end{align}
where here for simplicity the time dependence on the RHS of the jumps have been suppressed and we have, $\bar{\gamma}^{-2} = \big(  \dot{\xi}_{p}^{2} \Gamma(\xi_{p}) -  \dot{\xi}_{p} \upvarepsilon(\xi_{p}) - \upchi(\xi_{p})\big)$. 
Furthermore, it is noted that the initialising jumps $J_{0}(\tau), J_{1}(\tau)$, given explicitly in equations (\eqref{hyper_j0}, \eqref{hyper_j1}), are obtained through careful implementation of the chain rule and Dirac-$\delta$ distribution composition rules as given in equations (\eqref{appendix_a_dirac_compI}-\eqref{appendix_a_dirac_compositionII}) acting on the $J_{0}(t), J_{1}(t)$ jumps as given in equations (\eqref{ch3_j0}-\eqref{ch3_jt}) following the derivation in Appendices \ref{appA_frobs} and above.  We also note for the derivation of the jumps, as described thoroughly equations (\eqref{hyper_partialtau_rho}- \eqref{hyper_partialrhopartialtau_epsilon}), we work with the source term without dividing by the $\Gamma(\sigma)$, as we could be misled by equation \eqref{ch34_hyper_wave_equation}. This is then accounted for by incorporation of the source terms through eq.~\eqref{cha2_spatial_disc_generic} and explicitly given in eqs.~(\eqref{s1psi_iota} - \eqref{spi_epsilon}). 

\subsubsection{Initialising jumps $J_{0}(\tau), J_{1}(\tau)$}\label{app_initialising_jumps}
In \cite{da2023hyperboloidal} for convenience, we derived the initialising jumps directly from chain-rules transformations on the jumps in $(t,r)$ coordinates.\footnote{This have now been independently implemented, directly from the paper, in upcoming work using an independent fully-spectral time-domain \cite{macedo2014axisymmetric} algorithm \cite{paper2,paper3}. The only common factor in those algorithms numerically is the use of the same hyperboloidal chart.} Here one shows how these can be obtained by the correct application of the Dirac delta proprieties highlighted in equations (\eqref{appendix_a_dirac_compI}-\eqref{appendix_a_dirac_compositionII}).  Considering the RHS of equation \eqref{ch34_hyper_wave_equation} we have, 
\begin{align}
    &S(t,r_{p}) = F(t)\delta'(r-r_{p}) + G(t)\delta(r-r_{p}).
    \label{app_waveq_rhs}
\end{align}
The task now is to transform the source term to the new coordinate chart, where $t \rightarrow t(\tau,\sigma), x \rightarrow x(\sigma)$ as given by equations (\eqref{ch34_minimal_gauge_tx}, \eqref{ch34_heightfunction}). The new particle motion can be described by a function $f(t(\tau_{p},\sigma_{p}), x(\sigma_{p}))$ where $f(t,r_{p}):= x - r_{p}(t)$. Applying chain-rule is attained,  
\begin{align}
    \partial_{\sigma} f = x'(\sigma) + \dot{r_{p}} H'(\sigma), && \partial_{\tau}f = -\dot{r}_{p}, 
    \label{app_chainrule_correc}
\end{align}
For simplicity we use $\partial_{\sigma}f = \triangle_{c}(\xi_{p})$, and $\tau_{c} =t(\tau,\sigma_{p}(\tau))$, $\xi_{p} = x(\sigma_{p}(\tau)) $ to denote the new coordinate time when at the particle position $\xi_{p}$, whose motion will shortly be specified. Starting by applying equation \eqref{appendix_a_dirac_compI} we simple have, 
\begin{align}
    G(t) \delta(r-r_{p}) = \frac{G(\tau_{c})}{\triangle_{c}(\xi_{p})}\delta(\sigma-\xi_{p}). 
    \label{app_gtr_transf}
\end{align}
For simplicity we define $\zeta = \sigma - \xi_{p}$ as in \cite{da2023hyperboloidal}. The correction associated with the transformation on the derivative of the Dirac $\delta-$ distribution is more complex, 
\begin{align}
    & F(t) \delta'(r-r_{p})  = F(t) \bigg[\frac{\triangle_{c}(\xi_{p})}{|\triangle_{c}(\xi_{p})|^{3}}\delta'(\zeta)  +  \frac{\triangle_{c}'(\xi_{p})}{|\triangle_{c}(\xi_{p})|^{3}} \delta(\zeta) \bigg] \\
    &= \frac{\triangle_{c}(\xi_{p})}{|\triangle_{c}(\xi_{p})|^{3}} \bigg[F(\tau_{c}) \delta'(\zeta) - \frac{\partial F(\tau_{c})}{\partial \xi_{p}}\delta(\zeta)\bigg] +   F(\tau_{c}) \frac{\triangle_{c}'(\xi_{p})}{|\triangle_{c}(\xi_{p})|^{3}} \delta(\zeta), \\
    &=\frac{\triangle_{c}(\xi_{p})}{|\triangle_{c}(\xi_{p})|^{3}} \bigg[F(\tau_{c}) \delta'(\zeta) - \frac{\partial F(\tau_{c})}{\partial \xi_{p}} \frac{\partial t}{\partial t}\delta(\zeta)\bigg]  + F(\tau_{c}) \frac{\triangle_{c}'(\xi_{p})}{|\triangle_{c}(\xi_{p})|^{3}} \delta(\zeta), \nonumber \\
    &= F(\tau_{c})\frac{\triangle_{c}(\xi_{p})}{|\triangle_{c}(\xi_{p})|^{3}} \delta'(\zeta) \nonumber \\
    &+ \bigg[ \frac{\triangle_{c}'(\xi_{p})}{|\triangle_{c}(\xi_{p})|^{3}}F(\tau_{c})  - \frac{\triangle_{c}(\xi_{p})}{|\triangle_{c}(\xi_{p})|^{3}} \frac{\partial F(\tau_{c})}{\partial t} \frac{\partial t}{\partial \xi_{p}}  \bigg]\delta(\zeta), 
\end{align}
where the first equality is obtained by applying the composition rule in equation \eqref{appendix_a_dirac_compositionII} and the second from applying the selection propriety \eqref{appendix_a_dirac_selectionII} so as to ensure all is evaluated with respect to the particle worldline $\xi_{p}$ with carefully considering the coordinate chart transformation in question as evidenced in the third-to-fourth lines. Finally following the recipe highlighted in \ref{appA_frobs} the following initialising jumps are obtained, 
\begin{align}
    \label{hyper_j0}
    &J_{0} = \bar{\gamma}^{2} \frac{\triangle_{c}(\xi_{p})}{|\triangle_{c}(\xi_{p})|^{3}} F(\tau_{c})   \\
    &J_{1} = \bar{\gamma}^{2} \bigg[  \frac{\triangle_{c}'(\xi_{p})}{|\triangle_{c}(\xi_{p})|^{3}}F(\tau_{c})  - \frac{\triangle_{c}(\xi_{p})}{|\triangle_{c}(\xi_{p})|^{3}} \frac{\partial F(\tau_{c})}{\partial t} \frac{\partial t}{\partial \xi_{p}} +  \frac{G(\tau_{c})}{\triangle_{c}(\xi_{p})}\nonumber \\
    &+ \bigg( \ddot{\xi}_{p}\Gamma(\xi_{p})  + \dot{\xi}_{p}^{2}\Gamma'(\xi_{p}) - \dot{\xi}_{p}\varepsilon'(\xi_{p})  + \varrho'(\xi_{p})   \chi'(\xi_{p}) - \iota(\xi_{p}) \bigg)J_{0} \nonumber \\
    &+ \bigg( 2 \dot{\xi}_{p}\Gamma(\xi_{p})  - \varepsilon(\xi_{p}) \bigg)\dot{J}_{0}  \bigg]. 
    \label{hyper_j1}
\end{align}

Finally, we define the quantities describing the particle trajectory, $\xi_{p}$, in the new coordinate chart. Given the nature of the transformations we work with, the inverse function $\bar{\xi}_{p}$ is attained via
\begin{align}
    \label{hyper_trajectory_1}
     &x(\sigma) - v \ t(\bar{\xi}_{p}, \sigma) = 0, \\
     &\bar{\xi}_{p}(\sigma) = \frac{(1+v)\sigma \ln(1-\sigma) + (v-1)(\sigma \ln(\sigma) -1) }{2 v \sigma}.
    \label{hyper_trajectory_2}
\end{align}
The inverse equation, i.e. $\xi_{p}(\tau) = \sigma_{p}(\tau)$ of equation \eqref{hyper_trajectory_2} is then obtained numerically implicitly via interpolation on a given numerical domain. This is done after all discontinuous formalism terms have been computed in terms of the variable $\tau$ spanning the computational time interval, which, as stated in Section \ref{Sec34_DiscoTEX_for_wave_hyperboloidal} is $\tau \in [-1.52,4.50]$ with $ v=1/4$.

\subsection{First jumps in the $(\tau,\sigma)$ coordinate chart}\label{app_hyper_jumps}

The first initialising jumps in the hyperboloidal chart and a few of the higher order jumps as given by the recurrence relation are given by equation \eqref{cha34_rec_relation_njumps_RWZ_hyper}, 
\begin{align}
    \label{wave_hyper_j0}
    &J_{0} = -\frac{\iu \cos{\tau_{c}}}{v^{2}-1}, \\
    \label{wave_hyper_j1}
    &J_{1} = \frac{(v^{2}-1)(1-v + 2v \xi_{p}^{2})\cos{\tau_{c}}}{2(v^{2}-1)^{2} (\xi_{p} -1) \xi_{p}^{2}},  \nonumber \\
    &- \frac{\iu (-1(v-1)^{2} + 2(1+v^{2})\xi_{p}^{2} ) \sin{\tau_{c}}}{2(v^{2}-1)^{2} (\xi_{p} -1) \xi_{p}^{2}},\\
    \label{wave_hyper_j2}
    &J_{2} =  \frac{1}{2}\bigg(-\frac{1-\iu}{(-1+v)^{2}(\xi_{p}-1)^{2}} + \frac{1+(1+i) \xi_{p} (2+\xi_{p})}{(1+v)^{2} \xi^{4}_{p}} \bigg) \sin{\tau_{c}} \nonumber \\
    &+\frac{\iu (v-1)^{3} - 2(v-1)^{3} (1+v)^{2}\xi_{p} + (v-1)^{3}((3-2\iu)  }{2(v^{2} -1)^{3} (\xi_{p} -1)^{2}\xi^{4}_{p}} \cos{\tau_{c}} \nonumber \\
    &\times \frac{ +3v(2+v))\xi_{p}^{2} - 2 (\iu + v + 3 \iu v - 2v^{3} +v^{5})\xi_{p}^{4} }{2(v^{2} -1)^{3} (\xi_{p} -1)^{2}\xi^{4}_{p}}. 
\end{align}

\subsection{Time jumps for the implementation of \texttt{DiscoTEX} in the $(\tau,\sigma)$ \textit{minimal gauge} coordinate chart}\label{app_hyper_timejumps}
Similarly to \ref{app_timejumps}, the time jumps now prescribing $\textbf{J}_{\texttt{H4}}(\Delta t_{\times}, \Delta \tau )$ and equations \eqref{ch34_wavetr_timejumps_hyper} are explicitly, 
\begin{align}
    \label{hyper_wave_Jb_0}
   \mathbb{J}_{0} &= \frac{4}{15} \cos{\tau_{c}} + \frac{272}{225}  \iu \sin{\tau_{c}}  \\
    \label{hyper_wave_Kb_0}
    \mathbb{K}_{0} &= \frac{4864}{3375} \iu \cos{\tau_{c}} -  \frac{128}{225} \sin{\tau_{c}} , \\
    \label{hyper_wave_Lb_0}
    \mathbb{L}_{0} &= - \frac{3136}{3375} \cos{\tau_{c}} - \frac{90368}{50625} \iu \sin{\tau_{c}}, \\
    \label{hyper_wave_Mb_0}
    \mathbb{M}_{0} &= - \frac{1724416 }{759375} \iu \cos{\tau_{c}} + \frac{69632}{50625} \sin{\tau_{c}}. 
\end{align}
As we have done in Section \ref{Sec34_DiscoTEX_for_wave_tx} in equations (\eqref{d2x_timejump_gen1} - \eqref{d2x_timejump_gen2}) we need to incorporate the differential operators in $\textbf{L}$, as defined in (\eqref{ch34_red_1ode}, \eqref{ch34_hyperboloidal_L1_p} - \eqref{ch34_hyperboloidal_L2_p}), into the final time jump associated with the $\textbf{L}_{1,2}$ operators. We thus introduce the following,
\begin{align}
    \label{hyper_wave_jc_0}
    &\mathcal{J}_{0} =\mathcal{J}_{0,\chi} + \mathcal{J}_{0,\iota}  + \mathcal{J}_{0,\varepsilon} + \mathcal{J}_{0,\varrho},  \\
    \label{hyper_wave_Kc_0}
    &\mathcal{K}_{0} =\mathcal{K}_{0,\chi} + \mathcal{K}_{0,\iota}  + \mathcal{K}_{0,\varepsilon} + \mathcal{K}_{0,\varrho}, \\
    \label{hyper_wave_Lc_0}
    &\mathcal{L}_{0} = \mathcal{L}_{0,\chi} + \mathcal{L}_{0,\iota}  + \mathcal{L}_{0,\varepsilon} + \mathcal{L}_{0,\varrho}, \\
    \label{hyper_wave_Mc_0}
    &\mathcal{M}_{0} = \mathcal{M}_{0,\chi} + \mathcal{M}_{0,\iota}  + \mathcal{M}_{0,\varepsilon} + \mathcal{M}_{0,\varrho}.
\end{align}

Additionally, and unlike in the previous chapter, where $\textbf{L}$ only contained the term $\textbf{L}_{1} = -\partial^{2}_{x}$, i.e. with no $x$-dependent coefficient, we now, though have several differential operators which contain a coefficient given in terms of the $\sigma$ coordinate. We thus will need to apply the rule described in \ref{app2} in equation \eqref{final_proof_coefficents_handling}. To further elucidate this we thus define 
\begin{align}
    &\mathcal{J}(\tau) =\mathcal{J}_{(0,\chi), \bar{m}} +
    \mathcal{J}_{(0,\iota),\bar{m}}  + \mathcal{J}_{(0,\varepsilon), \bar{m}} + \mathcal{J}_{(0,\varrho), \bar{m}}, 
\end{align}
where $\bar{m} = \{\bar{k}, \cdots, m - \bar{k}\}$ and $\bar{k}$ is variable depending on the coefficient of the differential operator in question.  We now specify each of these terms,
\begin{align}
    \label{wave_dx2_jtau_0_exp}
    &\mathcal{J}_{0, \chi} =  \sum^{\bar{m}}_{k=0} {\bar{m}\choose k} \tilde{\chi}^{(k)}(\xi_{p}) J_{\bar{m}-k}(\tau),\\
    \label{wave_dx_jtau_0_exp}
    &\mathcal{J}_{0, \iota} =  \sum^{\bar{m}}_{k=0} {\bar{m}\choose k} \tilde{\iota}^{(k)}(\xi_{p}) J_{\bar{m}-k}(\tau),\\
    \label{wave_dx_jtau_0_exp}
   &\mathcal{J}_{0, \varepsilon} =  \sum^{\bar{m}}_{k=0} {\bar{m}\choose k} \tilde{\varepsilon}^{(k)}(\xi_{p}) \dot{J}_{\bar{m}-k}(\tau),\\
    \label{wave_jtau_0_exp}
    &\mathcal{J}_{0, \varrho} =  \sum^{\bar{m}}_{k=0} {\bar{m}\choose k} \tilde{\varrho}^{(k)}(\xi_{p}) \dot{J}_{\bar{m}-k}(\tau), \\
    \label{jjtau0}
    &\mathcal{J}_{0}(\tau) =  \mathcal{J}_{0, \chi}|_{\bar{k}=2} + \mathcal{J}_{0, \iota}|_{\bar{k}=1} + \mathcal{J}_{0, \varepsilon}|_{\bar{k}=0} + \mathcal{J}_{0, \varrho}|_{\bar{k}=1}.
\end{align}
The other terms namely, $\mathcal{K}_{0}, \mathcal{L}_{0}$ and $\mathcal{M}_{0}$ are determined as explained in Eqs.~(\eqref{tx_k_time}-\eqref{tx_m_time}) in Section \ref{Sec34_DiscoTEX_for_wave_tx}. Below we give the final results:

\begin{align}
   \label{wave_jc_0_exp}
    \mathcal{J}_{0} &= \frac{4\big( \big(225 + 1216\iu \big) \cos{\tau_{c}} - \big(480-1020\iu \big) \sin{\tau_{c}} \big)}{3375}, \\
    \label{wave_Kc_0_exp}
    \mathcal{K}_{0} &= \frac{64\big( \big(-735 + 1140\iu \big) \cos{\tau_{c}} - \big(450-1412\iu \big) \sin{\tau_{c}} \big)}{50625},\\
    \label{wave_Lc_0_exp}
    \mathcal{L}_{0} &= -\frac{64\big(\big(11025 + 26944\iu \big) \cos{\tau_{c}} - \big(16320-21180\iu \big) \sin{\tau_{c}} \big)}{759375},\\
    \label{wave_Mc_0_exp}
    \mathcal{M}_{0} &= \frac{(22133760 - 25866240\iu)\cos{\tau_{c}}}{11390625} \nonumber \\
    &+ \frac{(15667200 + 33492992 \iu) \sin{\tau_{c}}}{11390625}. 
\end{align}

Finally, we note all these equations and others are given before taking the limit as explained at the end of \ref{app_timejumps}. We have decided to do this, such that the match between the jumps in different coordinate charts can be recognised.

\begin{figure*}
\includegraphics[width=89mm]{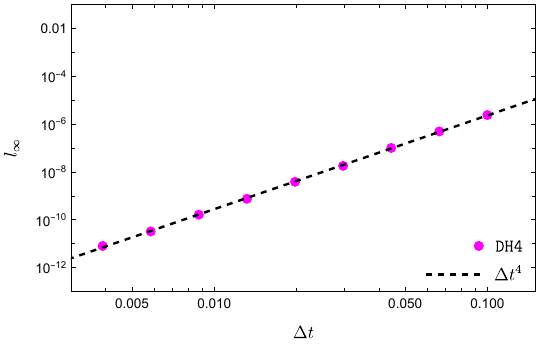}
\quad
\includegraphics[width=89mm]{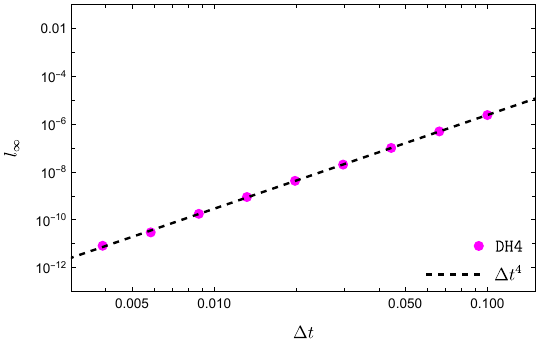}
\quad
\includegraphics[width=89mm]{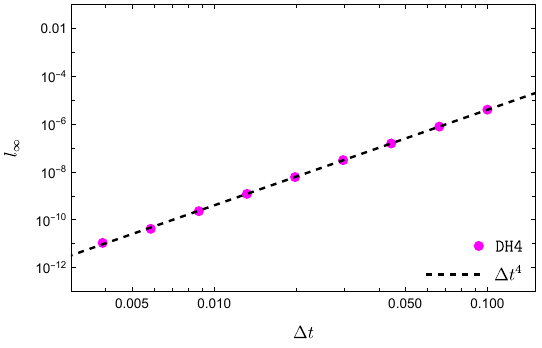}
\quad
\includegraphics[width=89mm]{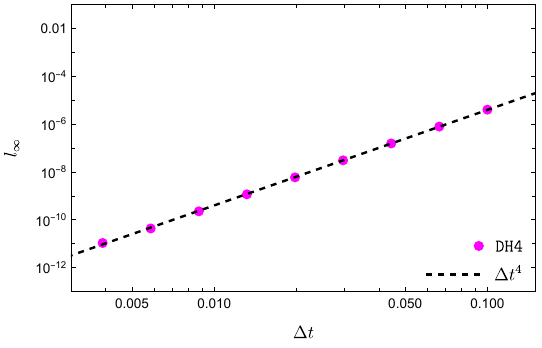}
\caption{Numerical error associated with computation of the numerical weak-form solution $\Psi(t,x_{f})$ against the exact solution in equation \eqref{ch2_distr_wavequation_delta0_sol} (plots on top) and equation \eqref{ch2_distr_wavequation_delta1} (plots on bottom) with both the discontinuous \texttt{IMP} implicit and \texttt{IMTEX} implicit-turned-explicit Hermite integrator order-4. \textbf{Left:} The two plots on the left give the implementation of \texttt{DiscoIMP} with an Hermite order-4 scheme as described by the evolution algorithm in equation \eqref{DiscoIMP_h4}. \textbf{Right:} The two plots on the right give the implementation of \texttt{DiscoTEX} with an Hermite order-4 scheme as described by the evolution algorithm in \eqref{ch3_discoTEX4_Wave_tr}.  \label{discoImp_vs_discotex}} 
\end{figure*}  
\subsection{Numerical weak-form solution to wave equations of Type II and III via \texttt{DiscoTEX} and \newline \texttt{DiscoIMP}}\label{app_discoimp}

To further complement our results in Section \ref{Sec34_DiscoTEX_for_wave} we give the solution to wave-like problems of Type II and Type III. Additionally, we will consider the implicit Hermite integration scheme, here called \texttt{DiscoIMP} at fourth-order and compare with \texttt{DiscoTEX}. In previous work, \cite{2014arXiv1406.4865M} have given an implementation of Type II equations with an implicit Hermite order-2 scheme and using radiation boundary conditions with $v=1/3$, $N=110$ Chebyshev collocation nodes and $J=30$ jumps. We find these computational control factors and those in \cite{o2022timeQMUL, o2022time} to not be optimal, particularly the factors in \cite{2014arXiv1406.4865M} just amount to extra computational cost with no accuracy improvements. The presentation follows very closely to that in Section \ref{Sec34_DiscoTEX_for_wave} and the terms that follow when not defined here, are as defined in that section. The final evolution algorithm describing \texttt{DiscoIMP} is, 
\begin{align}
    &\Pi^{n+1} = \Pi^{n} + \textbf{MF}\cdot \bigg[\textbf{MR} \cdot \Pi^{n} + \Delta t \textbf{L1} \cdot \Psi^{n} \nonumber \\
    &+ \bigg(\frac{\Delta t \textbf{L1} }{2} - \frac{\Delta t^{2}}{12}\textbf{L1}\cdot \textbf{L2}  \bigg) \cdot \textbf{MA}\cdot \bigg( \frac{\Delta t^{2}}{12} (s^{n} - s^{n+1}) - \textbf{MA} \cdot \big(  \Upsilon_{\Pi}\nonumber \\
    &  + \texttt{JH4}_{L2}(\Delta t_{\times}, \Delta t)\Xi \bigg)  + \frac{\Delta t^{2}}{12} \bigg( s^{n} - s^{n+1} \bigg) + \frac{\Delta t}{2}\bigg( s^{n}+s^{n+1}\bigg)   \nonumber \\
    & + \frac{\Delta t^{2}}{12}\bigg( st^{n} + st^{n+1} \bigg) + \Upsilon_{\Psi}  +  \texttt{JH4}_{L1}(\Delta t_{\times}, \Delta t)\Xi  \bigg) \bigg],  \nonumber \\
    &\Psi^{n+1} = \Psi^{n} + \textbf{MA} \cdot \bigg[ \bigg(\frac{\Delta t}{2} \textbf{I} + \frac{\Delta t^{2}}{12}\textbf{L2} \bigg)\cdot \bigg(\Pi^{n} - \Pi^{n+1}\bigg) \nonumber \\ 
    &  + \frac{\Delta t^{2}}{12} \big(s^{n} -s^{n+1} \big) + \Upsilon_{\Psi} + \texttt{JH4}_{L2}(\Delta t_{\times}, \Delta t)\Xi \bigg], 
    \label{DiscoIMP_h4}
\end{align}
where the matrices $\textbf{MF},\textbf{MR},\textbf{MA} $ are given as, 
\begin{align}
    &\textbf{MF} = \bigg[ \textbf{I} - \frac{\Delta t}{2} \textbf{L2} + \frac{\Delta t^{2}}{12}\bigg( \textbf{L1} + \textbf{L2}\cdot \textbf{L2} \bigg) - \bigg( \frac{\Delta t}{2} \textbf{L1} \nonumber \\
    &- \frac{\Delta t^{2}}{12}\textbf{L1}\cdot \textbf{L2}  \bigg) \cdot \textbf{MA} \cdot \bigg(\frac{\Delta t}{2}\textbf{I} - \frac{\Delta t^{2}}{12} \textbf{L2} \bigg)\bigg], \\
     &\textbf{MR} = \bigg[ \textbf{I} + \frac{\Delta t}{2} \textbf{L2} + \frac{\Delta t^{2}}{12}\bigg( \textbf{L1} + \textbf{L2}\cdot \textbf{L2} \bigg) + \bigg( \frac{\Delta t}{2} \textbf{L1} \nonumber \\
    &- \frac{\Delta t^{2}}{12}\textbf{L1}\cdot \textbf{L2}  \bigg) \cdot \textbf{MA} \cdot \bigg(\frac{\Delta t}{2}\textbf{I} - \frac{\Delta t^{2}}{12} \textbf{L2} \bigg) \bigg]\\
    &\textbf{MA} = \bigg[ \textbf{I} + \frac{\Delta t^{2}}{12}\textbf{L1}\bigg]. 
\end{align}
Additionally, here the terms $\texttt{JH4}_{L1/L2}(\Delta t_{\times}, \Delta t)$ refer to the time jumps associated with the operators in equation \eqref{ch34_waveTR_l_op_boundaryCondos}. As we can see from Figure \ref{discoImp_vs_discotex} the implementation of the algorithm with an \underline{imp}licit \texttt{IMP} scheme Hermite order-4 \texttt{DiscoIMP} and with an \texttt{IMTEX} Hermite order-4 integrator as \texttt{DiscoTEX} retrieves the same accuracy while retaining the same conservation geometric proprieties. In both our simulations the exact same numerical optimisation factors were used with $J=20$ jumps, $N=34$ Chebyshev nodes and a time-step of $\Delta t \approx0.003 $ in a computational domain of $x \in [-6,6]$ and $t\in[-1.52,6.60]$ time-interval. We have decided to use \texttt{DiscoTEX} due to its reduced-form easing computational implementations and because it reduces computational running time. 

\section{Black hole perturbation theory machinery}\label{app1_bhpt}
Note, unlike in \cite{da2023hyperboloidal}, the formalism of \cite{hopper2010gravitational} is now followed along with their conventions for the spherical harmonic decomposition formalism implementation.  
\subsection{Projections of the stress-energy tensor}\label{app4_bhpt-projections}
As applied in \cite{hopper2010gravitational} w use the decomposition of Martel and Poisson \cite{martel2005gravitational}, 
\begin{eqnarray}
    p_{ab}(x^{\mu}) = \sum_{l,m} h^{lm}_{ab}Y^{lm}_{B}, \ \ \ 
    p_{aB}(x^{\mu}) = \sum_{l,m} h^{lm}_{a}Y^{lm}_{B}, \\ 
    p_{AB}(x^{\mu}) = r^{2} \sum_{l,m} \bigg( K^{lm} \Omega_{AB}Y^{lm}  + G^{lm} \Omega_{AB}Y^{lm}_{AB} \bigg), 
\end{eqnarray}
where the tensor $\Omega_{AB}$ is the metric on the unit 2-sphere given as,
\begin{equation}
    ds^{2} = \Omega_{AB}dx^{A}dx^{B} = d\theta^{2} + \sin^{2}\theta \ d\varphi^{2}.  
    \label{app_unitsphere}
\end{equation}
\subsubsection{Polar/ZM projections of the stress-energy tensor}\label{app4_projections_even}
The source terms are given by the even-parity field equations described by equations (C3) in \cite{hopper2010gravitational}. 
\begin{align}
    &Q^{ab}(t,r) = 8\pi \int T^{ab}Y^{*}d\Omega,  \\
    &Q^{a}(t,r) = \frac{16 \pi}{l(l+1)}\int T^{aB}Y^{*}_{B}d\Omega, \\
    &Q^{\flat}(t,r) = 8\pi r^{2} \int T^{AB}\Omega_{AB} Y^{*}d\Omega, \\
    &Q^{\sharp}(t,r) = 32 \pi r^{4}\frac{(l-2)!}{(l+2)!}\int T^{AB}Y^{*}_{AB}d\Omega. \hspace{2cm}
    \label{appe_qterms_even}
\end{align}
Re-writing this in terms of time-dependent functions, 
\begin{align}
    &Q^{ab}(t,r) = q^{ab}(t)\delta(r-r_{p}), \ \ \ \ \  Q^{a}(t,r) = q^{a}(t)\delta(r-r_{p}), \\
    &Q^{\flat}(t,r) = q^{\flat}(t)\delta(r-r_{p}), \ \ \ \ \  Q^{\sharp}(t,r) = q^{\sharp}(t)\delta(r-r_{p}).
    \label{appe_qterms_dirac_exp}
\end{align}
The magnitudes of the $q$ terms are given as, 
\begin{align}
    &q^{tt}(t) = 8\pi \mu \frac{\mathcal{E}}{r^{2}_{p}f_{p} }Y^{*}, \ \ \ \ q^{rr}(t) = 8\pi \mu \frac{f_{p}}{\mathcal{E}r^{2}_{p}}(\mathcal{E}^{2} - U^{2}) Y^{*}, \ \ \ \ \\
    &q^{tr}(t) = 8\pi\mu \frac{u^{r}}{r^{2}_{p}},  \\
    &q^{t}(t) = 8\pi \mu \frac{\mathcal{L}}{r^{2}_{p}} Y^{*}_{\varphi}, \ \ \ \ q^{r}(t) = \frac{16 \pi \mu}{l(l+1)} \frac{\mathcal{L}}{\mathcal{E}} \frac{f_{p}}{r^{2}_{p}} u^{r} Y^{*}_{\varphi},  \\ 
    &q^{\flat}(t) = 8\pi \mu \frac{\mathcal{L}^{2}}{\mathcal{E}}  \frac{f_{p}}{r^{4}_{p}} Y^{*}, \ \ \ \ q^{\sharp}(t) = 32 \pi \mu \frac{(l-2)!}{(l+2)!} \frac{\mathcal{L}^{2}}{\mathcal{E}} \frac{f_{p}}{r^{2}_{p}} u^{r} Y^{*}_{\varphi\varphi}, 
    \label{appe_qterms_magnitude}
\end{align}
where the spherical harmonics are given as defined in Ref.\cite{hopper2010gravitational}.

\subsubsection{Axial/CPM projections of the stress-energy tensor}\label{app4_projections_odd}
For the axial perturbations, we have  \cite{hopper2010gravitational,martel2005gravitational}, 
\begin{align}
    &P^{a}(t,r) = \frac{16\pi r^{2}}{l(l+1)} \int T^{aB}X^{*}_{B}d\Omega, \\
    &P(t,r) = 16 \pi r^{4}\frac{(l-2)!}{(l+2)!} \int T^{AB}X^{*}_{AB}d\Omega,
    \label{appe_axial_gen_P}
\end{align}
as in \ref{app4_projections_even}, re-written it as time-dependent functions, 
\begin{eqnarray}
    P^{a}(t,r) &=& p^{a}(t)\delta(r-r_{p}), \\
    P(t,r) &=& p(t)\delta(r-r_{p}), 
\end{eqnarray}
where the magnitude of $p$ is given by, 
\begin{align}
    &p^{t}(t) = \frac{16\pi\mu}{l(l+1)}\frac{\mathcal{L}}{r^{2}_{p}} X^{*}_{\varphi}, \ \ \  
    p^{r}(t) = \frac{16\pi\mu}{l(l+1)}\frac{\mathcal{L}}{\mathcal{E}}\frac{f_{p}}{r^{2}_{p}} u^{r}X^{*}_{\varphi}, \nonumber \\  
    &p(t) = 16 \pi \mu \frac{(l-2)!}{(l+2)!}\frac{\mathcal{L}}{\mathcal{E}} \frac{f_{p}}{r^{2}_{p}}X^{*}_{\varphi\varphi}. \hspace{2cm}
\end{align}

\subsection{Axial/CPM source term}\label{app4_bhpt_axial_source}
The axial/CPM source term describing the perturbations of $\Psi^{a}_{lm}(t,r)$ given in equation \eqref{ch4_CPM_ZM_wavequation} is, 
\begin{eqnarray}
       S_{lm}^{a}(t,r_{p}(t)) =  \bar{F}^{a}_{lm}(t)\delta'(r-r_{p}(t))
      + \bar{G}^{a}_{lm}(t)\delta(r-r_{p}(t)). \ \ \ \ 
      \label{cha4_sourceterm_axial}
\end{eqnarray}
Following the notation of \cite{hopper2010gravitational} we have, 
\begin{eqnarray}
    \label{ch3_gaxial}
    \bar{G}_{lm}^{a}(t) &=&\mathcal{G}^{r_{1}}_{l}p^{r}_{lm}  + \mathcal{G}^{r_{2}}_{l}\frac{dp^{r}_{lm}}{dt} + \mathcal{G}^{t}_{l}p^{t}_{lm},\\
     \bar{F}_{lm}^{a}(t) &=&\mathcal{F}^{r}_{l}p^{r}_{lm} +  \mathcal{F}^{t}_{l}q^{t}_{lm},
    \label{ch3_faxial}
\end{eqnarray}
where, 
\begin{eqnarray}
    \label{ch3_gaxial_r_1}
    \mathcal{G}_{l}^{r_{1}}(t) &=& \frac{\dot{r}_{p}}{\lambda}, \ \ \ \  \mathcal{G}_{l}^{r_{2}}(t) = \frac{r_{p}}{\lambda}, \ \ \ \ \mathcal{G}_{l}^{t}(t) = -\frac{f_{p}}{\lambda}, \\
    \mathcal{F}_{l}^{r}(t) &=& -\frac{r_{p}\dot{r}_{p}}{\lambda}, \ \ \ \  \mathcal{F}_{l}^{t}(t) = \frac{r_{p}f^{2}_{p}}{\lambda}. 
    \label{ch3_faxial_terms}
\end{eqnarray}
\subsection{ZM/Polar source term}\label{app4_bhpt_polar_source}
The polar/ZM source term describing the perturbations of $\Psi^{p}_{lm}(t,r)$ given in equation \eqref{ch4_CPM_ZM_wavequation} is, 
\begin{eqnarray}
       S_{lm}^{p}(t,r_{p}(t)) =  \bar{F}^{p}_{lm}(t)\delta'(r-r_{p}(t))
      + \bar{G}^{p}_{lm}(t)\delta(r-r_{p}(t)). \ \ \ \ 
      \label{cha4_sourceterm_polar}
\end{eqnarray}
Following the notation of \cite{hopper2010gravitational} we have, 
\begin{align}
    \label{ch3_gpolar}
    \bar{G}_{lm}^{p}(t) =\mathcal{G}^{rr}_{l}q^{rr}_{lm}  + \mathcal{G}^{tt}_{l}q^{tt}_{lm}  + \mathcal{G}^{r}_{l}q^{r}_{lm}  + \mathcal{G}^{\flat}_{l}q^{\flat}_{lm} + \mathcal{G}^{\sharp}_{l}q^{\sharp}_{lm}, \\
     \bar{F}_{lm}^{p}(t) =\mathcal{F}^{rr}_{l}q^{rr}_{lm} +  \mathcal{F}^{tt}_{l}q^{tt}_{lm}, 
    \label{ch3_fpolar}
\end{align}
where, 
\begin{align}
    \label{ch3_gpolar_rr}
    &\mathcal{G}_{l}^{rr}(t) = \frac{1}{(\lambda + 1)r_{p}\Lambda^{2}_{p}} \bigg[ 
    (\lambda + 1)(\lambda r_{p} + 6M)r_{p} + 3M^{2} \bigg], \\
    \label{ch3_gpolar_tt}
    &\mathcal{G}_{l}^{tt}(t) = \frac{f_{p}^{2}}{(\lambda + 1)r_{p}\Lambda^{2}_{p}} \bigg[\lambda (\lambda + 1)r_{p}^{2} + 6\lambda M r_{p} + 15M^{2}\bigg],\\
    &\mathcal{G}_{l}^{r}(t) = \frac{2 f_{p}}{\Lambda_{p}}, \ \ \ \ \mathcal{G}_{l}^{\flat}(t) = \frac{r_{p} f_{p}^{2} }{(\lambda + 1)\Lambda_{p}}, \ \ \ \mathcal{G}_{l}^{\sharp}(t)=-\frac{f_{p}}{r_{p}},   \\
    &\mathcal{F}_{l}^{rr}(t) =-\frac{r^{2}_{p} f_{p}}{(\lambda +1)\Lambda_{p}}, \ \ \ \ \mathcal{F}_{l}^{tt}(t) = -\frac{r^{2}_{p} f_{p}^{3}}{(\lambda +1)\Lambda_{p}}. 
    \label{ch3_fpolar}
\end{align}

\subsection{Complementary results with \texttt{DiscoTEX} relatives: \texttt{DiscoIMP} \& \texttt{DiscoREX}}\label{app_discotex_family}

\begin{figure*}
\centering
\includegraphics[width=85mm]{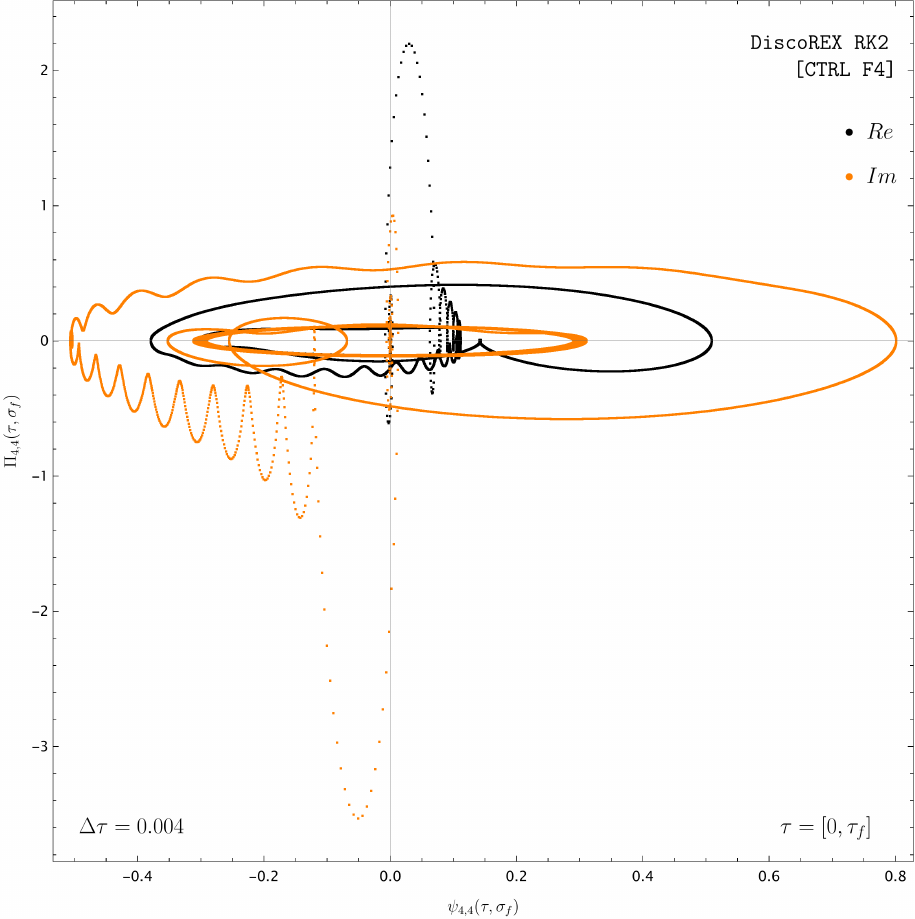}
\quad
\includegraphics[width=85mm]{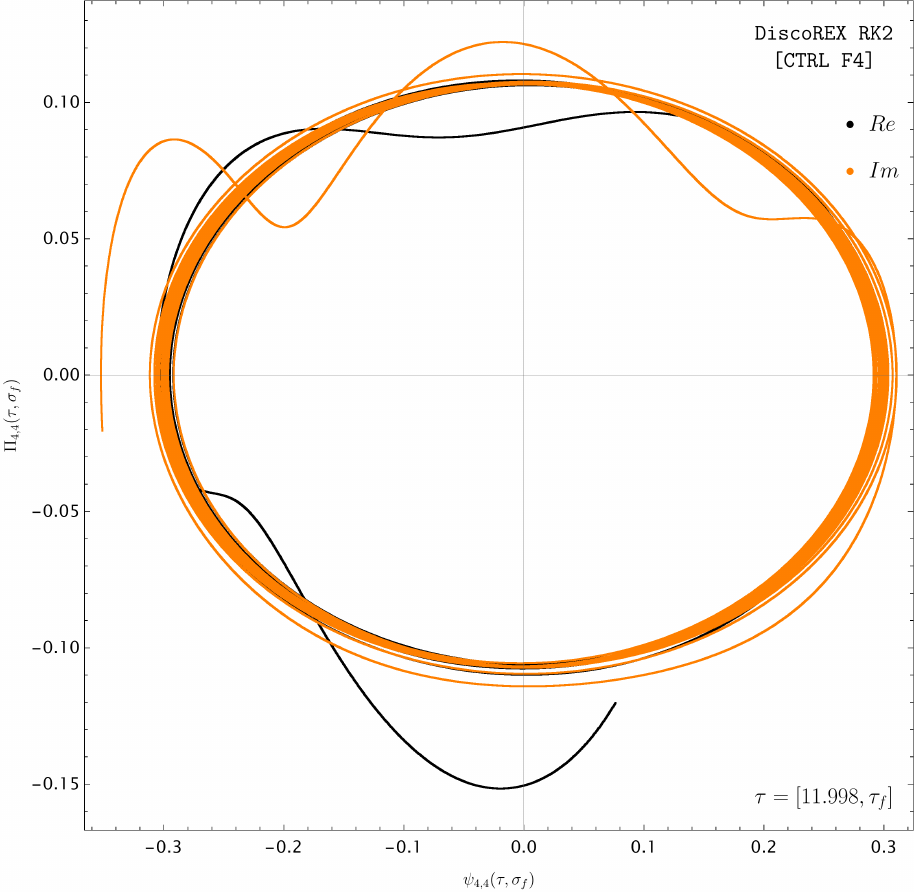}
\quad
\includegraphics[width=85mm]{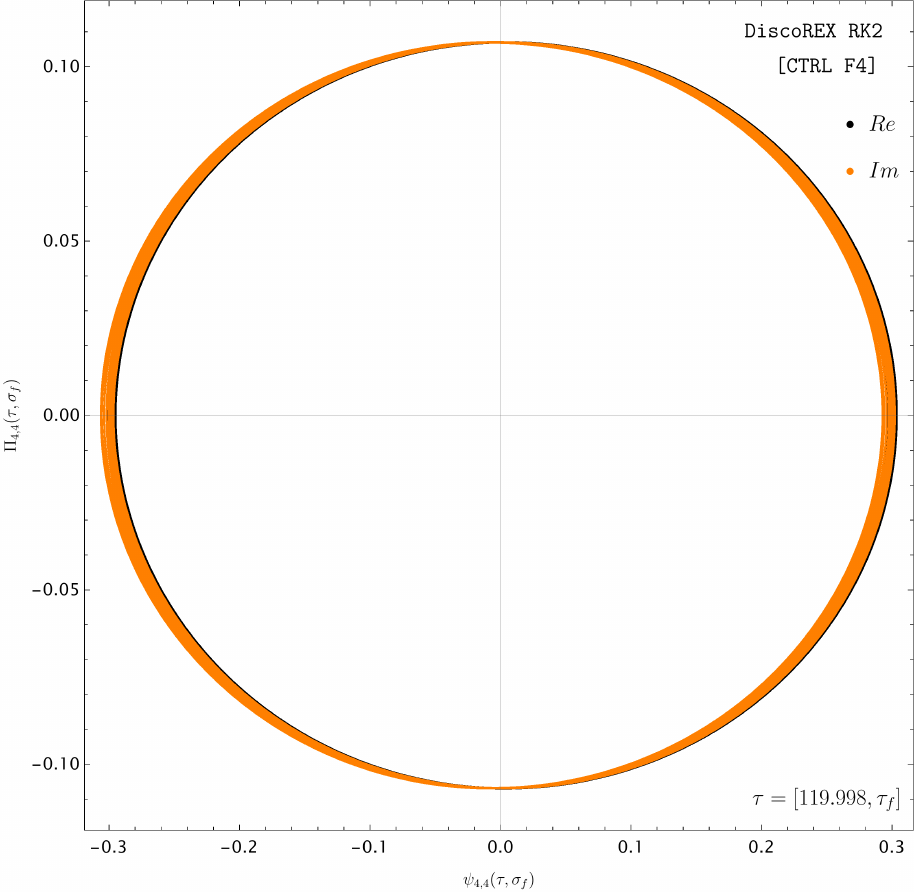}
\quad
\includegraphics[width=85mm]{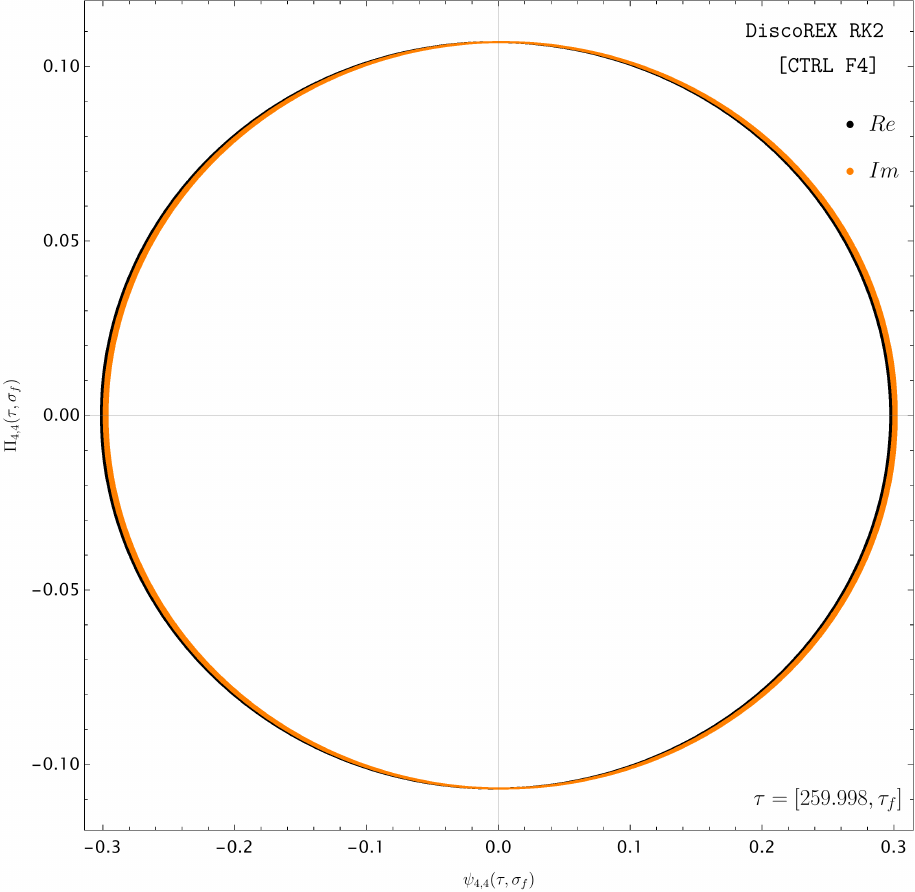}
\caption{Here we show the phase portrait of our numerical evolution for the field $\Psi_{2,3}(\tau,\sigma)$ at $\mathscr{I}^{+}$ obtained by implementing \texttt{DiscoREX RK2} as given by equation 
From left to right and then top to bottom, we include the evolution from an initial time of $\tau= \{0,11.98,119.98, 259.98\}$.\label{grav_pps_discoEX_RK2}}
\end{figure*}

\begin{figure*}
\centering
\includegraphics[width=85mm]{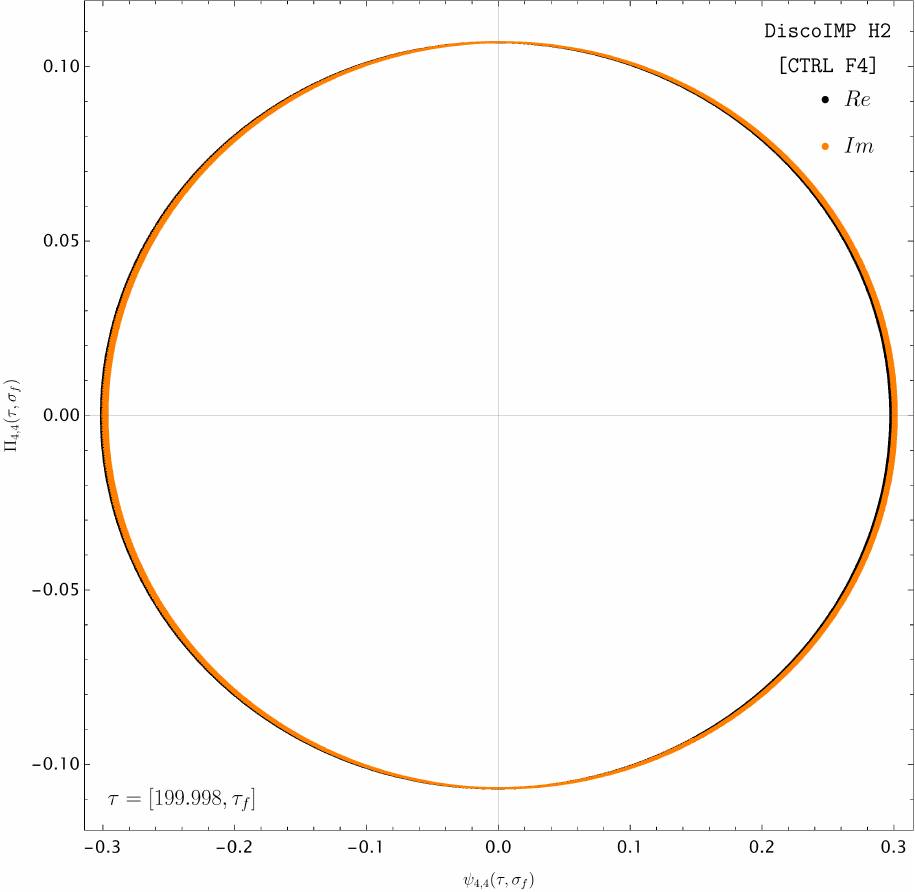}
\quad
\includegraphics[width=85mm]{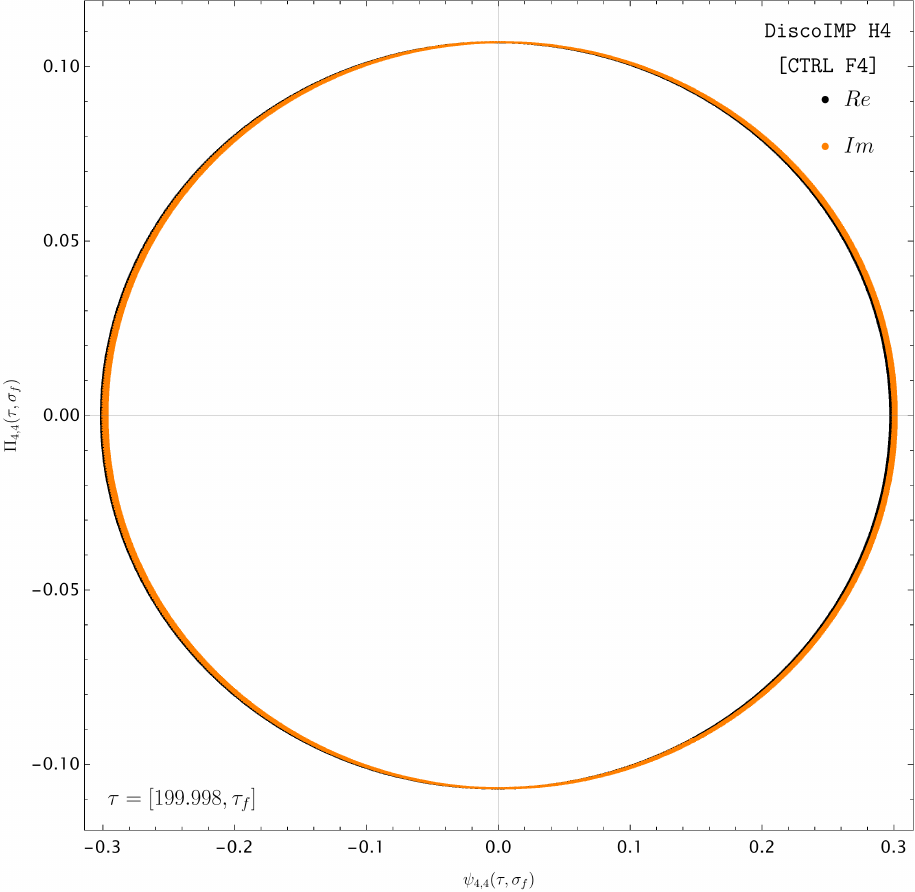}
\quad
\includegraphics[width=85mm]{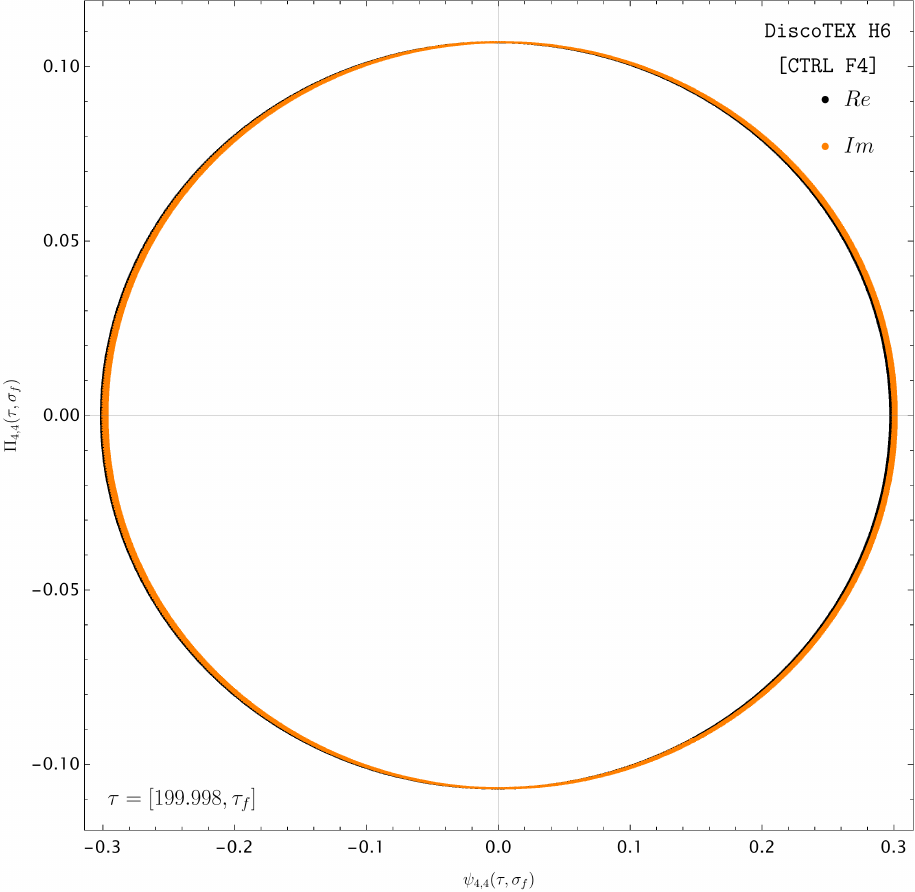}
\quad
\includegraphics[width=85mm]{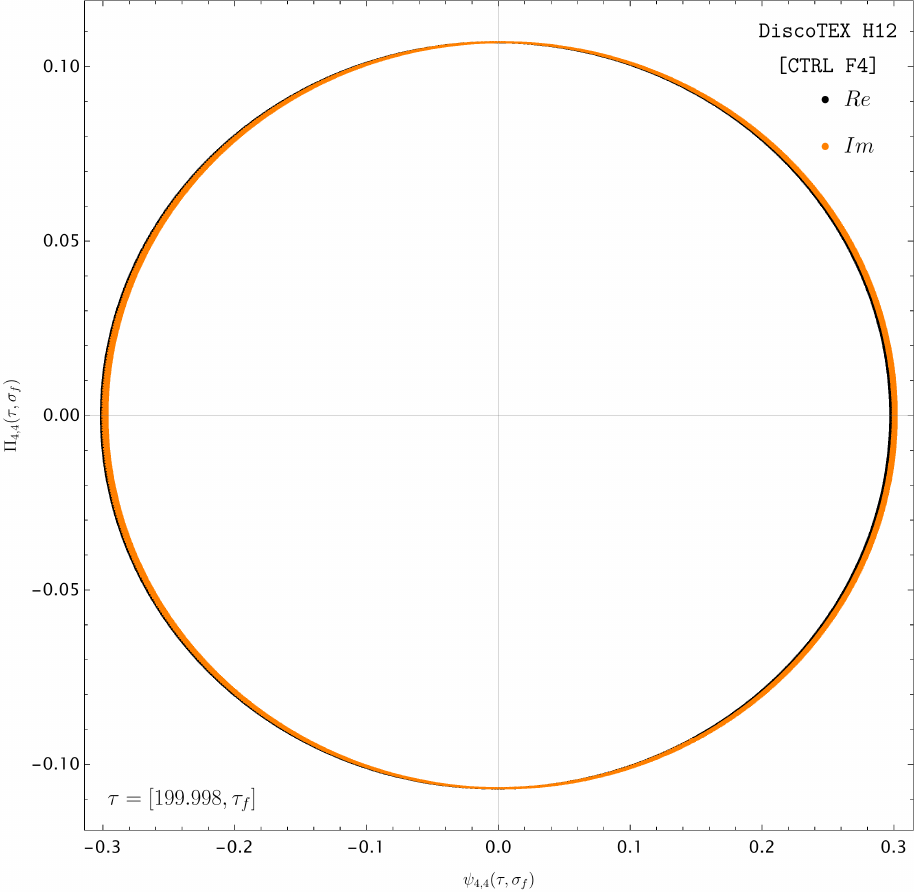}
\caption{Here we show the phase portrait of our numerical evolution for the field $\psi_{2,2}(\tau,\sigma)$ at $\mathscr{I}^{+}$ obtained by implementing \texttt{DiscoIMP H2-H4} (top) and from higher-order \texttt{DiscoTEX} algorithms (bottom). \label{grav_pps_parent_higher}}
\end{figure*}

To further complement Section \ref{sec_discotex_family} we include the plots assessing \texttt{[CTRL F\ref{controlfactor4}]} for \texttt{DiscoREX} in Figure \ref{grav_pps_discoEX_RK2}. Additionally, and finally, in Figure \ref{grav_pps_parent_higher}, we check whether numerically symplectic structure is preserved for higher-order numerical weak-form solutions of \texttt{DiscoTEX} and the parent algorithm \texttt{DiscoIMP}



\end{document}